# Pressure effects on the structural and electronic properties of $ABX_4$ scintillating crystals

Daniel Errandonea[1,*] and Francisco Javier Manjón[2]

[1] MALTA Consolider Team, Departamento de Física Aplicada – ICMUV, Universitat de València, c/Dr. Moliner 50, 46100 Burjassot (Valencia), Spain

[2] MALTA Consolider Team, Departamento de Física Aplicada – IDF, Universitat Politècnica de València, Cno. de Vera s/n, 46022 València, Spain

**Abstract**

Studies at high pressures and temperatures are helpful for understanding the physical properties of the solid state, including such classes of materials as, metals, semiconductors, superconductors, or minerals. In particular, the phase behaviour of $ABX_4$ scintillating materials is a challenging problem with many implications for other fields including technological applications and Earth and planetary sciences. A great progress has been done in the last years in the study of the pressure-effects on the structural and electronic properties of these compounds. In particular, the high-pressure structural sequence followed by these compounds seems now to be better

[*] Corresponding author, Email: daniel.errandonea@uv.es, Fax: (34) 96 3543146, Tel.: (34) 96 354 4475

understood thanks to recent experimental and theoretical studies. Here, we will review studies on the phase behaviour of different $ABX_4$ scintillating materials. In particular, we will focus on discussing the results obtained by different groups for the scheelite-structured orthotungstates, which have been extensively studied up to 50 GPa. We will also describe different experimental techniques for obtaining reliable data at simultaneously high pressure and high temperature. Drawbacks and advantages of the different techniques are discussed along with recent developments involving synchrotron x-ray diffraction, Raman scattering, and *ab initio* calculations. Differences and similarities of the phase behaviour of these materials will be discussed, on the light of the Fukunaga and Yamaoka´s and Bastide´s diagrams, aiming to improve the actual understanding of their high-pressure behaviour. Possible technological and geophysical implications of the reviewed results will be also commented.

**Contents**

**1. Introduction**

Scheelite-structured $ABO_4$ ternary oxides are important materials from both a theoretical and a technological point of view. In particular, $CaWO_4$, $SrWO_4$, $BaWO_4$, and $PbWO_4$ are promising materials for the next generation of cryogenic phonon-scintillation detectors [1]. This fact has motivated a renewed interest in the fundamental physical and chemical properties of the $AWO_4$ orthotungstates and related $ABX_4$ compounds (e.g. $YLiF_4$ and $BiVO_4$) as well as in their behaviour under compression. These compounds are in fact technologically important materials within a wider scope and have a long history of practical application since $CaWO_4$ (the mineral scheelite [2]) was first used by Thomas A. Edison in 1896 to detect x-rays [3]. In particular, scheelite-type $AWO_4$ compounds possess very attractive luminescence properties; e.g. they fluoresce bright bluish-white in ultraviolet radiation, a distinguishing feature utilized in prospecting and mining. Among other applications, these compounds have been used during the last years as solid-state scintillators [4, 5], laser-host materials [6], and in other optoelectronic devices like eye-safe Raman lasers [7 - 9]. They have several advantages over other scintillating materials due to its relatively large x-ray absorption coefficient and scintillation output, reason that have made them very popular for detecting x-rays and γ-rays in medical applications [10]. On top of that, recent studies on scheelite-type $ABO_4$ compounds have shown that these crystals have good prospects as heterogeneous catalysts [11], oxide ion conductors [12], and possible negative electrode materials to replace the graphite presently being used in the Li-ion batteries [13]. A significant amount of research work on the high-pressure behaviour of $ABX_4$ compounds has been performed in the last lustrum and this corpus forms a solid background for understanding the main physical properties of these materials [14 – 32]. Due to the importance of these high-pressure studies, several laboratories across the world are involved now in such studies.

Here we will review recent studies of the structural properties of scheelite-structured $ABX_4$ scintillating crystals under compression up to 50 GPa. We will comment on recent results regarding $AWO_4$ orthotungstates and $AMoO_4$ orthomolybdates obtained using different methods. These methods include angle-dispersive x-ray powder diffraction (ADXRD) [14 - 21] and x-ray-absorption near-edge structure (XANES) [17, 19, 22], using synchrotron radiation, *ab initio* total-energy calculations [17, 19, 23 - 25], Raman spectroscopy measurements [20, 26 - 30], and other experimental techniques, such as neutron diffraction [21] and shock-wave measurements [31]. The combination of the experimental measurements and *ab initio* calculations reveal the existence of complex high-pressure phase-diagrams in $ABX_4$ materials and in some cases the occurrence of a pressure-induced amorphization. A review of the above mentioned studies is timely, because the number of observed structural transformations driven by pressure in $ABX_4$ scintillating crystals has increased considerably. Among other things, here we will present a systematic of the structural sequence undergone by $ABX_4$ materials upon compression. According with this systematic, the structural sequences and transition pressures in these materials are related with the packing efficiency of the anionic $BX_4$ units around the A cations. In addition, a relationship between the charge density in the $AX_8$ bisdisphenoids of $ABX_4$ scheelite-related structures and their bulk modulus is discussed and used to predict the bulk modulus of other materials of technological interest, like ceramic $TiSiO_4$ and new scheelite oxynitrides [33]. Furthermore, a comparative analysis of the crystal chemistry of $ABX_4$ compounds under high-pressure will be done in order to present the whole body of structural studies available in the literature in a consistent fashion, and to suggest opportunities for future work. We will also review in detail, how the changes of the structural properties of scintillating scheelite-structured compounds affect their electronic properties, commenting recent optical-absorption and luminescence measurements and discussing the results in terms of the electronic structure. The conclusions extracted from the different studies here reviewed have important implications for the different technological applications described above. They may have also important geophysical and geochemical implications since $ABO_4$ ternary oxides are common accessory minerals in various kinds of rocks in the Earth's upper mantle, and have been also found in meteorite impact debris.

**2. Historical background**

The observation of pressure-driven phase transitions in $ABX_4$ compounds dates back to the early 1960s. Dachille and Glasser showed that $BAsO_4$ and $BPO_4$ crystallizing in the high-cristobalite structure at ambient conditions transform to the low-quartz structure at 5 GPa and 500ºC [34]. Young *et al.* found that $AlAsO_4$, crystallizing in the low-quartz structure at ambient conditions, transforms under pressure to the rutile structure at 9 GPa and 900ºC [35]. Seifert discovered that $InSbO_4$ with the rutile structure at ambient conditions transformed to the $\alpha$-$PbO_2$ structure around 11 GPa and 800ºC [36]. Snyman and Pistorius observed that many seleniates (e.g. $MgSeO_4$ and $ZnSeO_4$) isomorphous to the zincosite structure ($ZnSO_4$) transformed to the $CrVO_4$–type structure under compression [37]. Stubican reported that several vanadates and arsenates crystallizing in the zircon or in the monazite structures at ambient conditions transformed to the scheelite structure between 2 and 8 GPa [38]. Young and Schwartz found that many tungstates and molybdates with the wolframite structure transform to an unknown high-pressure form around 6 GPa and 900ºC [39]. In addition, in the 1970s, Muller *et al.* observed that CdCrO$_4$ with the monoclinic *C2/m* structure at ambient conditions transformed to the scheelite structure at 4 GPa and 200ºC [40]. Finally, Tamura found that FeTaO$_4$ with the rutile structure at ambient conditions transformed to the wolframite structure at 2.5 GPa and 1200ºC [41].

Regarding the scheelite-structured molybdates and tungstates, the first high-pressure studies were carried out also during the 1970s. In these studies, Nicol and Durana [42, 43] reported the vibrational Raman spectra for CaMoO$_4$, CaWO$_4$, and SrWO$_4$ at pressures as high as 4 GPa for the tetragonal scheelite phase [space group (S.G.): *I4$_1$/a*, No. 88, Z = 4] [44] and a previously unknown high-pressure phase. Based upon the changes found in the Raman spectra, they postulated that the high-pressure phases had a monoclinic wolframite-type structure (S.G.: *P2/c*, No. 14, Z = 2) [45]. In contemporary studies, the existence of high-pressure (HP) high-temperature (HT) phases in BaWO$_4$ and PbWO$_4$ was discovered [46, 47]. These phases were quenched from HP-HT conditions to ambient conditions and the metastable products were characterized by x-ray powder diffraction. The crystallographic structure of these HP-HT phases was solved by Kawada *et al.* [48] and Richter *et al.* [49], corresponding to a monoclinic structure (S.G.: *P2$_1$/n*, No. 14, Z = 8) named BaWO$_4$-II. Inspired by the high-pressure structural studies previously carried out by others [34 – 49], in the late 1970s, Fukunaga and Yamaoka tried to give a systematic explanation of the pressure-induced phase transitions in ABO$_4$ compounds [50]. In their study, they classified all ABO$_4$ compounds in a two-dimensional phase diagram on the basis of the mean cation to anion ionic radii ratio and the cation A to cation B ionic radii ratio.

The studies described above showed that scheelite was a high-pressure form of many other $ABO_4$ compounds with zircon, monazite or $CrVO_4$–type structures and that wolframite was a high-pressure form of rutile. These results motivated during the 1980s and 1990s an intensive effort to discover the high-pressure phases of the wolframite- and scheelite-structured compounds and to check whether or not wolframite could transform into scheelite under pressure or conversely. Jayaraman *et al.* performed high-pressure Raman studies of scheelite-type $ABO_4$ compounds like alkaline-earth tungstates ($CaWO_4$, $SrWO_4$, $BaWO_4$) [51], $PbWO_4$ and $PbMoO_4$ [52] up to 9 GPa. They found that the scheelite-structure remains stable in $CaWO_4$ and $SrWO_4$ up to the highest pressure reached in their experiments and reported the occurrence of pressure-driven phase transitions in $BaWO_4$, $PbWO_4$, and $PbMoO_4$ around 6.5 GPa, 4.5 GPa, and 9 GPa, respectively. They also suggested that these high-pressure phases could have an octahedral W-O coordination as in the $HgWO_4$-type structure (S.G.: *C2/c*, No. 15, Z = 4) [53]. They also performed high-pressure Raman studies of scheelite-type $CdMoO_4$ and wolframite-type $CdWO_4$ compounds up to 40 GPa finding two phase transitions near 12 GPa and 25 GPa in $CdMoO_4$ [54] and near 10 GPa and 20 GPa in $CdWO_4$ [55]. Other high-pressure Raman studies were conducted in the 1980s and 1990s on pseudoscheelite-structured compounds like alkaline perrhenates and periodates; e.g. $KReO_4$, $RbReO_4$, $CsReO_4$, $TlReO_4$, $KIO_4$, $RbIO_4$, and $CsIO_4$ [56 - 59].

In the 1980s, Hazen *et al.* performed by the first time x-ray diffraction studies under compression in scheelite-type tungstates and molybdates ($CaWO_4$, $PbWO_4$, $CaMoO_4$, $PbMoO_4$, and $CdMoO_4$) using a diamond-anvil cell (DAC) [60]. A DAC is a device that consists basically of two opposing cone-shaped diamond-anvils squeezed together. The resultant high pressures are produced when force is applied to the small areas of the opposing diamond culets, being possible to reach with the DAC pressures of up to 500 GPa [61]. Hazen *et al.* did not find in their single-crystal experiments any pressure-induced phase transition up to 6 GPa for the five studied compounds. However, they showed that the $BO_4$ ($WO_4$ and $MoO_4$) tetrahedra in scheelite oxides were rather incompressible and that the compressibility of scheelite-structured compounds was mainly due to the compression of the $AO_8$ polyhedra (e.g. $CaO_8$). These studies were followed by Macavei and Schulz who performed x-ray diffraction studies under compression in some wolframite-type tungstates ($MgWO_4$, $MnWO_4$, and $CdWO_4$) using a DAC up to 8 GPa [62]. They also showed that the $WO_6$ octahedra in wolframite oxides were rather incompressible and that the compressibility of wolframite-structured compounds was mainly due to the compression of the $AO_6$ octahedra. Finally, in the late 1980s, Bastide tried to give a systematic explanation of the pressure-induced phase transitions in $ABX_4$ compounds [63]. He classified all $ABX_4$ compounds in a two-dimensional phase diagram on the basis of their cation A and cation B to anion X ionic radii ratios and established a rule for the high-pressure phase transitions in $ABX_4$ compounds.

In the last decade, the development of the diamond-anvil cell technology [61] and of dedicated high-pressure facilities at synchrotron light sources [64] has brought a rebirth of the research on the high-pressure behaviour of $ABX_4$ compounds, in particular of the scheelite-type ones. Initially, Raman studies under pressure found a phase transition above 10 GPa in $CaWO_4$ and $SrWO_4$ [26, 65] and around 12 GPa in $SrMoO_4$ and $CdMoO_4$ [54, 66]. Besides, a second transition in $CdMoO_4$ was found at 25 GPa [54]. In parallel with these studies, it was done the first attempt to solve the crystalline structure of the high-pressure phases of scheelite-structured compounds. Jayaraman *et al.* performed energy-dispersive powder x-ray diffraction (EDXRD) on $CdMoO_4$ and concluded that the first high-pressure phase, observed above 12 GPa, had a wolframite-type structure and that the second high-pressure phase, observed beyond 25 GPa, had a $BaWO_4$-II-type structure [67]. Similar studies performed in $CaWO_4$ reached a similar conclusion for the first high-pressure phase [68]. Additional EDXRD studies on $CaWO_4$ extending the pressure range up to 65 GPa, observed that $CaWO_4$ becomes amorphous at pressures exceeding 40 GPa. These studies also found that a novel high-pressure phase of $CaWO_4$ can be obtained at 45 GPa after heating to 477 K and quenching to room temperature (RT) an amorphous sample [69].

In spite of all these efforts made to understand the high-pressure behaviour of $ABX_4$ compounds during three decades, it has been only in the last lustrum that the most significant contributions determining the high-pressure phases of scheelite and wolframite-structured compounds have been done. In particular, the combination of recent *ab initio* calculations [17, 19, 23 – 25], Raman spectroscopy studies [20, 26 – 30], and x-ray diffraction and absorption measurements [14 – 22] have allowed to clearly establish the sequence of the pressure-driven structural phase transitions undergone by scheelite-type compounds. These techniques are now also being used to explore the unknown high-pressure phases of wolframite and fergusonite structures [70]. A detail review of these studies and the main conclusions extracted from them will be done in the following sections.

**3. Crystal structure**

Metal orthomolybdates ($AMoO_4$) and orthotungstates ($AWO_4$) of relatively large bivalent cations (ionic radius > 0.99 Å; A = Ca, Ba, Sr, Pb or Eu, and Cd only for the molybdates) and many other $ABX_4$ compounds crystallize in the so-called scheelite structure [71]. Scheelite is the name of the mineral $CaWO_4$, which is used to describe the family of all the minerals isostructural to $CaWO_4$, like powellite (CaMoO$_4$), stolzite (PbWO$_4$), and wulfenite (PbMoO$_4$). Scheelite crystals have a tetragonal symmetry, appearing as dipyramidal pseudo-octahedra. They posses distinct cleavage planes and its fracture may be subconchoidal to uneven. Twinning is also commonly observed in natural crystals. In addition to the mineral sources, single-crystal scheelites have been synthesized using different methods since the late 1940s to satisfy the needs of different applications [72]. Nowadays, high-quality large crystals can be obtained via the Czochralski process [73].

The crystalline structure of scheelite was solved by Sillen and Nylander [74] who in the 1940s reviewed earlier works and by means of x-ray diffraction determined the oxygen atomic positions. More precise refinements of this structure were obtained during the 1960s by means of x-ray diffraction [44] and neutron diffraction [75] experiments. The scheelite crystal structure is characterized by the tetragonal space group *I4$_1$/a* listed as No. 88 in the International Tables of Crystallography. In this structure, the primitive unit cell has two ABX$_4$ units. The A and B sites have S$_4$ point symmetry, and the crystal has an inversion center. The X sites have only one trivial point symmetry. The scheelite crystals have three crystal parameters (x, y, z) which describe the location of the X (e.g. oxygen) sites at the 16f Wyckoff positions. There are a number of equivalent ways to describe the scheelite structure which have appeared in the literature [74]. As an example, **Table I** gives the Wyckoff positions of the atoms and lattice parameters of CaWO$_4$ in the conventional setting [60].

The scheelite crystal structure can be described as highly ionic with A$^{+2}$ cations and tetrahedral BX$_4^{-2}$ anions. Each B site is surrounded by four equivalent X sites in tetrahedral symmetry and the tetrahedral BX$_4^{-2}$ anions have short B-X bond lengths of approximately 1.78 Å which are quite rigid even under compression [11, 12, 60]. Each $A^{+2}$ cation shares corners with eight adjacent $BX_4$ tetrahedra and the $A^{+2}$ cations are surrounded by eight X sites at bond lengths of around 2.45 Å in approximately octahedral symmetry forming bisdisphenoids. **Figure 1** shows a perspective drawing of the crystal structure $CaWO_4$ at ambient conditions ($a = b = 5.2429$ Å, $c = 11.3737$ Å) in the conventional unit cell, indicating the *a*, *b*, and *c* axes and the nearly tetrahedral bonds between W and O and the dodecahedral bonds between Ca and O. In the scheelite structure, the $BX_4$ tetrahedra have the X atoms located at two different z values, with two X atoms lying at the same z and the other two at another z. In the tetragonal zircon structure the same is true and the X-X direction of the two X atoms lying in the same z in the $BX_4$ tetrahedra are aligned with the *a* and *b* axes. However, in the scheelite structure the X-X direction of the two X atoms lying in the same z is rotated with respect to the axes. Each scheelite structure has a characteristic setting angle ϕ (between 0 and 45º) which determines the minimum angle between the X-X direction and the *a*-axis.

On the other hand, metal tungstates ($AWO_4$) and molybdates ($AMoO_4$) of relatively small bivalent cations (ionic radius < 0.99 Å; A = Mg, Fe, Mn, Ni, Co, Zn), $CdWO_4$, and $MnReO_4$ crystallize in the so-called wolframite structure [71]. Some of these molybdates show an interesting polymorphism crystallizing also in different structures [76]. The wolframite structure belongs to the monoclinic space group *P2/c* and has two formula units per crystallographic cell [45]. It consists of infinite zig-zag chains, running parallel to [001] composed entirely of $AO_6$ octahedra and $WO_6$ octahedra which shares two corners with its neighbours. In particular, the $WO_6$ ($MoO_6$) octahedra are highly distorted since two of the W-O distances are much larger than the other four distances. As in the case of the scheelites, twining is also common in wolframites on {100}, usually as simple contact twins.

On top of that, the metal molybdates ($AMoO_4$; A = Mn, Mg, Fe) of some small bivalent cations also crystallize in the $CrVO_4$-type structure, whereas $CuWO_4$ and $CuMoO_4$ crystallize in the triclinic $P\bar{1}$ structure ($CuWO_4$-type) at ambient conditions. $ZnMoO_4$ could also crystallize in this structure. Although the $CuWO_4$-type structure is triclinic (S.G.: $P\bar{1}$, No. 2, Z = 2), it is also topologically similar to the wolframite structure [77]. Whereas in wolframite the $A^{2+}$ ions are present in axially compressed $AO_6$ octahedra, in the $CuWO_4$-type structure the $A^{2+}$ ions are responsible for a significant Jahn-Teller distortion and impose a large axial elongation upon the $AO_6$ octahedra. Regarding the orthorhombic $CrVO_4$-type structure (S.G.: *Cmcm*, No. 63, Z = 4) [78], it consists of $AO_6$ octahedra sharing edges to form chains that propagate in the *c* direction, with the chains linked to one another by $BO_4$ tetrahedral sharing corners with the octahedral of the chains. Finally, there are some molybdates like $CoMoO_4$, $MnMoO_4$, and $NiMoO_4$ that can also crystallize in a monoclinic structure with space group *C2/m* (S.G. No. 12, Z = 8) [76, 79], a structure that has been also observed under high-pressure and high-temperature conditions in $CaWO_4$ [69]. This structure is a close-packed arrangement of oxygen octahedra about both types of cations and can be thought of as a distorted modification of the CoO structure [76].

**4. High-pressure structural studies on $AWO_4$ orthotungstates: pressure effects on the scheelite structure**

**4.1. X-ray diffraction experiments**

After a first ADXRD study by Grzechnik *et al.* on the high-pressure structural behaviour of $YLiF_4$, where a scheelite-to-fergusonite phase transition was reported [80], similar studies have been carried out in $CaWO_4$, $SrWO_4$, $BaWO_4$, and $PbWO_4$ [14, 15, 17, 18, 19, 21]. They were performed up to pressures of approximately 20

GPa, but in the case of BaWO$_4$ the pressure range was extended up to 56 GPa [19]. In these experiments fine powder samples of these compounds were loaded into a DAC. The ADXRD patterns were measured using a micron-size beam of monochromatic x-ray radiation produced either by a synchrotron source or rotating anode x-ray generators at wavelengths ranging from 0.3679 Å to 0.7093 Å. More details about the experiments can be found in the above cited references.

**Figure 2** shows *in situ* ADXRD data measured at different pressures for CaWO$_4$. The x-ray patterns could be indexed with the scheelite structure from ambient pressure (0.0001 GPa) up to 9.7 GPa. A splitting and a broadening of the diffraction peaks are observed at 11.3 GPa together with the appearance of new reflections, in particular the weak peak located around $2\theta = 3.9°$ which is depicted by arrows in **Figure 2**. These facts are indicative of a structural phase transition which occurs at 10.5(8) GPa according with Ref. [17] and around 10 GPa in Ref. [14]. A similar behaviour has been found in SrWO$_4$, BaWO$_4$, and PbWO$_4$, being the phase transformation detected around 10 GPa [17, 18], 7 GPa [15, 19], and 5 – 9 GPa [19, 21], respectively. From the ADXRD data the evolution with pressure of the volume, lattice parameters, and axial ratios can be extracted. **Figure 3** shows the results obtained for CaWO$_4$ and SrWO$_4$. The data reported by different authors agree well within the uncertainty of the experiments [14, 17, 18, 44, 68, 81]. The pressure-volume (P-V) curves for the four orthotungstates have been analyzed in the standard way using a third-order Birch-Murnaghan equation of state (EOS) [82]. A selection of the obtained zero-pressure volume (V$_0$), bulk modulus (B$_0$), and its pressure derivative (B$_0$') are shown in **Table II**. The parameters of the EOS obtained from EDXRD data for EuWO$_4$ are also shown for comparison [32]. There is an excellent agreement among the values published for these parameters by different authors.

A comparison of the volume of all the scheelite-structured $AWO_4$ compounds shows that there is a direct relation between the ionic radii of the $A^{2+}$ cation and the equilibrium volume. A similar fact is also observed in the scheelite-type orthomolybdates [71]. In addition to that, it has been found that the compressibility of the orthotungstates increases following the sequence $CaWO_4 > SrWO_4 > EuWO_4 > PbWO_4 > BaWO_4$. This fact is a direct consequence of the different compressibility of the $c$-axis in the different compounds; see **Figure 3** for comparing $CaWO_4$ and $SrWO_4$. In **Figure 3**, it can be also seen that the linear compressibility of the $c$-axis is larger than that of the $a$-axis. As a consequence of this fact the anisotropy of the crystals decrease under compression, being this decrease also related with the size of the $A^{2+}$ cation. It is important to mention here that the presence of large uniaxial strains in the experiments seems to affect both the axial and bulk compressibility of the compounds of interest for this review. This can be seen in **Figure 3**, where the results obtained under non-hydrostatic conditions (diamonds) [68] tends to slightly underestimate the compressibility of $CaWO_4$.

It is very interesting to pay attention also to the evolution of the cation-anion distances in the scheelite-type $AWO_4$ compounds. It has been shown that the $AWO_4$ scheelites can be understood as made of hard anion-like $WO_4$ tetrahedra surrounded by charge compensating cations [17, 19, 60]. **Figure 4** shows the pressure evolution of the Pb-O and W-O distances in $PbWO_4$, which is in fully agreement with this picture. There, it can be seen that when pressure is applied the $WO_4$ units remain essentially undistorted. Therefore, the reduction of the unit-cell size should be mainly associated to the compression of the A (Pb) cation polyhedral environment. On the other hand, along the $a$-axis the $WO_4$ units are directly aligned, whereas along the $c$-axis there is an A cation between two $WO_4$ tetrahedra (see **Figure 1**). Thus, the different arrangement of hard $WO_4$ tetrahedra along the *c*- and *a*-axis accounts for the different compressibility of the two unit-cell axis. The different pressure behaviour of the two A-O distances (**Figure 4**) is also associated with the different compressibility of the unit-cell parameters. Effectively, the longest A-O distance has the largest projection along the *c*-axis. It is important to point out here that the asymmetric behaviour of the *c*- and *a*-axis is also revealed in their different thermal expansion [83] as well as in the evolution of the *c/a* ratio along the cationic A series.

**4.2. X-ray absorption measurements**

X-ray absorption spectroscopy is an experimental method used to determine the bonding of solids by analyzing oscillations in x-ray absorption versus photon energy that are caused by interference [84]. X-ray absorption measurements provide information about the geometrical arrangement of the atoms surrounding the absorbing atom. The measured spectra do not depend on long range order, so the information that they provide is complementary to the one yielded by x-ray diffraction. As a sensitive tool to the local atomic structure, x-ray absorption measurements can be used to obtain information about pressure-driven structural changes. In the materials we are interested, high-pressure x-ray absorption measurements have been performed in $CaWO_4$, $SrWO_4$, $BaWO_4$, and $PbWO_4$ in order to study the pressure effects on the coordination environment of W. Extended x-ray absorption fine structure (EXAFS) measurements were performed on $SrWO_4$ by Kuzmin *et al.* up to 30 GPa [22] working at the W $L_3$-edge (10.207 keV) using synchrotron radiation and a DAC at LURE. X-ray-absorption near-edge structure (XANES) measurements were performed by Errandonea *et al.* on $CaWO_4$, $SrWO_4$, $BaWO_4$, and $PbWO_4$ up to nearly 24 GPa using also a DAC and synchrotron radiation (W $L_3$-edge) [17, 19]. The last experiments were carried out using the energy dispersive set-up of the ID24 beamline of the European Synchrotron Radiation Facility (ESRF). More details on these experiments can be found elsewhere [17, 19, 22].

The experimental XANES spectra obtained for PbWO$_4$ under compression are shown in **Figure 5**. These results are representative of those obtained in the isostructural orthotungstates. As is shown in the figure, the spectra of the scheelite structure show five significant resonances labeled A, B, C, D, E. There is a clear tendency observed in the AWO$_4$ series: the intensity of the E resonance slightly increases whereas the resonances B, C, and D become less pronounced as the A cation becomes heavier. In the case of PbWO$_4$, the evolution of the spectra is smooth up to 9 GPa, suggesting that the W-O coordination environment is slightly modified by pressure, in good agreement with the conclusion extracted from the ADXRD studies [17, 19]. In **Figure 5**, it can be also seen that at 10.9 GPa the B resonance fades out and new weak resonances J, K, and L appear. These changes suggest that the W-O coordination changes and that a pressure-induced phase transition takes place. Additional changes corresponding to a second phase transition are observed at 16.7 GPa. These phase transitions will be discussed in detail in the next section of the manuscript. The experiments performed in CaWO$_4$, SrWO$_4$, and BaWO$_4$ provide similar results than those carried out in PbWO$_4$. According to the EXAFS and XANES experiments the range of stability of the scheelite phase extended from ambient pressure up to 11.3 GPa in CaWO$_4$, 11.2 – 12.4 GPa in SrWO$_4$, 9.8 GPa in BaWO$_4$, and 10.9 GPa in PbWO$_4$. These results are qualitatively in good agreement with those obtained from the ADXRD studies [14, 15, 17, 18, 19, 21]. However, as the resonances of the XANES spectra become less pronounced as the A cation atomic number increases, the transition is more difficult to detect in BaWO$_4$ than in SrWO$_4$ and CaWO$_4$. In fact, the XANES measurements detected the transition in CaWO$_4$ at the same pressure than the ADXRD measurements, but in SrWO$_4$ XANES detected the transition at a pressure 1 GPa higher than ADXRD, and the pressure difference is about 2 GPa in BaWO$_4$.

**4.3. Raman spectroscopy**

Raman scattering spectroscopy is an experimental technique that provides information about the vibrational properties of materials. This technique allows measuring the optical phonons of a material at the centre of the Brillouin zone, which are characteristic of both the structure and chemical composition of the material. Therefore, Raman scattering provides both chemical and structural information. The chemical information comes from the fact that the frequency of the vibrations is related to the atomic mass and bonding force between the atoms. The structural information comes from the fact that Raman scattering is sensitive to long-range order interactions between the atoms and it is subjected to strict selection rules imposed by the structural symmetry of the compound under study. Raman scattering is a technique complementary to ADXRD and XANES measurements for analyzing structural phase transitions because it is a subtle local probe capable of distinguishing small traces of various local phases coexisting in a compound. Thus in some cases, Raman scattering is a more sensitive technique for detecting pressure-induced phase transitions than ADXRD and XANES measurements.

The first high-pressure Raman scattering studies of scheelite-type alkaline-earth tungstates (CaWO$_4$ and SrWO$_4$) and molybdates (CaMoO$_4$) were performed by Nicol *et al.* up to 4 GPa [42, 43]. These studies were followed by the Raman experiments of Jayaraman *et al.* in CaWO$_4$, SrWO$_4$, BaWO$_4$, PbWO$_4$ and PbMoO$_4$ [51, 52] up to 9 GPa. Nicol *et al.* reported a pressure-induced splitting of some of the $E_g$ modes of the scheelite structure, which were interpreted as indications of the occurrence of a pressure-driven structural transformation [42]. In addition, Jayaraman *et al.* found that $BaWO_4$, $PbWO_4$, and $PbMoO_4$ underwent a phase transition near 6.5 GPa, 4.5 GPa, and 9 GPa, respectively. They suggested that the high-pressure phases of the three compounds could have an octahedral W coordination, as in the $HgWO_4$–type structure, but were not the same phase in the three compounds. Raman studies performed in the 1990s on $CaMoO_4$ up to 25 GPa [85] and $SrMoO_4$ up to 37 GPa [66] found evidence of phase transitions at 8 GPa and 15 GPa in $CaMoO_4$ and at 13 GPa in $SrMoO_4$ and suggested again the monoclinic nature of the high-pressure phases with an increase of the Mo-O coordination from 4 to 6. More recent Raman studies in $CaWO_4$ and $SrWO_4$ up to 20 GPa found a phase transition beyond 10 GPa [26, 65], and concluded that the fourfold W coordination in the scheelite phase was retained in the high-pressure phase, which is consistent with a transition towards the monoclinic M-fergusonite structure (S.G.: I2/*a*, No. 15, Z = 4) [86] (hereafter called fergusonite) reported by ADXRD and XANES experiments [17]. Finally, Raman studies under pressure in scheelite $BaMoO_4$ were recently conducted up to 8 GPa [27] and up to 15 GPa [20], and in $PbWO_4$ up to 15 GPa [30]. In $BaMoO_4$ a phase transition to the fergusonite structure was found at 5.8 GPa followed by a second phase transition around 8 GPa to an unknown phase. In $PbWO_4$, it was found a phase transition at 5.6 GPa to an unknown structure that did not revert completely to the original scheelite structure.

All the above cited studies of high-pressure Raman scattering in scheelites tungstates and molybdates performed during more than two decades formed a vast corpus of knowledge but failed at explaining the systematic of pressure-induced phase transitions in scheelites. This situation has recently changed and the nature and characteristics of the pressure-induced phase transitions in scheelite tungstates and molybdates have been finally clarified and explained. Manjón *et al.* have recently reported high-pressure Raman scattering studies in $BaWO_4$ and $PbWO_4$ up to 16 GPa and 17 GPa, respectively [28, 29]. In these two works, the Raman scattering of single crystals was excited with the 488 nm line of an $Ar^+$ laser and the scattered signal from samples loaded inside a DAC was collected and analyzed with a T64000 triple spectrometer equipped with a confocal microscope and a liquid $N_2$-cooled CCD detector. In these works, the Raman measurements were interpreted with the help of *ab initio* total-energy calculations performed within the framework of the density functional theory (DFT) using the Vienna *ab initio* simulation package (VASP) and *ab initio* lattice dynamics calculations carried out using the direct force constant approach (or supercell method) [28, 29].

Scheelite-structured $ABX_4$ compounds display 13 Raman-active modes corresponding to the following decomposition at the $\Gamma$ point:

$$\Gamma = 3A_g + 5B_g + 5E_g$$

The 13 Raman-active modes in the scheelite structure can be explained as internal or external modes of the $BX_4$ units, according to the following representation:

$$\Gamma = \nu_1(A_g) + \nu_2(A_g) + \nu_2(B_g) + \nu_3(B_g) + \nu_3(E_g) + \nu_4(B_g) + \nu_4(E_g)$$
$$+ R(A_g) + R(E_g) + 2T(B_g) + 2T(E_g)$$

where R and T modes denote the rotational and translational external modes observed in $ABX_4$ scheelites, and the internal modes of the $BX_4$ molecules are: $\nu_1$ (symmetric stretching), $\nu_2$ (symmetric bending), $\nu_3$ (asymmetric stretching) and $\nu_4$ (asymmetric bending).

**Figure 6** shows the Raman spectra of scheelite $PbWO_4$ at almost ambient pressure. The scheelite structure in $BaWO_4$ and $PbWO_4$ becomes unstable above 6

GPa, but scheelite Raman-active modes were followed under pressure up to 8.2 GPa (9 GPa) in $BaWO_4$ ($PbWO_4$). Manjón *et al.* compared the Raman spectra of $BaWO_4$ [28] and $PbWO_4$ [29] with those previously reported for scheelite $CaWO_4$ [45, 51], $SrWO_4$ [51, 65] $BaWO_4$ [51, 52], and $PbWO_4$ [51, 52] and explained satisfactorily some inconsistencies of the previous results of Raman studies under pressure. **Table III** summarizes the experimental frequencies, pressure coefficients, Grüneisen parameters and symmetry assignments of the Raman-active modes in the scheelite phase of the four tungstates. It can be observed that the frequencies and pressure coefficients of all scheelite-type Raman-active modes are rather similar in the three alkaline-earth tungstates and somewhat different in $PbWO_4$. A phonon gap between 400 $cm^{-1}$ and 750 $cm^{-1}$ is present in the scheelite structure for the four tungstates between the stretching internal modes and the rest of the modes. Manjón *et al.* demonstrated that the frequencies of the internal stretching modes of the $WO_4$ tetrahedra in the scheelite phase of the four tungstates are in agreement with the valence of the W cation according to Hardcastle and Wachs's formula [87].

As regards the phase transition pressures in $AWO_4$ scheelites, Raman scattering measurements reported by Manjón *et al.* in $BaWO_4$ ($PbWO_4$) found that a first high-pressure phase transition was observed at 6.9 GPa (6.2 GPa) and a second phase transition was observed at 7.5 GPa (7.9 GPa) [28, 29]. These phase-transition pressures were apparently in disagreement with the phase-transition pressures measured by ADXRD and XANES. We will be discuss these apparent discrepancies in detail in section 5.

## 4.4. Other experimental studies: neutron diffraction, shock-wave experiments, and Brillouin spectroscopy

Static high pressure can be generated not only with the use of diamond-anvil cells, but also with the so-called large-volume presses (LVP) [88, 89], like the Paris–Edinburgh pressure cell [90]. In this device, the sample is placed between two hard anvils made from either tungsten-carbide or sintered industrial diamond and a hydraulic ram is used to apply the pressure. This system has the advantage of allowing the study of large size samples but it is limited in the maximum pressure range achieved, which is an order of magnitude smaller than the routine pressure range of a DAC experiment. Using the Paris–Edinburgh pressure cell, Grzechnik *et al.* performed high-pressure time-of-flight neutron powder diffraction studies of $PbWO_4$ and $BaWO_4$ at ISIS (Rutherford Appleton Laboratory, UK) [21]. From these studies, structural information on the scheelite structure was extracted up to 4.9 GPa in $PbWO_4$ and 5.4 GPa in $BaWO_4$. Their main result is the confirmation that both compounds have similar pressure dependences of interatomic distances and bond angles. As from the ADXRD measurements [17, 19], the neutron studies showed that the compressibilities of $PbWO_4$ and $BaWO_4$ are primarily due to shortening of the Pb–O or Ba–O bonds rather than to the changes of the W–O bonds in the $WO_4^{-2}$ tetrahedral units. On the other hand, it was also found that the intratetrahedral bond angles O–W–O are not significantly sensitive to pressure, while the interpolyhedral W–O–A angles tend to converge upon compression.

In addition to the static high-pressure generated with the use of a DAC or a LVP, dynamic high-pressure can be also applied to a sample exposing it to shock-waves [91, 92]. A shock-wave is a strong pressure wave, propagating through an elastic medium, which is produced by a phenomenon that creates violent changes of pressure in the sample, like the impact of a high-velocity projectile. The wave front of the shock-wave in the sample where compression takes place is a region of sharp changes in stress, density, and temperature. In shock-wave experiments phase transitions are detected by the identification of discontinuities in the longitudinal sound speed. Zaretsky and Mogilevsky [31] investigated the dynamic response of natural, [001]-oriented, and synthetic, [201]-oriented, $CaWO_4$ in planar impact experiments with shock up to 20 GPa. These authors found that in both cases the plastic deformation is governed by the dislocation glide in {112} planes with resolved shear stress of about 0.6 – 0.7 GPa. However, the inelastic response and the dynamic strength in tension of the natural and synthetic materials are different. Zaretsky *et al*. also observed that the waveform obtained from the [001]-oriented sample contains the signature of the second-order scheelite–fergusonite transformation discovered with other experimental techniques [14, 17]. However, they did not observe it in the [201]-oriented sample. The absence of the transformation signature in the waveforms obtained, after strong impact in the [201]-oriented crystals, is probably due to faster transformation kinetics under loading in this direction.

Brillouin scattering [93] and ultrasonic measurements [94] are techniques that provide quite valuable information to constrain the independent elastic constants ($C_{11}$, $C_{33}$, $C_{44}$, $C_{66}$, $C_{12}$, and $C_{13}$) of a tetragonal crystal like scheelite, and therefore they can be used to determine its bulk and shear modulus. The occurrence of phase transitions under compression can be also corroborated by Brillouin experiments [95] through the observation of soft acoustic modes. Such experiments have been performed in $CaWO_4$ and in other $ABX_4$ scheelite-structured compounds like $YLiF_4$, $BiVO_4$, and $CaMoO_4$ [95 - 97]. The measured elastic constants for $CaWO_4$ are given in **Table IV**. From these constants a bulk modulus of 77 GPa is obtained for scheelite. This magnitude is in very good agreement with the value reported in the ADXRD studies; see **Table II**. The estimated values for the isotropic Young's modulus and the shear modulus are 143 GPa and 230 GPa, respectively.

**4.5.** *Ab initio* **calculations**

In the last decade *ab initio* calculations have become a quite useful tool for the study of matter at extreme conditions. They allow researchers to complement and check experimental observations and to predict properties of materials [98]. Several theoretical studies have been performed in $ABX_4$ scheelite-type compounds within the last years [17, 19, 23, 24, 25, 99, 100, 101]. These studies have been a great help to better interpret the results of the experimental studies described above, and in particular to determine the crystal structure of the high-pressure phases. Regarding the scheelite-structure, *ab initio* calculations also support the idea that in the orthotungstates the compression of the crystals is mainly due to the decrease of the A-O distances, behaving the $WO_4^{-2}$ tetrahedra as rigid units. In general, the agreement between calculations and experiments is quite good. Only in the case of the calculations performed by Li *et al.* a quite unrealistic bulk modulus of 185 GPa is reported for the scheelite phase of $PbWO_4$ [23].

The most extensive and systematic theoretical study on the structural behaviour of $AWO_4$ scheelites under pressure has been performed by A. Muñoz *et al.* [17, 19, 23, 25]. A. Muñoz and his group studied the structural stability of $CaWO_4$, $SrWO_4$, $BaWO_4$, and $PbWO_4$ by means of total-energy calculations performed within the framework of the density functional theory (DFT) with the Vienna *ab initio* simulation package (VASP) [102]. In these calculations the exchange and correlation energy was evaluated within the generalized gradient approximation (GGA) [103]. As an example, **Figure 7** shows the energy-volume curves obtained for each of the structures considered in BaWO$_4$. The relative stability and coexistence pressures of the phases can be obtained from this curve. **Figure 7** shows the scheelite phase of these compounds as being stable at zero and low pressure. The values of the volume, bulk modulus and pressure-derivative obtained for the four studied compounds are summarized in **Table V**. These values are in good agreement with the reported experimental results (see **Table II**), with differences within the typical reported systematic errors in DFT-GGA calculations. A similar agreement has been found for the unit-cell parameters and atomic positions for the oxygen atoms. In addition to that, a decrease of the axial anisotropy of scheelite orthotungstates under compression is predicted by the *ab initio* calculation for the low-pressure phase, in good agreement with the x-ray diffraction and absorption experiments [17, 19].

For PbWO$_4$ the *ab initio* calculations of Muñoz *et al.* [19] found the raspite structure (S.G.: *P2$_1$/a*, No 14, Z = 4) [104] to be very close in energy and equilibrium volume to the scheelite structure, with raspite slightly higher in energy by about 20 meV per formula unit. This fact is in perfect agreement with raspite-PbWO$_4$ (or PbWO$_4$-II) being found in Nature as a metastable polymorph of PbWO$_4$ under normal conditions [104]. This result disagrees with a previous theoretical calculation that obtained the raspite structure lower in energy than the scheelite structure [23]. In this sense, we have to note that the raspite structure has not been reported experimentally in CaWO$_4$, SrWO$_4$, and BaWO$_4$ in good agreement with its significantly higher total energy with respect to the scheelite structure in these three compounds [17, 19]. A similar disagreement exists for the PbWO$_4$-III phase which in Ref. [23] is shown much lower in energy than the scheelite phase. Calculations performed by Muñoz *et al.* have shown that this phase is considerably higher in energy than the scheelite phase in the four orthotungstates and is a competing post-scheelite phase, as will be discussed in the next section. As can be seen in **Figure 4**, from the *ab initio* calculations it can be also concluded that the compressibility of the WO$_4$ tetrahedra is much smaller than that of the AO$_8$ bisdisphenoids.

Regarding the pressure stability range of the scheelite structure, the theoretical calculations agree very well with the experiments in the cases of CaWO$_4$ and SrWO$_4$, extending it up to about 10 GPa. In the cases of BaWO$_4$ and PbWO$_4$ the pressure-driven phase transition is predicted to occur around 5 GPa in good agreement with Raman measurements [28, 29], but in somewhat less good agreement with ADXRD and XANES results. This apparent contradiction of Raman results and *ab-initio* calculations with ADXRD and XANES measurements is discussed in detail in the next section.

## 5. Pressure-induced phase transitions in AWO$_4$ orthotungstates

### 5.1. Scheelite-fergusonite transition

As we described above, the ADXRD spectra of CaWO$_4$ exhibit several changes around 11.3 GPa (see **Figure 2**). These changes are completely reversible upon pressure release. The observed splitting of peaks and the appearance of new reflections suggest the occurrence of a phase transition. The measured ADXRD patterns of the high-pressure phase can be indexed on the basis of the fergusonite structure (S.G.: *I2/a*, No. 15, Z = 4) [86], as shown by Grzechnik *et al.* [14] and confirmed by Errandonea *et al.* [17]. The new Bragg peaks observed at low angles correspond to the (020) reflection of the fergusonite structure. **Figure 8** shows the spectrum of CaWO$_4$ at 11.3 GPa and the refined profile obtained assuming the fergusonite structure. In order to perform the Rietveld refinement the starting Ca, W, and O positions were derived from the atomic coordinates in the scheelite structure using the *I4$_1$/a* → *I2/a* group-subgroup relationship [14, 17]. The agreement of the refined profiles with the experimental diffraction patterns is quite good as can be seen in **Figure 8** and Refs. [14] and [17]. **Table VI** summarizes the lattice parameters and atomic positions of fergusonite $CaWO_4$. There it can be seen that the agreement existent between the crystalline structures reported by different authors is quite good. In the diffraction studies of $SrWO_4$ [17, 18], $BaWO_4$ [15, 19], $PbWO_4$ [19, 21], and $EuWO_4$ [32], the crystal structure of the high-pressure phases has been also assigned to the fergusonite-type structure. According with these studies the scheelite-to-fergusonite transition pressures for these four compounds are 10 GPa, 7 GPa, 9 GPa, and 8.5 GPa, respectively. In the cases of $BaWO_4$ and $PbWO_4$, evidences of an additional phase transition were found at 10.9 GPa and 15.6 GPa, respectively.

The fergusonite structure of $PbWO_4$ is represented in **Figure 9** at two different pressures. There it can be seen that the fergusonite structure is a distorted version of scheelite. Indeed, at pressures close to the transition pressure is very difficult to distinguish the fergusonite structure from the scheelite structure (see **Figures 1** and **9(a)**). Regarding the mechanism involved in the transition, it is now accepted that the scheelite-to-fergusonite transition is caused by small displacements of the A and W atoms from their high-symmetry positions and large changes in the O positions and the consequently polyhedra distortion. In particular, all the A and W atoms of alternate layers of the scheelite structure need to shift in opposite directions along the *c*-axis. This axis correspond to the *b*-axis of the fergusonite structure according with the crystallographic setting generally used to describe the structures. The shift of the A and W atoms is accompanied by a shear distortion perpendicular to the *c*-axis of alternate O planes. Because of these atomic displacements, immediately after the transition the volume of $WO_4$ tetrahedra is enlarged by less than 10% and the volume of the $AO_8$ bisdisphenoids is reduced by a similar amount. It is interesting to note that immediately after the transition the fergusonite structure contains isolated $WO_4$ tetrahedra interlinked by A ions which have primarily an eightfold O coordination, like happens in the scheelite structure.

Now we would like to comment on the evolution of the lattice parameters of the fergusonite phase as a function of pressure. In **Figure 3**, it can be seen that after the scheelite-fergusonite phase transition, the difference between the *b/a* and *b/c* axial ratios of the fergusonite phase in $CaWO_4$ and $SrWO_4$ increases upon compression. A similar behaviour is observed for the monoclinic β angle. These two facts imply an increase of the monoclinic distortion with pressure. In **Figure 3**, it can be also seen that a volume discontinuity is not apparent at the transition pressure, which is consistent with the fact that fergusonite is a distorted and compressed version of scheelite which only implies a lowering of the point-group symmetry from *4/m* to *2/m*. In general, third-order Birch-Murnaghan fits to both the scheelite and the fergusonite pressure-volume data give EOS parameters that differ by less than one standard deviation from those obtained for the scheelite data only. Hence, the EOS parameters reported above in **Table II** can be assumed as valid parameters to describe the compressibility of both the scheelite and fergusonite phases.

It is also interesting to discuss the changes found in the bond distances at the phase transition. **Figure 4** shows that in $PbWO_4$ the four degenerate W-O bond distances inside the $WO_4$ tetrahedra split into two different ones, becoming the $WO_4$ tetrahedra distorted with two short distances (1.8 Å) and two long distances (1.87 Å). In addition to that, the $PbO_8$ dodecahedra distort in such a way that after the transition there are four different Pb-O distances in the range from 2.4 Å to 2.7 Å. In spite of these facts, it is interesting to see that in the fergusonite phase, again, the W-O bonds are much less compressible than the A-O bonds. As a consequence of this fact, there are virtually no changes between the bulk compressibility of the studied $AWO_4$ compounds between the scheelite (low-pressure) and fergusonite (high-pressure) phases. Therefore, the volume versus pressure data of both phases can be represented with the same EOS. Basically, in the high-pressure phase the linear compressibility of the different axis changes due to the monoclinic distortion of the crystal, but the bulk compressibility does not change since it is still governed only by the compression of the $AO_8$ dodecahedra. Another interesting fact we would like to point out here is that at the phase transition two of the oxygen atoms second neighbouring W approach considerable to the W atoms. In addition, this bond distance rapidly decreases with pressure after the phase transition; see **Figure 4**. Therefore, the W-O coordination gradually changes from 4 to 4+2 within the pressure-range of stability of fergusonite (as illustrated in **Figure 9**) acting this phase as a bridge phase between the scheelite phase and a second high-pressure phase (post-fergusonite phase), which has a W-O coordination equal to six.

X-ray absorption measurements give a similar picture of the structural changes undergone by $AWO_4$ scheelites than x-ray diffraction. The fading of the B resonance and the appearance of three new resonances, J, K, and L at 10.9 GPa are indicative of the occurrence of an structural change in $PbWO_4$. The absence of the B resonance in the high-pressure phase seems to contradict ADXRD measurements, since this resonance is characteristic of a W-O coordination number equal to four, like in the scheelite and fergusonite structure. However, this apparent discrepancy is not real since we should remember that in the fergusonite structure the internal parameters of the fergusonite structure change between 7.9 GPa and 9.5 GPa in such a way that the W coordination gradually changes from four oxygen atoms to 4+2 (see **Figures 4** and **9**). Thus, in the fourfold coordinated version of the fergusonite structure (close to the transition pressure) the B resonance is present, whereas in the quasi-sixfold coordinated version the B resonance is absent. Therefore, the XANES measurements indicate that at 9 GPa a phase transition takes place in $PbWO_4$ towards a fergusonite structure, being upon further compression the W environment distorted in such a way that the W atoms are surrounded by six O atoms in a 4+2 configuration at 10.9 GPa. At 16.7 GPa there are new, subtle changes in the XANES spectra that suggest the occurrence of a second phase transition, in agreement with the conclusions of the ADXRD studies [19]. In the cases of $CaWO_4$ and $SrWO_4$, the changes observed in the XANES spectra at 11 GPa and 12.4 GPa are also consistent with the occurrence of a scheelite-to-fergusonite phase transition. However, in the fergusonite phase of these compounds the gradual change of W coordination is not observed up to 20 GPa [17]. In $BaWO_4$, the situation is more similar to $PbWO_4$, two changes are observed in the XANES spectra at 7.8 GPa and 9 GPa. It seems that in this compound there is also a gradual change of W coordination taking place. Nevertheless, the spectra collected at 9.8 GPa already was attributed to the second high-pressure phase and not to fergusonite. The fact that the range of stability of the fergusonite phase is only 2 GPa added to the fact that some of the resonances of the spectrum become weaker for heavy elements like Ba are the reasons that make difficult the detection of the fergusonite phase in the XANES measurements of $BaWO_4$. However, the reported evidence that structural changes take place at 7.8 GPa and 9.8 GPa is in good agreement with the two phase transitions found in ADXRD experiments [15, 19].

In order to discuss the conclusions extracted from *ab initio* calculations we will refer to **Figure 7**. It shows the calculated energy-volume curves of $BaWO_4$, which are similar to that of $SrWO_4$, $PbWO_4$ and $CaWO_4$; see Refs. [17], [19], and [25]. In order to establish the sequence of stable phases, several structures were theoretically analysed because they were previously observed in these or related compounds. The name of the different structures considered in the calculations are given in the figure. To complement these static calculations and to help in the identification of the Raman modes observed in the high-pressure phase, lattice dynamics calculations were also performed for the same compounds by Muñoz *et al.* using the direct small-displacements method [28, 29].

For $CaWO_4$, $SrWO_4$, $BaWO_4$, and $PbWO_4$ the calculations indicate that on increasing pressure, a ferfusonite-type distortion of the scheelite structure becomes increasingly more noticeable from the structural point of view, and its energy becomes lower than that of the ambient pressure phase. Thus calculations found a second order, slow and continuous, phase transition from scheelite to fergusonite, in good agreement with the ADXRD and XANES experiments, being also the transition pressures in agreement with those found in the experiments. However, the theory also suggests that in $SrWO_4$, $BaWO_4$, and $PbWO_4$ there is a third competitive structure, the monoclinic $BaWO_4$-II-type [49]. This structure has indeed a lower enthalpy than the fergusonite structure, and thus from the theoretical point of view the transition should be from the tetragonal scheelite structure to this more compact monoclinic structure. It should be stressed here that the previous experimental observations of the $BaWO_4$-II-type phases in $PbWO_4$ and $BaWO_4$ required the application of both high pressure and high temperature to the respective scheelite phases, whereas the structural calculations of Muñoz *et al.* have been performed at 0 K. The scheelite-to-$BaWO_4$-II-type transitions are strongly first order with large density changes and involve extensive rearrangement of the crystal structure, in contrast to the second-order scheelite-to-fergusonite transitions. Thus, the barrier-less transition to the fergusonite structure may happen at pressures at which the first-order transition to the $BaWO_4$-II-type structure is kinetically hindered. The presence of kinetic barriers may also explain the need for high temperature in the previous experimental observations of the $BaWO_4$-II-type phase in $PbWO_4$ and $BaWO_4$. In support of this picture, x-ray experiments in $BaWO_4$ and $PbWO_4$ [19] found indications of the $BaWO_4$-II-type phases as post-fergusonite stable phases at higher pressures than the calculated coexistence pressures (see Section 5.2).

The above discrepancies between ADXRD and XANES measurements and total-energy *ab initio* calculations in $BaWO_4$ and $PbWO_4$ regarding the nature of the high-pressure phases and their phase transition pressures was solved in recent Raman measurements assisted with lattice-dynamics calculations for these two compounds [28, 29]. In these two works, Manjón *et al.* showed that there is a scheelite-to-fergusonite phase transition at 7.5 GPa and 7.9 GPa in $BaWO_4$ and $PbWO_4$, respectively [28, 29]; i.e., at slightly smaller pressures than those found in ADXRD and XANES measurements and in very good agreement with the theoretical estimations. However, Raman measurements revealed a previous scheelite-to-$BaWO_4$-II transition near 5 GPa in good agreement with theoretical estimations and found that fergusonite and $BaWO_4$-II coexisted in a range of several GPa in both compounds. This picture contrast with Raman scattering measurements in $CaWO_4$ and $SrWO_4$ up to 25 GPa that only observed a scheelite-to-fergusonite phase transition at 10 GPa and 12 GPa [45, 65], respectively, in good agreement with ADXRD and XANES measurements.

As regards the Raman modes of the high-pressure fergusonite phase, fifteen, twelve, twelve, and sixteen new Raman peaks have been observed in $CaWO_4$, $SrWO_4$, $BaWO_4$, and $PbWO_4$, respectively. These new Raman modes are compatible with the presence of the monoclinic fergusonite structure, which should display 18 Raman-active modes with the following mechanical representation:

$$\Gamma = 8A_g + 10B_g$$

Five additional Raman-active peaks appear in the fergusonite structure as compared to the scheelite structure due to the reduction of the tetragonal $C_{4h}$ symmetry of the scheelite structure to the monoclinic $C_{2h}$ symmetry of the fergusonite structure. This reduction in symmetry leads to a transformation of every $A_g$ and every $B_g$ scheelite mode into $A_g$ modes of the monoclinic symmetry, and of every doubly degenerate $E_g$ scheelite mode into two $B_g$ modes of the monoclinic symmetry. **Figure 6** shows the Raman spectra of $PbWO_4$ at 9.0 GPa where the fergusonite Raman peaks have been marked with arrows in good agreement with the theoretical frequencies obtained from *ab initio* lattice dynamics calculations (marks below the spectra). **Figure 10** shows the pressure dependence of the experimental Raman modes in the fergusonite phase of $PbWO_4$.

Manjón *et al.* compared the Raman modes of the fergusonite phase in $BaWO_4$ and $PbWO_4$ with those observed in the high pressure phase of $CaWO_4$ [45] and $SrWO_4$ [65] and with those observed in fergusonite-like $HgWO_4$ [105]. **Table VII** summarizes the experimental frequencies, pressure coefficients, and symmetry assignments of the Raman-active modes in the fergusonite phase for the four scheelite tungstates and fergusonite-like $HgWO_4$. It can be observed that the Raman modes of the fergusonite phase in $BaWO_4$ resemble very much those observed in $CaWO_4$ and $SrWO_4$ while $PbWO_4$ bear more resemblance with the Raman modes measured in fergusonite-like $HgWO_4$. In this sense, theoretical *ab initio* lattice dynamics calculations have been of great help to find the fergusonite Raman-active modes and assign their correct symmetries. The reason for the similar Raman spectra in the ferguson ite $CaWO_4$, $SrWO_4$, and $BaWO_4$ is that the coordination for the A cation is 8 and for the W cation is 4 in the three compounds. Furthermore, the ferguson ite Raman-active modes in the alkaline-earth tungstates have a similar frequency distribution than in the scheelite phase exhibiting a phonon gap between 500 cm$^{-1}$ and 800 cm$^{-1}$ (see **Table VII**). It was shown that the Raman-active modes in the ferguson ite phase can be correlated with the internal and external modes of the $WO_4$ units [28, 29], as in the scheelite structure, and that the internal stretching modes of the $WO_4$ tetrahedra in the ferguson ite phase can be calculated by knowing the frequencies of the Raman modes with highest frequencies and using Hardcastle and Wachs's formula [87]. On view of the success obtained in the calculation of the internal stretching frequencies of the ferguson ite phase in $BaWO_4$, Manjón *et al.* gave an estimation for the frequency of the only internal stretching mode of ferguson ite $CaWO_4$ and $SrWO_4$ that was not found in previous works [45, 51, 65]. This mode is proposed to have a frequency of 878 cm$^{-1}$ (890 cm$^{-1}$) in $CaWO_4$ ($SrWO_4$) at 15 GPa.

In ferguson ite $PbWO_4$ and in the structurally-related $HgWO_4$, the coordination for the A cation is 8 and for the W cation is 4 + 2 being the tendency to octahedral coordination for the W cation more pronounced in $PbWO_4$ than in $HgWO_4$. The increase of coordination of the W cation in the ferguson ite phase with respect to the scheelite phase is usually accompanied by an increase of the W-O bond distance. This fact explains that ferguson ite $PbWO_4$ and ferguson ite-like $HgWO_4$ exhibits some internal stretching modes, especially at frequencies above the phonon gap of the scheelite structure, at considerably lower frequencies than in the scheelite phase and than in the ferguson ite phase of alkaline-earth tungstates. The increase of coordination of the W cation in ferguson ite $PbWO_4$ is related to a strong decrease of the bond distance of two W-O second neighbours and a small increase of two W-O first neighbours as suggested by *ab initio* total energy calculations (see **Figure 4**). The latter structural changes caused a marked decrease of the frequency of the highest frequency stretching mode and the former structural change caused a marked increase of the frequencies of the two fergusonite modes derived from the $\nu_4(E_g)$ scheelite mode. Besides, these two fergusonite modes exhibit a very high positive pressure coefficient due to the strong tendency of the fergusonite phase in $PbWO_4$ towards the octahedral coordination for W. In practice, they could be considered as the two new internal stretching modes of the $WO_6$ octahedra in the fergusonite phase of $PbWO_4$, as found by applying Hardcastle and Wachs's formula [29]. Therefore, the fergusonite phase in $PbWO_4$ can be viewed as a bridge phase between the tetrahedral coordination of the W cation in the scheelite phase and the octahedral coordination of the W cation in a post-fergusonite phase. Furthermore, one can consider that the higher stability of the fergusonite phase in $PbWO_4$ with respect to $BaWO_4$ is due to the major ability of the fergusonite phase in $PbWO_4$ to accommodate the pressure-induced deformation, probably due to the smaller ionic radius of Pb with respect to Ba.

**5.2. Post-fergusonite structures**

X-ray diffraction and absorption measurements performed in $CaWO_4$ and $SrWO_4$ above 20 GPa did not find any evidence of any post-fergusonite phase [14, 17, 18, 26, 65]. However, similar studies performed in $BaWO_4$ and $PbWO_4$ found evidence of the existence of a more compact monoclinic post-fergusonite structure [15, 17, 19, 28, 29]. The quality of the x-ray diffraction data existent for the post-fergusonite phase does not permit an accurate Rietveld refinement of its structure. However, after a LeBail analysis [106] of the experimental data it was found that the most likely space group of the post-fergusonite phase is the monoclinic *P2$_1$/n*. Therefore, the post-fergusonite-phase corresponds to the $BaWO_4$-II-type structure reported as the most stable post-scheelite phase in the *ab initio* calculations. **Figure 11** shows the refined structure models and the residual of the refinement procedure for a diffraction pattern collected in BaWO$_4$ at 10.9 GPa. The calculated lattice parameters and atomic positions for this phase of BaWO$_4$ at 9.3 GPa are given in **Table VIII**. In BaWO$_4$ the post-fergusonite structure is named as BaWO$_4$-II and in PbWO$_4$ as PbWO$_4$-III. **Figure 12** gives a perspective drawing of the monoclinic BaWO$_4$-II structure. It is important to note that the transition from the fergusonite phase to the BaWO$_4$-II-type phase occurs in both compounds together with a large volume collapse of approximately 8%. This volume collapse reflects the fact that the structure of BaWO$_4$-II (PbWO$_4$-III) consists of densely packed networks of distorted WO$_6$ octahedra. XANES spectra collected beyond 9.5 GPa in BaWO$_4$ and 16.7 GPa in PbWO$_4$ were also compatible with the monoclinic *P2$_1$/n* structures. According to both ADXRD and XANES studies, other structures like LaTaO$_4$-type (S.G.: *P2$_1$/c*, No. 14, Z = 4) [107] and the BaMnF$_4$-type (S.G.: *A2$_1$/am*, No. 36, Z = 4) [108] structures were in close competence with the BaWO$_4$-II-type one. The combination of the experimental studies with *ab initio* calculations was crucial for the assignment of the structure of the post-fergusonite phase. Here it is important to mention that peak intensities can be slightly distorted in x-ray powder diffraction x-ray DAC experiments [109, 110] and thus an analysis based on residuals alone is in some cases not enough to discriminate between competing crystal structures. In such a case, the help of *ab initio* calculations is more than recommendable. The predictability shown for the theoretical calculations for these compounds give credibility for their predictions at higher pressure than those reached in the experiments.

Raman measurements in CaWO$_4$ and SrWO$_4$ found no evidence of a second phase transition in good agreement with the ADXRD and XANES experiments [26, 65]. However, Raman measurements in BaWO$_4$ (PbWO$_4$) found forty-one (thirty-seven) new peaks, corresponding neither to the scheelite nor to the ferguson phase [28, 29]. The new Raman modes appeared above 6.9 (6.2) GPa and were followed up to the maximum pressure attained in the experiments. These Raman results were in good agreement with the monoclinic $P2_1/n$ structure found in ADXRD and XANES measurements and predicted by *ab initio* total energy calculations. The new Raman peaks coexisted with those of the scheelite phase up to 9.0 GPa in both compounds, and coexisted with those of the ferguson phase up to 9.0 (14.6) GPa in BaWO$_4$ (PbWO$_4$). Manjón *et al.* demonstrated with the help of *ab initio* lattice dynamics calculations that the new peaks appearing above 6.9 GPa (6.2 GPa) in BaWO$_4$ (PbWO$_4$) do not correspond to the PbWO$_4$-II phase (raspite-type, S.G.: $P2_1/a$, No. 14, Z = 4) [104], but to the BaWO$_4$-II (PbWO$_4$-III) phase with monoclinic $P2_1/n$ space group symmetry [28, 29]. This structure should display 72 Raman-active modes with the following mechanical representation:

$$\Gamma = 36A_g + 36B_g$$

**Figure 6** shows the Raman spectra of the PbWO$_4$-III phase at 13.7 GPa, and **Figure 10** shows the pressure dependence of the Raman modes in the PbWO$_4$-III phase. In the monoclinic $P2_1/n$ structure, the W cation has an octahedral coordination and this correlates well with the appearance of Raman peaks in the phonon gap of the scheelite and ferguson structures. Manjón *et al.* showed that it was possible to identify the internal stretching modes of the WO$_6$ octahedra in the BaWO$_4$-II and PbWO$_4$-III phases [28, 29] by knowing the frequencies of the Raman modes with highest frequencies and using Hardcastle and Wachs's formula despite the objections raised by Daturi *et al.* [45]. In references [28] and [29] the authors noted that the Raman spectra of CaMoO$_4$ [85], SrMoO$_4$ [54], BaMoO$_4$ [20,27] and PbMoO$_4$ [52] at different pressures were similar to those observed in CaWO$_4$, SrWO$_4$ BaWO$_4$ and PbWO$_4$ in their different phases. Therefore, they suggested that scheelite tungstates (AWO$_4$) and molybdates (AMoO$_4$) with the same A cation (A= Ca, Sr, Ba, and Pb) should undergo the same pressure-induced phase transitions.

As regards the coexistence of the different phases in BaWO$_4$ and PbWO$_4$, it is interesting to note here that contrary to ADXRD and XANES measurements, Raman measurements showed that the BaWO$_4$-II and PbWO$_4$-III phases appeared before the fergusonite phase in good agreement with *ab initio* calculations [28, 29]. Therefore, strictly speaking they should not be classified as post-fergusonite, but as post-scheelite. In BaWO$_4$, Raman experiments detected the BaWO$_4$-II phase at 6.9 GPa and the fergusonite phase at 7.5 GPa, coexisting the three phases (scheelite, fergusonite, and BaWO$_4$-II) between 7.5 and 9 GPa. In PbWO$_4$, Raman experiments detected the PbWO$_4$-III phase at 6.2 GPa and the fergusonite phase at 7.9 GPa, coexisting the three phases (scheelite, fergusonite, and PbWO$_4$-III) between 7.9 and 9 GPa, while the fergusonite and PbWO$_4$-III phases coexist up to 14 GPa. In the Raman spectra of both compounds, it was observed that the Raman peaks of the $P2_1/n$ phase appear rather weak at the phase transition and that the Raman peaks of the fergusonite phase appear rather strong at the phase transition. On top of that, the fergusonite peaks decrease rather quickly in intensity with increasing pressure. All these results can be interpreted by considering that despite the fergusonite phase appeared after the $P2_1/n$ phase the former is the dominant phase when it appears and that afterwards the fergusonite phase becomes unstable rather quickly against the more compact $P2_1/n$ phase. The reason for the observation of the scheelite-to-fergusonite phase transition in Raman experiments after the observation of the scheelite-to-$P2_1/n$ phase transition is indicative of the kinetically-hindered nature of the latter phase transition and the second-order nature of the former one. Furthermore, the dominance of the scheelite phase near the scheelite-to-$P2_1/n$ phase transition, the dominance of the fergusonite phase after the scheelite-to-fergusonite phase transition, and the closeness of the two phase-transition pressures [less than 1 GPa (2 GPa) differences in $BaWO_4$ ($PbWO_4$)] allows to explain why the fergusonite phase has been detected in ADXRD and XANES measurements at lower pressures than the $P2_1/n$ phase in both $BaWO_4$ and $PbWO_4$.

*Ab initio* total-energy calculations agree with experimental measurements in concluding that both the fergusonite and $P2_1/n$ phases are competing post-scheelite phases with the fergusonite phase being more stable at high pressures for $CaWO_4$ and $SrWO_4$, and the $P2_1/n$ phase being more stable at high pressures in $BaWO_4$ and $PbWO_4$. The observation of the scheelite-to-fergusonite phase transition after the scheelite-to-$P2_1/n$ phase transition in Raman measurements supports the idea that there is a kinetic hindrance (energy barrier) that prevents the reconstructive $I4_1/a$-to-$P2_1/n$ phase transition for taking place, while the displacive second-order $I4_1/a$-to-$I2/a$ phase transition is not kinetically hindered. The observation of the theoretically-predicted sequence of phase transitions in Raman spectroscopy unlike in ADXRD and XANES is possible due to the ability of the former technique of distinguishing small traces of various local phases coexisting in a compound. It is important also to note that Raman measurements were performed using single crystals, while the ADXRD and XANES measurements were performed on powder samples.

On further pressure increase, beyond the limit reached by the experiments, the theoretical calculations found that for $BaWO_4$ the monoclinic $P2_1/n$ phases become unstable against the orthorhombic $BaMnF_4$-type structure (S.G.: $A2_1/am$, No. 36, Z = 4) [108] around 27 GPa. On further increase of pressure the orthorhombic $Cmca$ structure (S.G.: *Cmca*, No. 64, Z = 8) [17, 111] becomes favoured above around 56 GPa. In the case of PbWO$_4$, the *Cmca* structure becomes favoured over the *P2$_1$/n* phase around 35 GPa. In CaWO$_4$ and SrWO$_4$, the *Cmca* structure is predicted to become stable at 29 GPa and 32 GPa, respectively. An interesting further result of the theoretical calculations is the fact that at expanded volumes (and corresponding "negative" pressures) the zircon structure (S.G.: *I4$_1$/amd*, No. 141, Z = 4) [112] becomes stable against the scheelite structure in all the materials. This fact can be clearly seen in **Figure 7**. Mineral zircon (ZrSiO$_4$) transforms under pressure to a structural phase isomorphous with scheelite, named reidite [113], which is in agreement with the theoretical findings for the orthotungstates (though in this case of course the coexistence pressure is "negative").

**5.3. Pressure-induced amorphization**

In an EDXRD study performed in CaWO$_4$ [68], it was found that the diffraction lines gradually weaken and begin to broaden beyond 27 GPa. Under further compression at 40 GPa only some pronounce broad diffuse scattering, commonly observed in amorphous solids [114], was present in the x-ray patterns. These changes observed in the diffraction pattern were irreversible. A similar phenomenon was observed in ADXRD experiments on BaWO$_4$ beyond 47 GPa [19]. These facts have been interpreted as a pressure-induced amorphization. The occurrence of this kind of amorphization is sometimes related to a frustrated solid-solid phase transition. A similar amorphization has been also found in other scheelite-structured ABX$_4$ compounds, like NaLa(MoO$_4$)$_2$ [115], YCrO$_4$ [116], and YGdF$_4$ [117]. Precursor-effects of the amorphization have been observed in PbWO$_4$ too [19, 30]. Another possibility for the appearance of the broad features in the ADXRD patterns is the occurrence of a pressure-induced chemical decomposition as recently proposed for BaWO$_4$ [21]. This hypothesis is however in contradiction with the rest of experimental data available in the literature [15, 19, 28, 51] and with the fact that a thermal annealing of amorphous CaWO$_4$ at 45 GPa and 477 K during two hours led to the nucleation of a new crystalline structure of CaWO$_4$ [69]. Thus, most likely the disappearance of the Bragg peaks in the diffraction patterns of BaWO$_4$ and CaWO$_4$ near 47 GPa and 40 GPa, respectively, is the result of the frustrated transformation of their high-pressure crystalline phases into a non-crystalline solid. The irreversible nature of the amorphization implies that beyond a given pressure the polyhedra not only deform and interconnect probably differently, but also that the structural changes are significantly larger to hinder the reversal of deformations upon release of pressure.

Pressure-induced amorphization has been reported in many substances, and nowadays it is known that it differs from classical amorphization [118]. High-pressure amorphization can be understood in ABO$_4$ compounds in terms of the packing of the anionic BO$_4$ units around the A cations [119]. When the ionic radii of the BO$_4$ groups is small relative to that of the A cations, increasing repulsive and steric stresses induced by pressure can be accommodated by deformation of the A cation outer shell as opposed to significant changes in its average position, thereby favouring the transformation to a high-pressure crystalline phase. In contrast to this, a large ratio between the ionic radii (BO$_4$/A) will accommodate increased stresses through larger and more varied displacements from their average positions resulting in a subsequent loss of translational periodicity at high-pressure. This results in the transformation to a frustrated non-crystalline solid. The lower pressure for the onset of amorphization in CaWO$_4$ as compared to BaWO$_4$ is consistent with the larger WO$_4$/A ratio for the Ca compound (WO$_4$/Ca = 1.89 and WO$_4$/Ba = 1.47). A direct conclusion can be drawn for SrWO$_4$ and PbWO$_4$ for which the ratios WO$_4$/Sr = 1.76 and WO$_4$/Pb = 1.66 imply that they should become amorphous at pressures around 43 GPa and 45 GPa, respectively.

### 5.4. High-pressure high-temperature phases

New phases of $BaWO_4$ and $PbWO_4$ have been synthesized by different groups [47, 48] using piston-cylinder devices [120]. These new phases were quenched at ambient conditions after the high-pressure and high-temperature (HP-HT) treatment, being their structure assigned to the $BaWO_4$-II structure [47] and to the isostructural $PbWO_4$-III structure [48]. The structure of these HP-HT phases is very similar to that of the post-fergusonite structures found under compression at room temperature. To conclude whether the two phases are the same phase (like happen with the bcc HT phase of Ca and the bcc HP phase of Ca [121]) or just two independent phases with a similar crystalline structure an analysis of the HP-HT phase diagram of the scheelites is required. **Figure 13** shows the phase diagram of $BaWO_4$. To build it, it has been assumed that at RT the scheelite-to-fergusonite transition takes place around 7 GPa [19] and the fergusonite-to-$BaWO_4$-II transition occurs around 9.5 GPa. Further, it is known that the pressure for the synthesis of the HP-HT $BaWO_4$-II phase increases with temperature following the relation: P(GPa) = 2.67 + 0.00265 T (in ºC) [47]. This relation can be considered as the phase boundary between the HP-HT $BaWO_4$-II phase and the observed low temperature phase. It is draw in **Figure 13** as a solid line for the pressure range covered in Ref. [47] and extrapolated to higher pressure as a dotted line. It is also known that scheelites melt at low pressure directly from the scheelite phase [122] being the melting point 1775 K [123]. Therefore, a phase boundary could be drawn between the scheelite phase and the HP-HT $BaWO_4$-II phase. Since the latter phase is much denser than the former, according with the Clausius–Clapeyron equation ($dT/dP = T \Delta V/\Delta H$, where $\Delta V$ and $\Delta H$ are, respectively, the difference in molar volume and enthalpy between the two phases of interest) [124], the phase boundary should have a positive slope (dT/dP), as shown in **Figure 13**. The Clausius–Clapeyron equation can be also used to calculate the melting curve of $BaWO_4$ [125]. By taking the melting enthalpy as 913.8 J mol$^{-1}$ [123] and volume change at melting as 0.025 cm$^3$ mol$^{-1}$, a typical value for minerals [126], a melting slope of around 50 K/GPa is obtained, which is a quite reasonable value. The calculated melting curve of $BaWO_4$ is shown in **Figure 13** as a dashed line. At its interception with the scheelite-$BaWO_4$-II an abrupt change of the melting slope is expected due to the different densities of the solid phases [127]. This fact is also shown in **Figure 13**. Finally, the phase boundary between fergusonite and the RT $BaWO_4$ phase has been also drawn as a dotted line. This boundary should also have a positive slope due to the volume contraction that takes place at the transition. The phase diagram obtained for $BaWO_4$ resembles very much that calculated for $YLiF_4$ [128], which suggests that the assumptions made to build it are quite reasonable.

Let us discuss now the possible relation between the HT-HP $BaWO_4$-II phase and the RT-HP $BaWO_4$-II phase. The phase boundaries between both phases and the fergusonite phase are both positive. They are expected to intersect each other at very high temperatures. According with **Figure 13**, at the intersection point, a large change in the dT/dP value of the fergusonite-$BaWO_4$-II phase-boundary curve is expected to take place. However, this fact is in contradiction with the general principles of thermodynamics, which predict that there are no abrupt changes in the phase boundaries of any two-phase transformation. This conclusion strongly rules out the possibility that the two isostructural $BaWO_4$-II-type phases could be the same phase. One possible explanation for the different slopes of the phase boundaries between fergusonite and the HT-HP $BaWO_4$-II phase and between fergusonite and the RT-HP

BaWO$_4$-II phase is the existence of another phase at play. The existence of this new HP-HT phase beyond 7 GPa at high temperature, and between the fergusonite and the two BaWO$_4$-II phases, will imply the existence of two additional triple points in the HP-HT phase boundary. It will also mean that in spite of having a similar crystal structure the two BaWO$_4$-II phases are not exactly the same phase. The possible location of the proposed HP-HT phase is schematically shown in **Figure 13**. It is important to note here that recent HP-HT x-ray diffraction experiments performed in BaWO$_4$, showed that when heating it at 10 GPa the BaWO$_4$-II phase transforms to a new phase around 1000 K, being the structure of the new phase not yet characterized [129]. This fact confirms the hypothesis above discussed about the possible existence of extra phases in the phase diagram of scheelite orthotungtates. The existence of previously unknown phases was also observed in CaWO$_4$, but around 40 GPa [69]. In this case, a monoclinic structure has been proposed for it ($a$ = 10.38 Å, $b$ = 6.11 Å, $c$ = 7.41 Å, $\beta$ = 104.28°, S.G.: *C2/m*, No. 12, Z = 8). This structure is very similar to that of $\alpha$-MnMoO$_4$ [79] and was recently found as a HT phase of AlWO$_4$ [130]. On it, the W environment can be assumed as a highly distorted octahedron, being therefore the W-O coordination similar to that of the BaWO$_4$-II phase.

All these conclusions suggest that the HT-HP phase diagram of scheelite-structured compounds is more complicated that believed up to very recently. Further studies are needed to better understand it. Its accurate knowledge may have important geophysical and geochemical implications since the scheelite-structured orthotungstates are common accessory minerals in various kinds of rocks in the Earth's upper mantle since the pressures and temperatures limits of this section are found at a depth of < 100 km in the Earth's upper mantle [131].

### 5.5. Wolframite-to-$\beta$-fergusonite phase transition

It is known that $AWO_4$ orthotungstates with A = Cd, Co, Fe, Ni, and Zn crystallize in the wolframite structure [45, 71]. There are very few studies about the high-pressure behaviour of $AWO_4$ orthotungstates with wolframite structure. Only two high-pressure Raman studies have been published in $CdWO_4$ up to 40 GPa [54] and $ZnWO_4$ up to 24 GPa [132]. In the last work no phase transition was reported; however, in $CdWO_4$ two phase transitions were found at 12 GPa and 20 GPa. Jayaraman *et al.* suggested that the first phase transition in $CdWO_4$ was of ferroelastic character and the second one was to a phase with pure octahedral coordination for W.

Recently, a new work reporting the Raman spectra of $ZnWO_4$ up to 45 GPa has been published [70]. In this work also some changes in the Raman spectra have been found around 12 GPa and 30.6 GPa in close resemblance to those occurring in $CdWO_4$ at 12 GPa and 20 GPa. The changes occurring at 12 GPa in $ZnWO_4$ were considered to be due to the formation of different wolframite-type domains in the sample, as a consequence of the introduction of defects in $ZnWO_4$ single crystals, a phenomenon also known as twinning. However, the changes occurring at 30.6 GPa were attributed, with the help of *ab initio* lattice dynamics calculations, to the onset of a phase transition to a distorted β-fergusonite-type structure, similar to the structure of $YNbO_4$ (S.G.: *C2/c*, No. 15, Z = 4) [133]. In the HP structure of $ZnWO_4$ the Zn atoms replace Y and the W atoms replace Nb. The HP structure of $ZnWO_4$ is closely related to the high-pressure monoclinic M-fergusonite structure found in scheelite-structured orthotungstates. Indeed, both structures consist of zigzag chains of W-O polyhedra with eight coordinated A cations. The characterization of the reported high-pressure phase of $ZnWO_4$ lacks to be confirmed by ADXRD measurements but it is in agreement with the Fukunaga and Yamaoka's and Bastide's diagrams discussed in section 7.

**6. High pressure phase transitions in other ABX$_4$ materials: molybdates, vanadates, chromates, silicates, germanates, niobates, tantalates, perrhenates, periodates, phosphates, and fluorides.**

Many AMoO$_4$ orthomolybdates crystallize in the same structures than AWO$_4$ orthotungstates since the Mo cation has a similar ionic radius than the W cation. In this sense, the above considerations regarding the phase transitions occurring in AWO$_4$ orthotungstates are also valid for AMoO$_4$ orthomolybdates. This means that we expect a scheelite-to-fergusonite phase transition in CaMoO$_4$ and SrMoO$_4$ and a scheelite-to-*P2$_1$/n* phase transition in PbMoO$_4$ and BaMoO$_4$. In fact, SrMoO$_4$ was recently studied under compression up to 25 GPa by angle-dispersive x-ray diffraction and a phase transition was observed from the scheelite phase to a monoclinic M-fergusonite phase [134] as also observed in SrWO$_4$ [17]. As already commented, the Raman spectra of these molybdates at high pressures are very similar to those of the corresponding tungstates in good agreement with the hypothesis described above.

Similarly, we expect that wolframite-type orthomolybdates like ZnMoO$_4$, NiMoO$_4$, MgMoO$_4$, MnMoO$_4$ and FeMoO$_4$ behave in a similar way to CdWO$_4$ and ZnWO$_4$ exhibiting a pressure-induced phase transition to a distorted β-fergusonite phase. A special case is that of CdMoO$_4$ which crystallizes in the scheelite structure while CdWO$_4$ crystallizes in the wolframite structure. Raman measurements up to 40 GPa [52] and EDXRD measurements up to 52 GPa [67] in CdMoO$_4$ suggest that there is a phase transition from scheelite-to-wolframite at 12 GPa and another phase transition from wolframite-to-BaWO$_4$-II at 25 GPa.

Germanates, like ZrGeO$_4$, HfGeO$_4$, and ThGeO$_4$, are also compounds that can be synthesized in the scheelite structure at temperatures higher that 1200 K from the appropriate amounts of ZrO$_2$/HfO$_2$/ThO$_2$ and GeO$_2$. Synthetic ZrGeO$_4$ and HfGeO$_4$ have been recently studied by means of ADXRD measurements up to 20 GPa [135]. These studies show that these compounds do not undergo any phase transition or decomposition in this pressure range and that they are one of the less compressible $ABO_4$ compounds ($B_0$ = 230 – 240 GPa). Their incompressibility can be understood on the basis of the empirical model proposed in Ref. [17] to predict the bulk moduli of scheelite-structured $ABX_4$ compounds as we will show in section 7.5. As we will see later, this fact is related to the high cation charge density of the $ZrO_8$ and $HfO_8$ bisdisphenoids.

As regards to the vanadates, chromates, and silicates, many of these compounds crystallize in the zircon structure. Prototype structures of vanadates, chromates, and silicates are $YVO_4$, $YCrO_4$, and $ZrSiO_4$. There are several studies under pressure of these compounds. It has been found that, like other zircon-structured compounds, $ZrSiO_4$ undergoes a phase transition towards the scheelite structure around 22 GPa [136]. Recently, Raman spectra in $HfSiO_4$ show that it undergoes a phase transition from the zircon structure to the scheelite structure around 20 GPa [137]. Also ADXRD, photoluminescence, and Raman studies under pressure in $YVO_4$ show that this compound undergoes the zircon-to-scheelite phase transition at 7.5 GPa [138 - 140], and Raman studies in $YCrO_4$ show the onset of a zircon-to-scheelite phase transition at 3 GPa, being the transition completed at 15 GPa [116]. The zircon-to-scheelite phase transition in $CaCrO_4$ has been also measured by EDXRD at 5.7 GPa [141]. Recently, *ab initio* calculations of Marqués *et al.* have shown that the zircon-to-scheelite phase transition in $ZrSiO_4$ is of reconstructive type [142], as already suggested by Kusaba et al. [143], despite the group-subgroup relationship between both structures points towards a transition of displacive-type. The reconstructive mechanism for the zircon-to-scheelite phase transition explains why scheelite phases are metastable at ambient pressure and do not return to the zircon phase on release of pressure [144].

There are other molybdates, tungstates, vanadates, and chromates that crystallize in the monoclinic α-MnMoO$_4$ structure (S.G.: $C2/m$, No. 12, Z = 8) or in structures belonging to the orthorhombic $Cmcm$ structure (S.G. No. 63, Z = 4) [40], like AlWO$_4$, ZnMoO$_4$, CrVO$_4$, and CdCrO$_4$. There are very few studies of these compounds under pressure, but it is known that CdCrO$_4$ transforms under pressure to the scheelite structure [40], and that many molybdates with the α-MnMoO$_4$ structure crystallize in the wolframite structure when synthesized at high pressures [145].

Perrhenates and periodates are compounds whose structures are closely related to scheelite. AgReO$_4$, KReO$_4$, RbReO$_4$, KIO$_4$, and RbIO$_4$ crystallize in the scheelite phase while TlReO$_4$, CsReO$_4$, and CsIO$_4$ crystallize in the orthorhombic $Pnma$ structure (S.G.: $Pnma$, No. 62, Z = 4), also known as pseudoscheelite. Similarly to the scheelite structure, the pseudoscheelite structure also consists of isolated BO$_4$ (ReO$_4$ or IO$_4$) tetrahedra linked together by highly coordinated Tl or Cs polyhedra (e.g. TlO$_8$ and TlO$_9$ polyhedra). Several x-ray diffraction and Raman studies under pressure have been made in perrhenates showing several phase transitions up to 30 GPa. Raman studies under pressure in KReO$_4$, RbReO$_4$, TlReO$_4$, and CsReO$_4$ suggested the sequence of pressure-induced phase transitions scheelite → pseudoscheelite → wolframite → denser monoclinic or triclinic structure [56, 146]. X-ray diffraction studies in TlReO$_4$ under pressure reported a sequence of phase transitions up to 25 GPa. TlReO$_4$ goes from the pseudoscheelite to another orthorhombic phase at 1 GPa. This transition is followed by a second transition to a wolframite-related monoclinic phase at 2 GPa, and then by a third transition to a BaWO$_4$-II-type phase around 10 GPa [57]. On the other hand, Raman and x-ray diffraction in AgReO$_4$ under pressure up to 20 GPa showed a phase transition near 13 GPa to an unknown structure which seems to be neither the BaWO$_4$-II nor the pseudoscheelite structure [147]. Regarding periodates, Raman studies under pressure up to 12 GPa in KIO$_4$, RbIO$_4$, and CsIO$_4$ found several phase transitions to unknown structures [58, 59, 148].

As regards to phosphates and arsenates, they crystallize in either the berlinite or quartz structure, like AlPO$_4$, or in the *Cmcm* structure, like InPO$_4$. Berlinites were intensively studied in the 1980's because of their relation with quartz (SiO$_2$) in order to understand the polymorphism and transformations occurring in the Earth's mantle. It was considered that berlinites undergo a pressure-induced amorphisation at relatively low pressures like happens in SiO$_2$. However, Raman scattering measurements in AlPO$_4$ suggested that they could be a poorly crystallized solid [149]. Recent ADXRD measurements have shown that there are two solid-solid high-pressure phase transitions in AlPO$_4$ that can be well observed after laser heating a AlPO$_4$ sample loaded in a DAC [150]. The first phase transition in AlPO$_4$ is to the *Cmcm* structure where Al is sixfold-coordinated and P is fourfold-coordinated [151]. This pressure-induced phase transition is also observed in GaPO$_4$ [152]. The second phase transition in AlPO$_4$ is from the *Cmcm* structure to a monoclinic structure with space group *P2/m* (S.G. No. 10, Z = 4) where both Al and P are sixfold-coordinated [153]. This last work has found a long-waiting material with P in octahedral coordination and paves the way to obtain new materials with P with this coordination.

Niobates and tantalates crystallize in a number of monoclinic structures related to scheelite. The fergusonite mineral (YNb$_{1-x}$Ta$_x$O$_4$), YNbO$_4$, LaNbO$_4$, NdNbO$_4$, and NdTaO$_4$ crystallize in the monoclinic space group *I2/a* (S.G. No. 15, Z = 4). YTaO$_4$ also crystallizes in the monoclinic wolframite structure [45], and LaTaO$_4$ crystallize in a monoclinic structure with *P2$_1$/c* symmetry (S.G. No. 14, Z = 4) [107]. There are very few structural studies under pressure in these compounds reporting phase transitions. Mariathasan *et al.* noted that increasing pressure in these compounds was analogous to reducing temperature [154]. They also observed that on increasing temperature ferrgusonite-related compounds undergo a phase transition to the scheelite phase which is in agreement with the scheelite-to-ferrgusonite pressure-induced phase transition observed in scheelite tungstates. It has been also shown that $NdTaO_4$ undergoes a phase transition from the ferrgusonite to the $LaTaO_4$-type structure on increasing pressure [107]. This finding quantitatively agrees with the fact that the $LaTaO_4$-type structure becomes energetically favoured against the ferrgusonite structure beyond 20 GPa in $AWO_4$ tungstates [19].

Finally, we want to mention $ABX_4$ fluorides. There are many $ABF_4$ fluorides with different structures. $YLiF_4$, $GdLiF_4$, and $CaZnF_4$ crystallize in the scheelite structure. $BaMnF_4$ crystallizes in the *A2$_1$/am* orthorhombic structure (S.G. No. 36, Z = 4). $NaTiF_4$ and $BaZnF_4$ crystallize in the *Pbcn* space group (S.G. No. 60, Z = 4). $KAlF_4$ and $RbAlF_4$ crystallize in the orthorhombic *P4/mbm* structure (S.G. No. 127, Z = 2) [155]. $KBF_4$ crystallizes in the *Pnma* structure (S.G.: *Pnma*, No. 62, Z = 4) [156]. Finally, $KLaF_4$ and $RbBiF_4$ crystallize in a structure closely related to that of $KBF_4$ where the cations occupy randomly the cation sites [157]. There are very few studies under pressure in fluorides except in $YLiF_4$, $CaZnF_4$, and $GdLiF_4$. Phase-transition studies in fluorides under pressure have been performed by Raman scattering in $YLiF_4$ and $CaZnF_4$ [158 - 160], by ADXRD in $YLiF_4$ and $GdLiF_4$ [80, 117], and by luminescence measurements in $YLiF_4$ doped with $Nd^{3+}$ and $Eu^{3+}$ [140, 160, 161]. Phase transitions in scheelite-type $YLiF_4$ and $CaZnF_4$ have been observed around 10 GPa [158 - 160]. ADXRD measurements suggest that $YLiF_4$ undergoes a phase transition from the scheelite to the M-ferrgusonite phase [80]. However, molecular dynamics calculations and *ab initio* total-energy calculations have suggested that the phase transitions is from scheelite to a M'-ferguson ite phase (S.G.: $P2_1/c$, No. 14, Z = 2) with similar structural parameters than those of the fergusonite phase [162]. Also Raman studies in $KAlF_4$ and $RbAlF_4$ under pressure have been performed showing that a pressure-induced phase transition takes place at 0.2 GPa in $KAlF_4$ to an unknown phase [163]. More studies are needed in order to achieve for the fluorides an understanding of the high-pressure structural properties similar to that achieved for the tungstates and other oxides.

# 7 Towards a systematic understanding

## 7.1. Crystal chemistry of scheelites and zircons

A number of crystal structures, some of them corresponding to $ABX_4$ compounds, consist of $BX_4$ tetrahedra and $AX_8$ eight-coordinated polyhedra, which can be seen as two interpenetrating tetrahedra, known as bidisphenoids or dodecahedra. Among these structures some important mineral structures as scheelite or zircon are included. It is known that in many cases, these structures are related via simple crystallographic operations [164]. In particular, a bidisphenoid sharing an edge with a tetrahedron is easily deformed to a pair of edge-sharing octahedra, which allows the establishment of a relation between $ABX_4$ structures and $MX_2$ octahedral structures as rutile (S.G.: $P4_2/mnm$, No. 136, Z = 2) [165]. This fact also gives the possibility to draw a parallelism between the high-pressure structural behaviour of $ABX_4$ and $MX_2$ ($MMX_4$) compounds. Therefore, crystal chemistry can be used not only to improve the understanding of the already observed pressure-induced phase transitions in $ABX_4$ compounds, but also to make predictions for future studies.

In order to present the results reviewed in the previous sections in a consistent fashion, we will make here an analogy between the high-pressure structural behaviour of scheelite-structured and zircon-structures $ABO_4$ compounds and rutile-structured $MO_2$ compounds. It is known that upon compression zircon-structured silicates transform into the scheelite structure (e.g. zircon-to-reidite transition in $ZrSiO_4$) [113]. Another interesting fact is that, according to *ab initio* calculations, the zircon structure becomes stable against the scheelite structure in $AWO_4$ materials at expanded volumes (i.e. "negative" pressures) [19]; e.g. $PbWO_4$ and $CaWO_4$. Thus, including the "negative" pressures in the picture for the orthotungstates, the following systematic arise for the structural sequence undergone by the materials of interest to this review upon pressure increase: *$I4_1/amd$* (zircon) → *$I4_1/a$* (scheelite) → *$I2/a$* (fergusonite) → *$P2_1/n$* ($BaWO_4$-III-type) → orthorhombic phases ($BaMnF_4$-type, *Cmca*) → amorphous. The zircon structure transforms by means of a translationgleiche transformation into the scheelite structure. This transformation consist of twining zircon on (200), (020), and (002). The first operation yields an anhydrite-type structure [166], the second operation produces the $AgClO_4$-type structure [167], and the third one generates the scheelite-type structure. Regarding the scheelite-to-fergusonite transition, it also consists in a translationgleiche transformation, since the scheelite structure can be transformed into the fergusonite structure by means of another translationgleiche transformation that involves a lowering of the point-group symmetry from *4/m* to *2/m*. Finally, the $BaWO_4$-II-type structure is naturally obtained by a klassengleiche transformation from fergusonite. These structural relations may have important implications for a number of $ABX_4$ structures, especially those with a large difference between the sizes of the A and B atoms, which include some important minerals in addition to zircon and scheelite.

Let us compare now the zircon structure with the rutile structure. The latter structure consists of infinite rectilinear rods of edge-sharing $MO_6$ octahedra parallel to the *c*-axis, linked by corner sharing to the octahedra in identical corner rods. If M is alternatively substituted by bigger and smaller cations A and B and the cations are shifted in each rod then the zircon structure is obtained. The chains of edge-sharing $MO_6$ octahedra in rutile become chains of alternating $AO_8$ bidsphenoids and $BO_4$ tetrahedra in the zircon. Under compression rutile transforms to the α-$PbO_2$-type structure [168] and both scheelite and fergusonite can be thought as distorted superstructures of α-$PbO_2$. In particular, scheelite is related to the α-$PbO_2$ structure in a way that is analogous to the relationship between zircon and rutile. Additionally, the monoclinic post-fergusonite structure that has been observed in $BaWO_4$ and $PbWO_4$ is related to the baddeleyite, a post-α-$PbO_2$ structure of rutile-type $TiO_2$ [169]. Thus the crystal chemistry systematic of $MO_2$ compounds provides additional support to the structural sequence that can be extracted for $ABO_4$ compounds from previous studies. It is worth to mention here that based upon similar crystallochemical arguments it can be concluded that the scheelite-to-wolframite phase transition is not expected in $ABX_4$ compounds [14].

It is interesting to note here that in rutile-type dioxides like $PbO_2$, $ZrO_2$, and $SiO_2$ it was found that kinetics has a large effect on pressure-induced phase transitions [168, 170]. The high-pressure behaviour of these compounds depends upon the starting material and pressure–temperature history. In addition, a coexistence of different phases under compression has been observed in them, existing large pressure ranges of two-phase intergrowth. A similar phenomenon was observed in scheelite-structured $BaWO_4$ and $PbWO_4$ [19, 28, 29], confirming that the analogy between $ABX_4$ and $MX_2$ is an efficient tool for analyzing the high-pressure behaviour of zircon-structured and scheelite-structured $ABX_4$ compounds. The analogies in the crystal chemistry of the $MX_2$ compounds and their $AMX_4$ superstructures under pressure warrant further investigations to elucidate pressure-induced post-scheelite (post–ferguson ite) structures in many unstudied compounds. It could be also applied to compounds like monazite $CePO_4$ (S.G.: $P2_1/n$, No. 15, Z = 4) [171], which is also a superstructure of rutile. Huttonite (monazite-structured $ThSiO_4$) and thorite (zircon-structured $ThSiO_4$) are two metastable forms of $ThSiO_4$ [172]. According with the high-pressure behaviour of $SiO_2$, both huttonite and thorite can be expected to transform under compression to the scheelite structure.

**5.2. Size criterion: A correlation between the packing ratio and transition pressures in scheelite $ABX_4$ compounds**

One important empirical result regarding the pressure stability of the scheelite structure is the size criterion. It has been shown in the past that in $ABX_4$ scheelite compounds it was possible to correlate the packing ratio of the anionic $BX_4$ units around the A cations with the phase-transition pressures of these compounds [173]. Based upon the data on 16 different scheelite $ABX_4$ compounds, it was established by Errandonea *et al.* [173] that the transition pressure ($P_T$) for these compounds can be estimated following the relationship: $P_T = (1 \pm 2) + (10.5 \pm 2)\left(\frac{BX_4}{A} - 1\right)$, where the $BX_4$/A represents the radii ratio of $BX_4$ units and A cations, being the sum of the X/A and the B/A effective ionic ratios. The $BX_4$/A values can be calculated using the data of the ionic radii of A, B, and X atoms available in the literature [174 – 177]. This relationship indicates that for $BX_4$/A = 1 the scheelite structure is hardly stable even at ambient pressure. To understand the physics underlying this relationship, we have to remember that both the effective ionic radii decrease in cations and anions with increasing pressure, the radius decrease being larger for the larger anionic radii. Therefore, the B/A ratio is almost constant with increasing pressure while the X/A ratio decreases considerably. Consequently, it is expected that the $BX_4/A$ ratio decreases with increasing pressure and that those compounds showing a smaller $BX_4/A$ ratio should exhibit lower transition pressures. This has already been empirically found in scheelite compounds. The above hypothesis for the instability of the scheelite compounds with $BX_4/A$ radii ratios being near or below 1 is also supported by the transition pressures found in the alkaline-earth perrhenates and periodates families [56, 58, 59, 147, 148]. It has been shown that $KReO_4$, $RbReO_4$, $KIO_4$ and $RbIO_4$ crystallize in the scheelite structure. However, $TlReO_4$, $CsReO_4$, and $CsIO_4$, showing smaller $BX_4/A$ ratios near 1, crystallize in a pseudoscheelite structure at ambient pressure.

The size criterion can be used to predict the pressure range of stability of different $ABX_4$ scheelite compounds. It has been shown to be very effective for $BaMoO_4$, $GdLiF_4$, $LuLiF_4$, $TlReO_4$, and $NaAlH_4$. For these five compounds the estimated transition pressures are 5.8 GPa, 11.6 GPa, 10.8 GPa, 9.8 GPa, and 16.5 GPa, respectively, while experimentally very recently the transition pressures have been found to be 5.8 GPa [20], 11 GPa [117], 10.7 GPa [178], 10 GPa [179], and 14 GPa [180], respectively. The same criterion has been probed to be also efficient to estimate the transition pressures of double tungstates like $NaSm(WO_4)_2$, $NaTb(WO_4)_2$, and $NaHo(WO_4)_2$, which become unstable near 10 GPa [181]. For these compounds assuming the A cation radii as the average of the ionic radii of Na and the trivalent rare earth (e.g. Sm) the estimated transition pressures are 10.2 GPa, 10.5 GPa, and 10.7 GPa, respectively. In addition to that, it has been shown recently that the size criterion can also be applied to other $ABX_4$ compounds with structures different than scheelite if these compounds are also made or isolated $BX_4$ tetrahedra and $AX_8$ dodecahedra, like the zircon structure [182]. Including the new data available in the literature, the relation given in Ref. [173] can be recalculated. Twenty-four scheelite compounds known to undergone pressure-induced phase transitions as summarized in **Table IX**. As can be seen in **Figure 14**, these data can be fit with a linear function which gives the following relation: $P_T = (1.9 \pm 1.5) + (9.5 \pm 1.2)\left(\frac{BX_4}{A} - 1\right)$. The only data point that is not within the standard deviations of the fit is that corresponding to $RbReO_4$, which suggest that its transition pressure could have been probably underestimated. On the other hand, the accuracy of the size criterion for calculating the transition pressures in the other 23 scheelites recommend its use for predicting the occurrence of pressure-driven instabilities in additional scheelite compounds like, e.g., $CdCrO_4$, $NaReO_4$, and $KRuO_4$. For these compounds the new relationship predicts the occurrence of pressure-driven phase transitions at 7.9 GPa, 10.4 GPa, and 6.9 GPa, respectively, i.e. at accessible pressures for DAC experiments. In the case of scheelite-structured $HfGeO_4$, the size criterion predicts a transition pressure beyond 15 GPa. This suggests that a phase-transition should be expected to take place in this compound very close to maximum pressure attained in previous experiments [135].

It is worth to comment here that some double-molybdates like $NaBi(MoO_4)_2$ and $NaLa(MoO_4)_2$ apparently do not follow the size criterion. In these two compounds the experimentally determined transition pressure is beyond 25 GPa [183, 184], while the predicted transition-pressure using the size criterion is only around 11 GPa. However, the same empirical criterion seems to work properly with double-tungstates [181]. Therefore, more studies are needed in order to determine whether the size criterion applies only to simple scheelite-structured $ABX_4$ compounds or also to more complex compounds like the double-molybdates and double-tungstates.

**7.3. Common trends of pressure-induced phase transitions in ABX$_4$ compounds: The Fukunaga and Yamaoka's diagram and the Bastide´s diagram.**

One of the characteristics of pressure-induced phase transitions in ABX$_4$ compounds is the tendency to increase the coordination number for both A and B cations. As already commented, in the late 1970s, Fukunaga and Yamaoka tried to give a systematic explanation of the pressure-induced phase transitions in ABO$_4$ compounds [50]. In their study, they classified most of the ABO$_4$ compounds in a two-dimensional phase diagram with (t, k) coordinates. Coordinate t = $(r_A+r_B)/2r_X$ represented the average cation to anion radii ratio, and coordinate k= $r_A/r_B$ represented the cation A to cation B radii ratio. In the Fukunaga and Yamaoka's (FY) diagram, the ordering of the compounds is based on the following rules: 1) similarities between ABO$_4$ compounds and AO$_2$ compounds; 2) A cation with higher coordination (or lower valence) than B cation; and 3) possibility of structural changes without increase of cation coordination. In this way, the compounds with the same structure locate along a diagonal path along the (1,1) direction of the FY diagram. The pressure-induced phase transitions can be understood if we consider that anions are usually larger and more compressible than cations and that compounds under compression undergo phase transitions to more compacted structures, provided that steric stresses between cations are small [71], that is the case when anions are larger than cations. With these considerations in mind, Fukunaga and Yamaoka suggested that pressure-induced phase transitions follow the east (E) rule in the FY diagram; i.e., t increases while k remains relatively constant with increasing pressure.

The FY diagram allows us to understand the following pressure-induced phase transitions: high cristobalite → quartz, quartz → rutile, rutile → α-PbO$_2$ , α-PbO$_2$ → fluorite, quartz → *Cmcm*, ZnSO$_4$ → zircon, zircon → scheelite, monazite → scheelite, and the *Cmcm* → wolframite or scheelite phase transitions. It also predicts that the wolframite-structured compounds can undergo a phase transition to a ferregusonite-type structure but not to the scheelite structure. However, there were several facts not explained by the FY diagram. In this sense, some compounds do not crystallize in the corresponding structure according to their location in the FY diagram. However, the most striking feature is that $BAsO_4$ and $BPO_4$ crystallizing in the high-cristobalite structure undergo a pressure-induced phase transition to quartz [34]. This phase transition in both compounds seems not to follow the E rule for pressure increase in the FY diagram. Manjón *et al.* have reasoned that the inconsistency arises from the wrong location of the boron compounds with high cristobalite structure in the FY diagram [185]. The wrong location of these two compounds in the FY diagram comes from considering that the valence of the A cation must be lower than the valence of the B cation. However, the boron anomaly in the FY diagram disappears if the second rule of the FY diagram is changed by the following: 2) A cation is always the cation with larger ionic radius ($r_A > r_B$). In this case, all the compounds in the FY diagram basically remain at the same locations but the arsenate and phosphate compounds $BAsO_4$ and $BPO_4$ must be renamed to become "borates" $AsBO_4$ and $PBO_4$ because the radius of B is very small and leads boron to have the smaller coordination despite having smaller valence than As or P. This result is related to the reversal of the cation and anion role in BAs and BP recently reported and discussed in the literature [186, 187]. Therefore, assuming that the compounds $AsBO_4$ and $PBO_4$ are not arsenates and phosphates but "borates", both compounds have $k > 1$ and the phase transition from high-cristobalite to quartz is compatible with the E rule in FY diagram. Another anomaly is that the $ZnSO_4$ → *Cmcm* phase transition does not follow the E rule in the FY diagram but seems to follow the south-east (SE) rule. This anomaly can be understood if we consider that on the basis of the new second rule ($r_A > r_B$), the ionic radius of the cation A is slightly more compressible than that of cation B and consequently $k = r_A/r_B$ tends to decrease slightly with increasing pressure, especially for large k ($k \gg 1$). With this consideration in mind, we can establish that pressure-induced phase transitions in the FY diagram should follow a path between the E rule and the SE rule.

In summary, we think that all $ABX_4$ compounds can be properly located in the FY diagram provided that $r_A > r_B$; i.e., with $k \geq 1$ and the FY diagram can help in understanding the ambient pressure structures and high-pressure phase transitions in a number of $ABX_4$ compounds. We want to stress that there are two striking features in the FY diagram. First, the FY diagram predicts that some $ABX_4$ compounds crystallizing in the scheelite structure should undergo a phase transition to the wolframite structure under pressure. This has not been observed in the most recent experiments with scheelite orthotungstates and orthomolybdates with the only exception of $CdMoO_4$, a compound near the border between the scheelite and wolframite structure as we will discuss later in the light of the Bastide's diagram. Note that $CdWO_4$ crystallizes in the wolframite phase and that ionic radii of W and Mo are very similar. Second, the FY diagram predicts a transition from the wolframite structure either to the fergusonite or to the rutile structure and its high-pressure forms. A phase transition from the wolframite to a distorted β-fergusonite structure has been found in $ZnWO_4$ [70] and likely also for $CdWO_4$ [134]. The pressure-induced phase transition from wolframite to rutile is unlikely as we will discuss later using the Bastide's diagram, but a phase transition from wolframite to a high-pressure phase of rutile, like α-$PbO_2$ seems to be likely for the wolframites with smaller A cation, like $NiWO_4$. The only pitfall of the FY diagram is that it does not give clues for the post-ferguson­ite and most post-scheelite high-pressure phases [50].

Another attempt to found a systematic in the crystal structures and pressure-induced phase transitions in $ABX_4$ compounds was done by Bastide in the late 1980s [63]. Bastide located the $ABX_4$ compounds in a diagram with two axes that represent the cation-to-anion radii ratios ($r_A/r_X$, $r_B/r_X$), and where the $ABX_4$ compound are formulated with $r_A > r_B$ and using the revised ionic radii of Shannon [174 – 177, 188]. The Bastide's diagram is divided in different [$c_A$–$c_B$] regions with $c_A$ and $c_B$ being the coordination of cations A and B, respectively. **Figure 15** shows an updated version of the Bastide´s diagram with the location of the structures which are relevant for the present discussion. Again, taking into consideration that anions are usually larger and more compressible than cations and that compounds under compression undergo phase transitions to more compacted structures, provided that steric stresses between cations are small [71], Bastide suggested that pressure-induced phase transitions follow the north-east (NE) rule in the Bastide's diagram; i.e., both $r_A/r_X$ and $r_B/r_X$ increase with increasing pressure.

The NE rule of the Bastide's diagram allows us to understand the following phase transitions: high cristobalite → quartz, quartz → rutile, rutile → α-$PbO_2$, α-$PbO_2$ → fluorite, quartz → *Cmcm*, $ZnSO_4$ → zircon, zircon → scheelite, monazite → scheelite, *Cmcm* → wolframite or scheelite, xenotime → scheelite, scheelite → fergusonite, wolframite → fergusonite, fergusonite → *P2$_1$/c* or *P2$_1$/n*, *P2$_1$/c* or *P2$_1$/n* → *Cmca*, and *Pnma* → *P2$_1$/c* or *P2$_1$/n* phase transitions. It also allows to predict that: 1) wolframite cannot go to scheelite structure with increasing pressure because the scheelite structure is not in NE direction with respect to wolframite in the Bastide's diagram.; 2) wolframite structure is not a candidate for a post-scheelite phase; and 3) a pressure-induced phase transition from wolframite to rutile is unlikely.

Manjón *et al.* have discussed that the Bastide's diagram has some anomalous features [185]. For instance, scheelite $YLiF_4$ and other scheelite fluorides have [8-4] cation coordination but their $r_A/r_X$ and $r_B/r_X$ lye above the normal stability region of scheelite structures (with $r_A/r_X$ between 0.6 and 1.1 and $r_B/r_X$ around 0.3) and even above the limits of stability of the fergusonite structure (with $r_A/r_X$ between 0.55 - 0.6 and 1.0 and $r_B/r_X$ around 0.4). However, the pressure-induced phase transitions in $YLiF_4$ are different than those found in alkaline-earth tungstates and in good agreement with its position in the Bastide's diagram and the NE rule [185]. It is curious that despite scheelite $YLiF_4$ has an anomalous position in the Bastide's diagram, it has a cation A to anion $BX_4$ radius ratio $BX_4/A$ [$(r_B+r_X)/r_A= 2.11$] inside the limit of stability of zircon and scheelite structures (with $BX_4/A$ ratios between 1.2 and 2.2). A similar case is that of $CdCrO_4$ which crystallizes in the *Cmcm* structure [189] while its position in the Bastide's diagram is more prone to crystallize in the zircon structure. Again, $CdCrO_4$ undergoes a phase transition to the scheelite phase on increasing pressure in good agreement with the NE rule [40]. Another anomaly of the Bastide's diagram is that the compressed structure of compounds with [12-4] cation coordination like those with the barite ($BaSO_4$) structure tend to undergo a phase transition increasing $r_B/r_X$ but with constant (or even decreasing $r_A/r_X$); i.e., following a N (or even a NW) path in the Bastide's diagram. This fact can be understood if we consider that the ionic radius of the cation A is slightly more compressible than that of cation B and that large A cations can be as compressible as the X anion. This means that for $ABX_4$ compounds with large A cations $r_A/r_X$ can be almost pressure independent (or even decrease) what leads to the N (or NW) path in the Bastide's diagram. With this consideration in mind, we can establish that pressure-induced phase transitions in the Bastide's diagram usually should follow a path between the NE rule and the N rule. In this respect, we think that the experimental observation of the scheelite-to-wolframite transition [67, 68] can only be justified under non-hydrostatic conditions that enhance steric stresses leading to the NW rule in the Bastide's diagram. An exception could be $HfGeO_4$ and $ZrGeO_4$ for whom the wolframite structure is in the NE direction in the Bastide's diagram.

In the following we will show, that unlike the FY diagram, the Bastide's diagram is very useful for predicting the high-pressure phases of $ABX_4$ compounds especially post-wolframite, post-scheelite, and post-fergusonite phases that are not covered by the FY diagram. The coordinates of $CaWO_4$ in Bastide's diagram (0.8, 0.3) allow us to predict, according to the NE rule for increasing pressure, a scheelite → fergusonite transition, as has been experimentally observed and theoretically predicted [17, 18]. The NE rule with increasing pressure drives $CaWO_4$ to the region of stability of the fergusonite structure. In $SrWO_4$, $PbWO_4$ and $BaWO_4$, whose Bastide's coordinates are (0.9, 0.3), (0.92, 0.3), and (1.01, 0.3), respectively, the NE rule points towards the limit of the region of stability of the fergusonite structure. Recent *ab initio* calculations provide evidence that the fergusonite and $P2_1/n$ structures are in strong competition in $SrWO_4$ while the fergusonite structure is much more stable in $CaWO_4$ [25]. The competition between the fergusonite and the $P2_1/n$ structures experimentally observed in $PbWO_4$ and $BaWO_4$ is stronger than in $SrWO_4$ and is directly related to the $r_A/r_X$ ratio in the series $CaWO_4$, $SrWO_4$, $PbWO_4$, and $BaWO_4$. For smaller $r_A/r_X$ ratios the fergusonite phase dominates over the $P2_1/n$ phase while the latter phase dominates over the former in the compounds with larger $r_A/r_X$ ratios in good agreement with the NE rule in the Bastide's diagram.

Further transitions have been theoretically predicted in the four $AWO_4$ compounds (A = Ca, Sr, Ba and Pb), ultimately to the C*mca* orthorhombic structure (S.G. No. 64, Z = 8) which, with a cation coordination [9-6] or higher, are located close to the orthorhombic $SrUO_4$ (S.G.: P*bcm*, No. 57, Z = 4) [190] and $BaMnF_4$ (S.G.: *A2₁/am*, No. 36, Z = 4) [108] structures in the Bastide´s diagram. These structures are again in agreement with the NE rule for pressure increase with respect to the scheelite phase in the Bastide's diagram. On the basis of the similar radii of Eu and Sr, we expect similar phase transitions for $SrWO_4$ and $EuWO_4$. Indeed, in both compounds the appearance of the ferguson phase was observed at similar pressures [17, 32]. Also due to the similar radii of W and Mo and to the similar location in the Bastide's diagram, the same phase transitions of the Ca, Sr, Pb and Ba scheelite tungstates could be expected for the Ca, Sr, Pb, and Ba scheelite molybdates. Therefore, the Bastide's diagram allows us to predict the pressure-induced structural transformations from tetragonal symmetry to monoclinic symmetry and to orthorhombic symmetry following the structural sequence: (zircon or monazite or xenotime) → scheelite → (fergusonite or $P2_1/n$) → ($BaMnF_4$ or C*mca*). The zircon → scheelite → fergusonite sequence has been also recently predicted theoretically for $YCrO_4$ [182] and $CaCrO_4$ [141] and has been recently observed experimentally in $YCrO_4$ [116] and $ZrSiO_4$ [191].

On the other hand, scheelite $YLiF_4$ (0.78, 0.45) is a compound in the limit of 4-to-6 coordination for Li, that has been predicted to undergo a phase transition from the scheelite to the M'-fergusonite phase (S.G.: *P2₁/a*, No. 14, Z = 4) and afterwards to a C*mca* phase in good agreement with the NE rule in the Bastide's diagram [63]. In fact, above the fergusonite region there is a region of phase stability for some monoclinic compounds with structures belonging to the monoclinic space group No.

14 and some orthorhombic compounds (see **Figure 15**). In particular, NaCrF$_4$ (0.91, 0.48) and SrUO$_4$ (0.97, 0.69) are compounds crystallizing in the monoclinic M'-fergusonite structure and orthorhombic P*bcm* structure at ambient conditions, respectively, both in NE direction with respect to YLiF$_4$. Similar pressure-induced phase transitions to those of scheelite YLiF$_4$ are expected in other scheelite lithium fluorides, like GdLiF$_4$ and NdLiF$_4$.

Recently, a high-pressure phase of AlPO$_4$ with monoclinic structure belonging to the space group *P2/m* (S.G. No. 10) has been observed after a phase transition to the *Cmcm* phase [153]. This transition is in agreement with the Bastide's diagram since the compounds with the *Cmcm* structure can undergo a phase transition to a monoclinic structure belonging to space group numbers 10, 12, or 13 or to the scheelite structure following the NE rule depending on their location in the Bastide's diagram. In this sense, the *P2/m* structure (observed in ZrTiSe$_4$ and ZrTiTe$_4$ at ambient pressure) is a structure located near the AO$_2$ border in the Bastide's diagram between quartz and rutile in NE direction with respect to AlPO$_4$.

The unknown high-pressure phases of other ABX$_4$ compounds can be estimated from the Bastide's diagram. For instance, the high-pressure phases of fergusonite-like HgWO$_4$ are likely to have subsequently a monoclinic *P2$_1$/c* or *P2$_1$/n* structure (S.G. No. 14) and an orthorhombic *Cmca* or *Pbcm* structure. On the same basis, we think that berlinites AlAsO$_4$ and GaAsO$_4$ could also undergo a phase transition to the *P2/m* phase; whereas FePO$_4$, GaPO$_4$, and InPO$_4$ may have a completely different pressure behaviour than AlPO$_4$ due to their different location in the diagram. In FePO$_4$ and GaPO$_4$ a phase transition to the *Cmcm* structure and afterwards to a monoclinic *C2/m* structure (S.G. No. 12) or *P2/c* (S.G. No. 13) could be observed, while for InPO$_4$ a phase transition to the zircon structure and afterwards to the scheelite structure is expected.

Pressure-induced phase transitions from the *Cmcm* structure are expected to the rutile, wolframite, zircon, or to the scheelite phase depending on the location of the compounds in the Bastide's diagram. CrVO$_4$ crystallizing in the *Cmcm* structure at high temperatures transforms to the rutile-type structure and afterwards is expected to go to the wolframite phase. On the other hand, InPO$_4$ and CaSO$_4$ would likely transform to the zircon and scheelite or even to the monazite structure in the latter compound. As regards the pseudoscheelite structure, its high-pressure phases depend on the location of the compound in the Bastide's diagram. For PbCrO$_4$ a transition to the scheelite phase is expected. For TlReO$_4$ and CsReO$_4$ a phase transition to the scheelite phase is unlikely but they can undergo a phase transition either to the tetragonal *P4/mbm* structure or to the monoclinic *P2$_1$/c* and *P2$_1$/n* structures. In summary, the phenomenological FY and Bastide's diagrams have proved to be useful tools for understanding the pressure-induced phase transitions found experimentally and predicted by theoretical calculations. Therefore, these two diagrams, despite being far from being complete, are very convenient in order to predict high-pressure phases of ABX$_4$ compounds.

To close this section, we want to note that Depero *et al.* have reported that most ABO$_4$ compounds crystallize in structures with space groups Nos. 13, 14, 15, 60, 62, 88, 141 and 216 [192]. These space groups correspond to the wolframite, M'-fergusonite, fergusonite, *Pbcn*, pseudoscheelite, scheelite, zircon and cubic $F\bar{4}3m$ structures, respectively. On one hand, the coordination of the A cation is around 6 in the wolframite, around 8 in the fergusonites, zircon, and scheelite structures, around 10 in the orthorhombic *Pbcn*, and around 12 in the pseudoscheelite and the $F\bar{4}3m$ structure. On the other hand, the coordination of the B cation is around 4 in the zircon, scheelite, pseudoscheelite, and $F\bar{4}3m$ structures, around 4+2 in the wolframite and ferguson­ite structures, around 6 in the M'-ferguson­ite structure, and around 10 in the orthorhombic *Pbcn* structure. Therefore, from the viewpoint of the cation A and B coordination, it is reasonable that with increasing pressure the ABX$_4$ structures follow the sequences: A) wolframite $\rightarrow$ zircon, scheelite, ferguson­ite, M'-ferguson­ite $\rightarrow$ *Pbcn* $\rightarrow$ pseudoscheelite, cubic or B) zircon, scheelite, pseudoscheelite, cubic $\rightarrow$ wolframite, ferguson­ite $\rightarrow$ M'-ferguson­ite $\rightarrow$ *Pbcn*. Since some of these phase transitions are incompatible because an increase of one cation coordination could lead to the decrease of the coordination of the other cation one can join both sequences together to infer the structural sequence of ABX$_4$ compounds. Thus, the most probable pressure-induced phase transitions are wolframite $\rightarrow$ ferguson­ite $\rightarrow$ M'-ferguson­ite $\rightarrow$ *Pbcn* and zircon $\rightarrow$ scheelite $\rightarrow$ ferguson­ite $\rightarrow$ M'-ferguson­ite $\rightarrow$ *Pbcn*. These sequences are in good agreement with the results summarized in this review, being useful to predict phase transitions in as yet unstudied ABX$_4$ compounds. **Figure 16** summarizes the most likely pressure-induced phase transitions to occur in ABX$_4$ compounds depending on their $r_A/r_X$ ionic ratios. More high-pressure studies are needed to complete the diagrams shown in **Figures 15 and 16** in order to better understand the structural relations among ABX$_4$ compounds and also among ABX$_4$ compounds and their related AX$_2$ counterparts.

**7.4. Spontaneous strain and the ferroelastic nature of the scheelite-to-ferguson­ite phase transition**

The pressure-driven transition from the tetragonal scheelite to the monoclinic ferguson­ite phase has been reported to occur not only under compression in orthotungstates and orthomolybdates [14, 15, 17, 19] but it can be also temperature-induced in other compounds that belong to the $ABO_4$ family like $LaNbO_4$ [154]. In this case, the scheelite-to-fergusonite transition has been characterized as a ferroelastic second-order transition [193]. At the phase transition, and upon further compression of the fergusonite phase, one pair of parallel unit-cell edges of the tetragonal phase contract and another pair elongates, while the angle between them gradually increases. There are two possible ways to achieve this situation and these options can be illustrated by two choices of direction in the tetragonal cell. These two choices are crystallographically identical, and related through the fourfold rotation symmetry of the tetragonal system. We will call these two monoclinic orientation states $S_1$ and $S_2$. They are identical in structure, but different in orientation. These orientation states are crystallographically and energetically equivalent, being impossible to distinguish one from the other if they appear separately. The similitude between the pressure-induced and temperature-induced scheelite-to-fergusonite phase transition strongly suggest that the pressure-driven transition is also a second-order transformation with a ferroelastic nature. One possibility to probe this hypothesis is to analyze the spontaneous strains of the monoclinic phase applying the Landau theory [194].

In a ferroelastic transformation the $S_1$ and $S_2$ states can be seen as a small distortion caused by slight displacements of the atoms of the parent tetragonal phase. The spontaneous strain characterizes the distortion of each orientation state relative to the prototype structure (i.e. the scheelite-type structure). Following Schlenker *et al.* [195] the elements of the strain tensor for a crystal can be calculated based upon the lattice parameters. In the case of the transition we are dealing with, the strain elements of one of the orientation states are:

$$\varepsilon_{11} = \frac{c_M \sin \beta_M}{a_T} - 1 \quad (1),$$

$$\varepsilon_{22} = \frac{a_M}{a_T} - 1 \quad (2),$$

$$\varepsilon_{33} = \frac{b_M}{c_T} - 1 \quad (3),$$

$$\varepsilon_{12} = \varepsilon_{21} = \frac{1}{2}\frac{c_M \cos \beta_M}{a_T} \quad (4),$$

where the subscripts T and M refers to the tetragonal and monoclinic phases. The remaining tensor elements are reduced to zero by the cell parameters. The strain tensor for a single orientation state ($S_1$) is then:

$$e_{ij}(s_1) = \begin{pmatrix} \varepsilon_{11} & \varepsilon_{12} & 0 \\ \varepsilon_{12} & \varepsilon_{22} & 0 \\ 0 & 0 & \varepsilon_{33} \end{pmatrix} \quad (5)$$

and $e_{ij}(S_2)$ is related to $e_{ij}(S_1)$ by $e_{ij}(S_2) = \mathbf{R}\, e_{ij}(S_1)\, \mathbf{R}^T$, where $\mathbf{R}$ and $\mathbf{R}^T$ are the 90° rotation matrix around the *b*-axis of the monoclinic unit cell and its transpose. According to Aizu [196] in the present case the spontaneous strain tensor can be expressed as:

$$e_{ij}^s(s_1) = e_{ij}(s_1) - \frac{1}{2}\left[e_{ij}(s_1) + e_{ij}(s_2)\right] = \begin{pmatrix} -u & v & 0 \\ v & u & 0 \\ 0 & 0 & 0 \end{pmatrix} \quad (6)$$

$$\text{and} \quad e_{ij}^s(s_2) = \begin{pmatrix} u & -v & 0 \\ -v & -u & 0 \\ 0 & 0 & 0 \end{pmatrix} \quad (7),$$

where $u = \frac{1}{2}(\varepsilon_{22} - \varepsilon_{11})$ is the longitudinal spontaneous strain and $v = \varepsilon_{12}$ is the shear spontaneous strain. The scalar spontaneous strain $\varepsilon_s$ is defined as [193, 196, 197]:

$$\varepsilon_s^2 = \sum_{i=1}^{3}\sum_{j=1}^{3}(\varepsilon_{ij}^s)^2 = 2(u^2 + v^2) \quad (8).$$

Following eq. (8) the spontaneous strains tensor as well as the scalar spontaneous strain for any scheelite-structured $AWO_4$ and $AMoO_4$ compound can be calculated using the pressure dependences of the lattice parameters reported in the literature [17, 19, 32, 134]. The obtained results for $CaWO_4$ are plotted in **Figure 17** as a function of $\sqrt{(P-P_T)/P_T} = \eta'$, where $P_T$ is the transition pressure.

The deviation of the fergusonite structure from the $I4_1/a$ symmetry can be expressed by the magnitude of the order parameter η. According to the Landau theory [194], for a second-order transition η is small close to the critical value of the relevant thermodynamic variable. In the present case to $P_T$. Thus, the Gibbs free energy ($G$) of the fergusonite phase relative to the scheelite phase can be expressed as a Taylor expansion in terms of η, yielding the following relation: $G = k_1(P-P_T)\eta^2 + k_2\eta^4$ [197]. From this equation, the relation between pressure and the order parameter can be found by minimizing $G$. This condition is only satisfied if the order parameter has the form: $\eta \propto \sqrt{(P-P_T)/P_T} = \eta'$, which can be defined as the phenomenological order parameter in Landau's theory [194]. In a ferroelastic transition $\varepsilon_s$ can be considered as being proportional to the primary order parameter η [198] and consequently also to η'. Therefore, if the studied transition is a second-order ferroelastic transition, $\varepsilon_s$ should be a linear function η' as indeed happen in **Figure 17**. The same kind of dependence for $\varepsilon_s$ has been observed in $SrWO_4$, $BaWO_4$, $PbWO_4$, $EuWO_4$, and $SrMoO_4$ [134, 197]. This fact strongly suggests that the pressure-induced scheelite to fergusonite transition is a ferroelastic second-order phase transition. A similar ferroelastic transformation has been also found at low temperature in the scheelite-structured $CaMoO_4$ [199]. On top of that, the detection of a soft acoustic mode in Brillouin scattering measurements in scheelite-structured $BiVO_4$ [200] is another conclusive evidence of the accuracy of the ferroelastic interpretation of the scheelite to fergusonite transition.

**7.5. Equation of state of ABX$_4$ compounds and the relationship between bulk compressibility and polyhedral compressibility**

In Section 4.4, we have commented that the volume compression undergone by scheelite-structured AWO$_4$ compounds is mainly due to shortening of the A–O bonds rather than to changes of the W–O bonds in the WO$_4^{-2}$ tetrahedral units. Indeed, as a first approximation, it can be assumed that under compression these tetrahedra behave as rigid uncompressible units. The same behaviour has been observed in other ABO$_4$ compounds. This fact has led to high-pressure scientists to think that there could be a relation between the bulk and polyhedral compressibility of ABO$_4$ compounds. First, Hazen *et al.* found that the bulk modulus of certain ternary oxides and silicates can be directly correlated to the compressibility of the A-cation coordination polyhedra [201]. In particular, they proposed that the bulk compressibility (1/B$_0$) in these compounds is proportional to the average volume of the cation polyhedron divided by the cation formal charge; i.e., B$_0$ is proportional to the cation charge density per unit volume inside the cation polyhedron. They also found that A$^{2+}$B$^{6+}$O$_4$ scheelite-structures tungstates and molybdates compress in an anisotropic way with the WO$_4$ and MoO$_4$ tetrahedral behaving as rigid units [60]. Furthermore, they ordered the compressibility of scheelite compounds according to the A-cation formal charge and, on this basis, suggested that the compressibility of ABO$_4$ scheelites could be given mainly by the compressibility of the softer AO$_8$ polyhedron. For the compounds analysed by Hazen *et al.* [60], one finds that B$_0$ (in GPa) is equal to *750 Z$_i$/d$^3$*, where Z$_i$ is the cationic formal charge and *d* is the mean value of the A–O bond distance (in Å). More recently, Errandonea *et al.* [17] updated

Hazen´s idea. Among other things, these authors analysed the experimentally determined bulk modulus of approximately 25% of the $ABO_4$ compounds with scheelite and scheelite-related structures that can be found in the Inorganic Crystal Structure Database. They fitted the $B_0$ data as a function of the A cationic charge per unit volume of the $AO_8$ polyhedra ($Z_i/d^3$) and found that these two magnitudes are correlated. This correlation follows a linear relationship, namely $B_0$ (in GPa) = *610 $Z_i/d^3$*. The difference in the prefactor between the equation of Hazen *et al.* and that of Errandonea *et al.* arises because of the fact that the last authors have included in the fit only those compounds which have scheelite or scheelite-like structures, unlike Hazen *et al.*, who had included 38 oxides and silicates having various (non-scheelite) structures. As it was shown by Panchal *et al.* [135], the fit given by Errandonea *et al.* [17] is the most appropriate for the compounds discussed in this review. Indeed the empirical relation reported in Ref. [17] has been shown to be very effective for predicting the bulk modulus of scheelite-structured $BaMoO_4$, $YCrO_4$, $ZrGeO_4$, $HfGeO_4$, and $NaAlH_4$, among other compounds. For these five materials the predicted bulk modulus is 59 GPa, 135 GPa, 229 GPa, 232 GPa, and 25 GPa, respectively, while the measured values are 57 GPa [20], 142 GPa [182], 238 GPa [135], 242 GPa [135], and 27 GPa [180], respectively. **Figure 18** and **Table X** summarizes the bulk modulus of different scheelite-structured $ABO_4$ compounds and structurally related compounds as a function of $Z_i/d^3$, including recently published results. $NaAlH_4$ and $YLiF_4$ are also included in the plot and the table, showing that the compressibility of these compounds can be also described by the phenomenological equation obtained for the oxides. There it can be clearly seen that both magnitudes ($B_0$ and $Z_i/d^3$) are linearly correlated. The linear fit obtained by Errandonea *et al.* [17], $B_0$ (in GPa) = *610 $Z_i/d^3$*, is also shown to illustrate that the empirical model proposed by these authors so far can be used to predict the bulk modulus of many $ABX_4$ compounds. In particular, since in wolframite-structured tungstates the $WO_6$ octahedra are as rigid as the $WO_4$ tetrahedra in scheelite-structured tungstates, the model of Ref. [17] apparently also works for wolframite-structured oxides, like $CdWO_4$ and $ZnWO_4$, which are also included in **Figure 18** and **Table X**.

In addition to the relation between bulk and polyhedral compressibility established by Errandonea *et al.* [17], other empirical relations have been proposed to estimate the bulk modulus of $ABO_4$ compounds. Ming *et al.* [145] have suggested that their bulk modulus can be related to their ambient pressure molar volumes ($V_0$). Assuming this correlation and the formulation of Cohen for diamond-like semiconductors [214], Scott *et al.* [191] have proposed a semi-theoretical relationship for the bulk modulus of scheelite-structured compounds. On the other hand, Westrenen *et al.* [136] have suggested that the bulk modulus of $ABO_4$ compounds varies as a function of the molar volume and the product of the formal charges on the A and B cations. However, it has been shown that these two empirical approaches cannot predict the bulk modulus of scheelite-structured compounds with the same accuracy than the model presented in Ref. [17]. In particular Scott's and Wastrenen's models show significant deviations for the vanadates, chromates, and hydrides.

To close this section, we would like to stress that from the analysis of the literature the simple rule $B_0$ (in GPa) = *610 $Z_i/d^3$* appears to be the most effective empirical criterion for predicting the bulk modulus of any scheelite or scheelite related $ABO_4$ compound. The linear relationship between $B_0$ and the A-cation charge density of the $AO_8$ polyhedra is consistent with the fact that $AO_8$ polyhedra exhibiting a large A-cation charge density result in a larger electronic cloud inside the polyhedra than those $AO_8$ polyhedra with a low A-cation charge density. In the $AO_8$ bisdisphenoids with a high $Z_i$ the electrons around the cation are highly localized and the bond distances cannot be highly deformed upon compression. On the contrary, in $AO_8$ bisdisphenoids with a low $Z_i$ the density of electrons around the cation is highly delocalized and the bond distances can be considerably deformed under pressure. Then, since the compressibility of $ABO_4$ compounds is mainly given by the compression of the $AO_8$ polyhedra, the above-described facts explain why $B_0$ is proportional to $Z_i$. In addition to that, they also explain why $AO_8$ polyhedra with A valence +1, +2, and +3 are highly deformed upon compression as compared to $BO_4$ polyhedra with B valence +7, +6, and +5 in $ABO_4$ scheelites and scheelite-related structures, being the compounds with A- and B-cation valence equal to +4 the most uncompressible $ABO_4$ materials.

The effectiveness shown by the relation proposed in Ref. [17] for predicting the bulk modulus of several compounds recommends its use for predicting the bulk modulus of additional scheelite-structured and scheelite-related compounds. Some interesting predictions of this empirical rule are: $B_0$ = 42 GPa and 27 GPa for scheelite-structured $NaReO_4$ and $KIO_4$, $B_0$ = 87 GPa and 245 GPa for zircon-structured $EuCrO_4$ and $TiGeO_4$, $B_0$ = 120 GPa and 160 GPa for wolframite-structured $MnWO_4$ and $NiWO_4$, and $B_0$ = 300 GPa for cotunnite-structured $ZrTaO_3N$, suggesting that this last oxynitride is a potential super-hard material.

## 8. Studies of the electronic properties

### 8.1. Optical absorption measurements

High pressure is not only an efficient tool to investigate the specificity of chemical bonds of materials but also to study the electronic structure of semiconductors which can be strongly affected through the tuning of their chemical bonds [215]. Because of this fact, several studies of the pressure effects in the optical properties of semiconductors have been performed in the last two decades [216, 217]. In the case of the scheelite-structured $ABO_4$ compounds, the first studies of the pressure-effects on the optical properties have been performed only in the last years [218]. These materials are wide-gap semiconductors with band-gap energies ($E_g$) larger than 3.7eV [219]. In particular, recent experiments have shown the band-gap energy of these scintillating materials to decrease following the sequence $BaWO_4$ (5.26 eV) > $SrWO_4$ (5.08 eV) > $CaWO_4$ (4.94 eV) > $HgWO_4$ (4.55 eV) > $CdWO_4$ (4.15 eV) > $MgWO_4$ (3.92 eV) > $ZnWO_4$ (3.90 eV) > $PbWO_4$ (3.77 eV). This fact makes optical-absorption measurements in scheelite-structured $ABO_4$ compounds very challenging experiments when performed using a DAC. According with electronic band-structure calculations in these materials the valence-band maxima and conduction-band minima are located at the $\Gamma$ point, so that they are direct-gap materials [219, 220]. These calculations also indicate that $2p$ O states dominate the character of the valence bands and that the bottom of the conduction bands is dominated by Mo and W $4d$ or $5d$ states. These conclusions are also supported by x-ray photoemission spectroscopy measurements [219]. Regarding the optical absorption coefficient of $ABO_4$ compounds, it is known that the absorption coefficient at the band edge is an exponential function of the photon energy, which is called the Urbach´s tail [221].

Optical absorption spectra have been measured in single crystals of $PbWO_4$ upon compression up to 15 GPa [218]. The obtained results are shown in **Figure 19**. The spectra measured at low pressures resemble those reported in literature at ambient conditions [221]. Under compression the absorption spectra red-shift, indicating that the band gap of $PbWO_4$ decreases with pressure. The pressure dependence obtained for $E_g$ from the experimental data is given in **Figure 20**. There it can be seen that up to 6.1 GPa the reduction of $E_g$ is linear, closing at a rate of −71(2) meV/GPa. A similar behaviour has been observed for the scheelite-structured $PbMoO_4$ [52]. The results can be explained in terms of the crystal chemistry and electronic structure of $PbWO_4$ and $PbMoO_4$ [218]. The observed decrease of the band gap upon compression is a direct consequence of the increase of the strength of the crystal field acting on the $4d$ and $5d$ states of Mo and W [52, 218]. Recent band-structure calculations corroborate this hypothesis. Additional changes on the band-structure of $ABO_4$ compounds are caused by pressure-induced structural changes. For example, at 6.3 GPa in $PbWO_4$ a collapse of the band-gap energy from 3.5 eV to 2.75 eV was observed [218]. This collapse is a direct consequence of the structural transformation induced by pressure in $PbWO_4$ [19, 29]. This transformation induces changes on the tungsten-oxygen tetrahedron, which are expected to affect the band structure of $PbWO_4$. From 6.3 GPa to 11.1 GPa the absorption edge of $PbWO_4$ also moves to lower energies upon compression (as does in the scheelite structure), but with a pressure coefficient of −98(3) meV/GPa. At 12.2 GPa, additional changes in the absorption edge of $PbWO_4$ are observed (see **Figure 19**). It has been argued that these additional changes can be related to the conclusion of the phase transition to the $PbWO_4$-III phase. Beyond 12.2 GPa the absorption edge also red-shifts but with a pressure coefficient of −26(2) meV/GPa. In the scheelite-structured $BaMoO_4$, two transitions also take place upon compression [20]. The second transition takes place from the ferguson ite phase to a phase related to PbWO$_4$-III. The pressure evolution of the absorption edge in this phase of BaMoO$_4$ has been measured [222] being found that it red-shifts at −21 meV/GPa, a value very close to the one observed for the PbWO$_4$-III phase. Very recently, high-pressure optical measurements have been also performed in BaWO$_4$. In this compound for the high-pressure BaWO$_4$-II phase (isomorphic to PbWO$_4$-III) the same behaviour than in BaMoO$_4$ and PbWO$_4$ was found [223]. On the other hand, for the low-pressure scheelite phase a small increase of the band-gap energy with pressure has been observed. The difference in the behaviour of E$_g$ in the pressure range of stability of the scheelite phase for BaWO$_4$ and PbWO$_4$ has been attributed to a different hybridization of the Ba and Pb states with 2$p$ O states and 5$d$ W states [223]. In spite of the difference of the pressure effects on the optical properties of the scheelite phase of PbWO$_4$ and BaWO$_4$, at the end of the pressure range of stability of this phase in both compounds a collapse of E$_g$ of about 0.7 - 1 eV has been reported [218, 223]. This fact gives additional support to the idea that the band-gap collapse is related to the occurrence of a phase transition at a similar pressure.

**8.2. Luminescence studies**

In addition to the optical-absorption measurements, other optical studies have been also performed in scheelite-structured compounds under compression. High-pressure luminescence studies have been performed by Martin *et al.* in rare-earth doped SrWO$_4$ [224, 225]. These studies are very interesting since the variation of interatomic distances by pressure-tuning allows a better understanding of the ion-ion coupling mechanisms (e.g. the Nd$^{3+}$-Nd$^{3+}$ magnetic interaction). On doping, rare-earth ions substitute for the A$^{2+}$ cations in ABO$_4$ compounds, so they are located in the polyhedra most sensitive to pressure of these compounds. Because of this fact, the luminescence properties of rare-earth doped $ABO_4$ are considerable affected by pressure.

The emission spectra for the $^4F_{3/2} \rightarrow {}^4I_{9/2}$ transition have been obtained under pressure until 13 GPa in $Nd^{3+}$:$SrWO_4$ crystals [225]. Under compression the wavelength number of the multiplets $^4F_{3/2} \rightarrow {}^4I_{9/2}$ transition increases within the range of stability of the scheelite structure due to the decrease of the Nd-Nd distances. Additional changes observed in the emission spectra confirm the occurrence of scheelite-to-fergusonite phase transition in $SrWO_4$ which is observed around 10 GPa, in good agreement with the conclusion obtained from x-ray diffraction experiments [17]. On top of that, from the analysis of the luminescence decays as function of pressure from the $^4F_{3/2}$ level, it was found that the decay curves follow a non-exponential dependence, being concluded that the energy transfer processes are enhanced under compression. These processes are probably the most important factor which explains the reduction of the effective lifetime when the pressure is increased. Very similar conclusions were extracted from studies carried out on $Eu^{3+}$:$SrWO_4$ crystals [224].

Studies on the pressure dependence of the $^4F_{3/2} \rightarrow {}^4I_{9/2}$ and the $^4F_{3/2} \rightarrow {}^4I_{11/2}$ transitions of $Nd^{3+}$ have been also performed in scheelite-structured $YLiF_4$ [160]. The high-pressure behaviour of the $^4F_{3/2} \rightarrow {}^4I_{9/2}$ satellites indicates a significant increase of the $Nd^{3+}$-$Nd^{3+}$ ferromagnetic exchange interaction as the distance between pairs decreases. The observation of satellites of the $^4F_{3/2} \rightarrow {}^4I_{11/2}$ transition suggests that the splitting of the $^4F_{3/2}$ multiplet as a consequence of the exchange interaction. The luminescence spectra of $Nd^{3+}$:$YLiF_4$ indicate that slight changes in the scheelite structure occur around 5.5 GPa. Discontinuous spectral changes near 10 GPa, also observed for Eu$^{3+}$:YLiF$_4$, were attributed to a structural phase transition [161]. Similar high-pressure effects have been observed in the zircon-structured YVO$_4$. The effect of pressure on the $^4F_{3/2}\rightarrow{}^4I_{9/2}$ and the $^4F_{3/2}\rightarrow{}^4I_{11/2}$ emissions of Nd$^{3+}$ in zircon-type YVO$_4$ were also investigated [140]. These studies show that chemical composition and hydrostatic pressures affect differently the crystal field parameters in the case of the zircon structure. This behaviour observed in scheelite- and zircon-structured compounds is in contrast to the one of the spinels and garnets. In particular, the increase of the crystal field interaction with pressure is similar in Nd$^{3+}$:YVO$_4$ and Nd$^{3+}$:YLiF$_4$. The larger pressure coefficients for photoluminescence transitions in YVO$_4$ as compared to those of YLiF$_4$ can be attributed to a larger decrease of the Slater and spin-orbit parameters, which may be related to a stronger increase of the covalency with increasing pressure on the oxide compounds. Regarding YVO$_4$, the reported studies also indicate a significant increase of the Nd$^{3+}$-Nd$^{3+}$ ferromagnetic exchange interaction as the distance between the pairs decreases.

**9. Technological and geophysical implications**

The studies under high pressure in ABX$_4$ compounds may have important technological and geophysical implications. From the technological point of view there is a current interest in producing nanocrystals of different ABX$_4$ compounds in order to increase their luminescent properties. Recently, CaWO$_4$ nanocrystals with controlled size have been prepared by a hydrothermal method [226]. Moreover, CdWO$_4$ nanocrystals have been synthesized in a tetragonal scheelite structure despite bulk CdWO$_4$ crystallizes in the monoclinic wolframite structure [227]. Additionally, high-pressure studies in combination with heavy ion bombardment at relativistic velocities in ZrSiO$_4$ have shown the formation of nanocrystals and of the scheelite phase at pressures smaller than that of the zircon-to-scheelite phase transition in the bulk [228]. On top of that, the understanding of the effects of irradiation during long time in $ABO_4$ scintillating crystals can be now better understood thanks to the improvement of knowledge of the effects on their electronic and optical properties caused by pressure-induced structural changes [229]. Another important technological application is the possible development of ultra-hard $ABO_4$ compounds by means of high-pressure and high-temperature treatments and the development of synthesis methods that could be easily transferable to the industry [230]. In particular, the potential use of these compounds as thoughener for oxide ceramic composites seems to be a very promising application [231]. Also the development of high-dielectric thin films of compounds like $TiSiO_4$ [232] can be improved with the information obtained from high-pressure studies on $ABO_4$ compounds. Finally, a new class of materials with phosphor in octahedral coordination has been recently devised [153] and new materials with boron in octahedral coordination are pursued [233].

Additionally, the information obtained from high-pressure studies has contributed to improve the knowledge of the dielectric properties of scheelite-structured materials. These materials have been shown to be high-quality microwave dielectric ceramics, being such material required for to the rapid development of mobile telecommunication systems, such as mobile phones [234]. There are many other research topics related to scheelite-structured materials that have been indirectly benefited by high-pressure studies, they include: the development of photoluminescence enhanced disordered thin films [235], the study of the layering effect of water on the structure of scheelites [236], and the study of $NaBH_4$, $NaAlH_4$ and similar metal hydrides [180] as promising hydrogen storage systems among others. In particular, the structural and electronic information obtained on $AWO_4$ compounds can be very helpful for the developing of highly-sensitive energy-resolving cryogenic detectors. The development of these cryogenic detectors is fundamental in particle physics for the detection of rare events and low background experiments [1].

From the geological point of view, many $ABO_4$ compounds, like zircon ($ZrSiO_4$), thorite ($ThSiO_4$), and hafnon ($HfSiO_4$), are part of the crust and lower mantle of the Earth, therefore it is important to understand the behaviour of these compounds under high pressure and also at high temperatures. A recent study of amorphized zircon due to natural radiation under high pressure has shown that there are phase transitions between different amorphous structures that are different than those found in crystals [237]. Conclusions drawn from high-pressure studies of scheelite-structured compounds could also have important geophysical and geochemical implications since the scheelite-structured orthotungstates are common accessory minerals in various kinds of rocks in the Earth's upper mantle. Pressures around 10 GPa and temperatures higher than 700 K (i.e. within the P-T range covered by present high-pressure studies) are found at a depth of 100 km in the upper mantle. On top of that, the results here reviewed could be important for extracting information from meteorite debris found in the Earth, which have been submitted to high P-T conditions when they impact with the Earth [238]. In the particular case of zircon, the high-pressure transformed phase with scheelite structure has been found in meteorite impact debris and has been named reidite [239]. The presence of reidite and its transformation pressure from zircon can be used as a geobarometer. Finally, the information obtained from high-pressure studies in zircons and scheelites is very important to decipher element mobility in ultrahigh-pressure eclogite-facies metamorphic rocks during subduction and exhumation of continental crust [240].

**10. Future prospects**

The Fukunaga and Yamaoka's (FY) and Bastide's diagrams as well as the crystallochemical arguments above presented can be used not only to predict high-pressure phases, but also to understand the behaviour of $ABX_4$ compounds with increasing or decreasing temperature. Therefore, it is interesting to try to predict high-pressure phases at different temperatures. As commented previously, Mariathasan *et al.* considered that the effect of temperature in $ABX_4$ compounds was contrary to that of pressure provided that the A-X polyhedra are more compressible than the B-X polyhedra [154]. This phenomena has been observed not only in $LaNbO_4$ but also in $SrMoO_4$ [134]. The NE or N rules for pressure increase in the Bastide's diagram shows that both cation polyhedra could compress in a different way depending on their cation to anion ionic radii ratios. Therefore, one cannot assume in general that temperature is opposite to pressure for all $ABX_4$ compounds. The study of the phases of $ABX_4$ compounds at different temperatures shows that the increase of temperature follows a south-west (SW) or even a west (W) path along the Bastide's diagram. For instance, $CrVO_4$ is a compound usually attributed to the orthorhombic structure *Cmcm*, however at ambient conditions crystallizes in the monoclinic *C2/m* phase [241]. At high temperature, $CrVO_4$ undergoes a phase transition above 675K to the *Cmcm* phase [242], which can be quenched at ambient conditions and it is in SW direction with respect to the *C2/m* phase. On the other hand, $LaTaO_4$ crystallizes in the monoclinic *P2$_1$/n* structure at ambient conditions [243] and undergoes a phase transition above 450 K to an orthorhombic *A2$_1$/am* phase [244], which is in W direction with respect to the *P2$_1$/n* phase. Similarly, $CsReO_4$ crystallizes in the orthorhombic *Pnma* structure at ambient conditions [245] and undergoes a phase transition above 400 K to the zircon *I4/amd* phase [246] which is in W-SW direction with respect to the *Pnma* phase of CsReO$_4$ in the Bastide's diagram. Therefore, unexplored phases at different pressures and temperatures can be estimated with the help of the systematic here presented, giving a new perspective for the high-pressure high-temperature synthesis of ABX$_4$ materials.

Furthermore, the FY and Bastide's diagrams may also allow us to understand the metastability of some phases in certain ABX$_4$ compounds at ambient conditions. For instance, the mineral crocoite (PbCrO$_4$) is observed both in the orthorhombic *Pnma* phase [247] and in the monazite monoclinic *P2$_1$/n* phase [248]. This can be understood because PbCrO$_4$ is located near the stability border of both structures in the Bastide's diagram. A similar case is that of ThGeO$_4$ which is found in both zircon and scheelite [249] phases since its location in the Bastide's diagram is near the border of stability of both structures. Other examples are those of FeVO$_4$ that crystallizes either in the triclinic $P\bar{1}$ phase or in an orthorhombic *Cmcm* phase, and of AlTaO$_4$ with several metastable phases, one of them being the orthorhombic *Pbcn* structure and another the cubic rutile-type structure. Both phases are reasonable on the light of the location of AlTaO$_4$ in the Bastide's diagram. In this sense, we have to mention that the ABX$_4$ compounds with higher polymorfism or metastable phases are those located near the borderline corresponding to AX$_2$ compounds and this opens the possibility for the search of metastable phases in this region and in the borderline region among several structures.

Finally, we want to stress that the FY and Bastide's diagrams allows us to understand why some nanocrystals can be synthesized in structures that are not stable in bulk materials. For instance, nanocrystals of CdWO$_4$ have been recently grown in the tetragonal scheelite phase despite bulk CdWO$_4$ crystallizes in the monoclinic wolframite *P2/c* phase [227]. This result can be understood by considering that the decrease of the nanocrystal size is sometimes equivalent to apply a "negative" pressure on the crystal which expands the lattice in all directions and leads to a decrease of the $r_A/r_X$ and $r_B/r_X$ ionic radii ratios. In this sense, we can understand that the nanocrystals of $CdWO_4$ crystallize in the scheelite phase since there is a small region of stability of the scheelite structure in SW direction with respect to monoclinic $CdWO_4$ in the Bastide's diagram. This example of "negative" pressure in nanocrystals opens the possibility to search for new compounds with smaller cation coordinations than those present in bulk materials. For instance, in principle it is possible to synthesize niobates, antimonates, tellurates, tantalates, and even uraniates with cation B coordination of 4 while the bulk materials usually they have a cation B coordination of 6 or even a higher coordination.

## 11. Concluding Remarks

In this contribution, we reviewed recent studies of the high-pressure effects on the structural and electronic properties of scintillating $ABX_4$ materials. In particular, the occurrence of pressure-induced phase transitions in scheelite-structured compounds and related materials have been discussed in detail. The different experimental and theoretical techniques used for obtaining reliable HP–HT data have been described. Drawbacks and advantages of the different techniques have been discussed including recent developments. Other techniques, like inelastic neutron scattering, have not been included in the review, but recent studies in $LuPO_4$ and $YbPO_4$ have shown that they could be very helpful to improve the knowledge of the physical properties of $ABX_4$ scintillating compounds [250]. Furthermore, a comparative analysis of the crystal chemistry of $ABX_4$ compounds under high-pressure was done in order to present the whole body of structural studies available in the literature in a consistent fashion, and to suggest opportunities for future work.

From the review results, it has been concluded that in the studied compounds the following systematic arises for their structural sequence: zircon → scheelite → fergusonite → denser monoclinic phases → orthorhombic phases. This conclusion has been confirmed by a recent work by Mittal *et al.* in $LuVO_4$ [251]. This authors reported by the first time the occurrence of the zircon → scheelite → fergusonite transition sequence in a zircon-structured compound. The implications of the reviewed results for technological applications of $ABX_4$ compounds and in geophysics were also discussed and probable trends for the future research on $ABX_4$ compounds presented. An issue of particular interest for the future could be the study of pressure effects on rare-earth compounds like scheelite(zircon)-like $HoCrO_4$ [252]. In these compounds magnetic transitions can be induced by pressure and interesting physical phenomena can be triggered by the pressure-driven *f*-electron delocalisation induced in the rare-earth metals [253].

**Acknowledgments**

This study was made possible through financial support from the Spanish government MCYT under Grants Numbers MAT2007-65990-C03-01 and MAT2006-02279 and the project MALTA-Consolider Ingenio 2010 CSD-2007-00045. D.E. acknowledges the financial support from the MCYT of Spain through the "Ramón y Cajal" program. He is also indebted to the Fundación de las Artes y las Ciencias de Valencia for granting him the IDEA prize. The authors are grateful to all the collaborators with whom they have the pleasure to interact along the years. In particular, they acknowledge the collaboration in their research on the reviewed subject of: J. Pellicer-Porres, A. Segura, D. Martínez-García, Ch. Ferrer-Roca, J. Ruiz-Fuertes, R. Lacomba-Perales, J. López-Solano, P. Rodríguez-Hernández, A. Muñoz, S. Radescu, A. Mujica, N. Garro, O. Tschauner, R. Kumar, M. Somayazulu,

D. Häusermann, I. R. Martin, U. R. Rodrıguez-Mendoza, F. Lahoz, M. E. Torres, V. Lavın, P. Lecoq, C. Y. Tu, P. Bohacek, J. M. Recio, and A. Vegas.

**Table I:** Structural parameters of the scheelite $CaWO_4$ [60].

|     | Site | $x$    | $y$    | $z$    |
|-----|------|--------|--------|--------|
| Ca  | 4b   | 0      | 0.25   | 0.625  |
| W   | 4a   | 0      | 0.25   | 0.125  |
| O   | 16f  | 0.2414 | 0.0993 | 0.0394 |

**Table II:** Equilibrium volume, bulk modulus, and first pressure derivate of the bulk modulus all at normal conditions for different scheelite-structured $AWO_4$ compounds. These values were fitted from experimental volume-pressure data.

| Compound      | $V_0$           | $B_0$      | $B_0$'   |
|---------------|-----------------|------------|----------|
| $CaWO_4$[a]   | 312(1) Å$^3$    | 74(7) GPa  | 5.6(9)   |
| $SrWO_4$[a]   | 347.4(9) Å$^3$  | 63(7) GPa  | 5.2(9)   |
| $EuWO_4$[b]   | 348.9(8) Å$^3$  | 65(6) GPa  | 4.6(9)   |
| $PbWO_4$[c]   | 357.8(6) Å$^3$  | 66(5) GPa  | 5.6(9)   |
| $BaWO_4$[c]   | 402.8(9) Å$^3$  | 52(5) GPa  | 5(1)     |

[a] Reference [17], [b] Reference [32], and [c] Reference [19]

**Table III:** Frequency, pressure coefficient, Grüneisen parameter, and symmetry of the Raman modes observed in scheelite-type AWO$_4$ (A = Ca, Sr, Ba, Pb) at room temperature, as obtained from linear fits to experimental data. Grüneisen parameters, $\gamma = B_0/\omega(0) \cdot d\omega/dP$, have been calculated with the zero-pressure bulk moduli, $B_0$, reported in Table II.

| | CaWO$_4$ [a] | | | SrWO$_4$ [b] | | | BaWO$_4$ [c] | | | PbWO$_4$ [d] | | |
|---|---|---|---|---|---|---|---|---|---|---|---|---|
| Peak /mode | $\omega(0)$ cm$^{-1}$ | $d\omega/dP$ cm$^{-1}$/GPa | $\gamma$ | $\omega(0)$ cm$^{-1}$ | $d\omega/dP$ cm$^{-1}$/GPa | $\gamma$ | $\omega(0)$ cm$^{-1}$ | $d\omega/dP$ cm$^{-1}$/GPa | $\gamma$ | $\omega(0)$ cm$^{-1}$ | $d\omega/dP$ cm$^{-1}$/GPa | $\gamma$ |
| T(B$_g$) | 84 | -0.4 | -0.35 | 75 | -0.4 | -0.34 | 63 | -0.8 | -0.67 | 58 | -1.1 | -1.30 |
| T(E$_g$) | 116 | 1.7 | 1.08 | 102 | 1.3 | 0.80 | 74 | 1.0 | 0.73 | 65 | 1.8 | 1.90 |
| T(B$_g$) | 227 | 4.7 | 1.53 | 171 | 3.4 | 1.25 | 133 | 4.1 | 1.58 | 77 | 3.3 | 2.80 |
| T(E$_g$) | 196 | 3.7 | 1.40 | 135 | 2.9 | 1.35 | 101 | 3.3 | 1.70 | 90 | 2.3 | 1.60 |
| R(A$_g$) | 212 | 3.8 | 1.33 | 190 | 4.4 | 1.46 | 150 | 4.2 | 1.47 | 178 | 3.3 | 1.20 |
| R(E$_g$) | 276 | 7.0 | 1.88 | 238 | 6.8 | 1.80 | 191 | 6.3 | 1.71 | 193 | 4.2 | 1.40 |
| $\nu_2$(A$_g$) | 334 | 2.5 | 0.55 | 337 | 3.3 | 0.62 | 331 | 2.5 | 0.40 | 323 | 1.9 | 0.40 |
| $\nu_2$(B$_g$) | 334 | 2.5 | 0.55 | 337 | 3.3 | 0.62 | 332 | 3.0 | 0.46 | 328 | 2.1 | 0.40 |
| $\nu_4$(B$_g$) | 402 | 4.1 | 0.75 | 371 | 4.1 | 0.70 | 344 | 2.0 | 0.30 | 357 | 2.8 | 0.50 |
| $\nu_4$(E$_g$) | 406 | 4.6 | 0.84 | 380 | 4.6 | 0.76 | 352 | 3.4 | 0.50 | 362 | 2.7 | 0.50 |
| $\nu_3$(E$_g$) | 797 | 3.0 | 0.28 | 799 | 3.0 | 0.24 | 795 | 3.2 | 0.21 | 752 | 2.4 | 0.20 |
| $\nu_3$(B$_g$) | 838 | 1.9 | 0.17 | 837 | 2.1 | 0.16 | 831 | 2.0 | 0.12 | 766 | 0.9 | 0.08 |
| $\nu_1$(A$_g$) | 911 | 1.5 | 0.12 | 921 | 2.2 | 0.15 | 926 | 2.7 | 0.15 | 906 | 0.8 | 0.06 |

[a] Ref. 51, [b] Ref. 26, [c] Ref. 28, [d] Ref. 29.

**Table IV:** Elastic constants of CaWO$_4$. Data taken from Ref. [96].

| $C_{11}$ | $C_{33}$ | $C_{44}$ | $C_{66}$ | $C_{12}$ | $C_{13}$ |
|---|---|---|---|---|---|
| 145 GPa | 127 GPa | 33.5 GPa | 39.9 GPa | 60 GPa | 41 GPa |

**Table V:** Calculated equilibrium volume, bulk modulus, and first pressure derivate of the bulk modulus all at normal conditions for different scheelite-structured AWO$_4$ compounds.

| Compound | $V_0$ | $B_0$ | $B_0$' |
|---|---|---|---|
| CaWO$_4$[a] | 318.3 Å$^3$ | 72 GPa | 4.3 |
| SrWO$_4$[a] | 362.2 Å$^3$ | 62 GPa | 4.9 |
| PbWO$_4$[b] | 376 Å$^3$ | 66 GPa | 4.7 |
| BaWO$_4$[b] | 402 Å$^3$ | 52 GPa | 5 |

[a] Reference [17] and [b] Reference [19]

**Table VI:** Structural parameters of the fergusonite-type CaWO$_4$. (a) Ref. [14], P = 11.2 GPa, $a$ = 5.0706(5) Å, $b$ = 10.8528(8) Å, $c$ = 5.0812(9) Å, and β = 90.082(13)°, pressure medium: helium. (b) Ref. [17], P = 11.3 GPa, $a$ = 5.069(2) Å, $b$ = 10.851(5) Å, $c$ = 5.081(7) Å, and β = 90.091(9)°, pressure medium: silicone oil.

| (a) | Site | $x$ | $y$ | $z$ |
|---|---|---|---|---|
| Ca | 4e | 0.25 | 0.6098(10) | 0 |
| W | 4e | 0.25 | 0.1320(5) | 0 |
| O$_1$ | 8f | 0.9214(44) | 0.9661(17) | 0.2327(29) |
| O$_2$ | 8f | 0.4807(23) | 0.2164(16) | 0.8529(51) |
| (b) | Site | $x$ | $y$ | $z$ |
| Ca | 4e | 0.25 | 0.6100(8) | 0 |
| W | 4e | 0.25 | 0.1325(3) | 0 |
| O$_1$ | 8f | 0.9309(39) | 0.9684(23) | 0.2421(24) |
| O$_2$ | 8f | 0.4850(35) | 0.2193(31) | 0.8637(37) |

**Table VII:** Frequency, pressure coefficient, Grüneisen parameter, and symmetry of the Raman-active modes observed in the high-pressure fergusonite phase in $AWO_4$ (A= Ca, Sr, Ba, Pb), as obtained from linear fits to experimental data. Raman frequencies of fergusonite-like $HgWO_4$ are also shown for comparison.

| Peak /mode | $CaWO_4$ (15 GPa)[a] ω cm$^{-1}$ | dω/dP cm$^{-1}$/GPa | $SrWO_4$ (15 GPa)[b] ω cm$^{-1}$ | dω/dP cm$^{-1}$/GPa | $BaWO_4$ (7.5 GPa)[c] ω cm$^{-1}$ | dω/dP cm$^{-1}$/GPa | $PbWO_4$ (9 GPa)[d] ω cm$^{-1}$ | dω/dP cm$^{-1}$/GPa | $HgWO_4$ (1 bar)[e] ω cm$^{-1}$ |
|---|---|---|---|---|---|---|---|---|---|
| $T(B_g)$ | 120 | 2.2 | 110 | 3.4 | 37.5 | -0.6 | 44 | 0.8 | |
| $T(A_g)$ | | | | | 59 | 0.8 | 53 | 0.4 | |
| $T(B_g)$ | | | | | 67 | 0.6 | 85 | 1.4 | |
| $T(B_g)$ | 180 | 2.0 | | | 93 | 2.2 | 136 | 3.7 | |
| $T(A_g)$ | 225 | 1.8 | 180 | 2.5 | 118 | 1.2 | 144 | 5.5 | |
| $T(B_g)$ | 250 | 2.0 | 220 | 2.0 | 161 | 1.8 | 158 | 4.0 | |
| $R(B_g)$ | 270 | 2.7 | | | 192 | 3.3 | | | 200 |
| $R(B_g)$ | 288 | 1.8 | | | | | | | 235 |
| $R(A_g)$ | 367 | 1.6 | 260 | 2.7 | | | 261 | 1.9 | 285 |
| $\nu_2(A_g)$ | 368 | 3.1 | 361 | 2.5 | 338 | - | 320 | 1.9 | 300 |
| $\nu_2(A_g)$ | 392 | 2.7 | 400 | 2.5 | | | 352 | 1.7 | 335 |
| $\nu_4(A_g)$ | 460 | 3.0 | 460 | 1.0 | | | 396 | 0.9 | 380 |
| $\nu_4(B_g)$ | 477 | 3.4 | 480 | 3.0 | | | 446 | 3.2 | 515 |
| $\nu_4(B_g)$ | 500 | 5.0 | 500 | 6.5 | | | 471 | 5.6 | 540 |
| $\nu_3(A_g)$ | 800 | 1.2 | 790 | 0.0 | 826 | 1.8 | 693 | 1.0 | 705 |
| $\nu_3(B_g)$ | | | | | 839 | 4.1 | 725 | 3.3 | 815 |
| $\nu_3(B_g)$ | 865 | 3.3 | 860 | 3.9 | 859 | 0.5 | 779 | 3.6 | 850 |
| $\nu_1(A_g)$ | 950 | 3.1 | 950 | 2.9 | 940 | 0.5 | 872 | 0.9 | 930 |

[a]Obtained from Ref. 51, [b]Obtained from Ref. 26, [c]Ref. 28, [d]Ref. 29, [e]Ref. 105.

**Table VIII:** Calculated structural parameters of BaWO$_4$-II phase at 9.3 GPa. Space group *P2$_1$/n*, Z = 8, *a* = 12.7173 Å, *b* = 6.9816 Å, *c* = 7.4357 Å, and β = 91.22º. Data reproduced from Ref. [19].

|       | Site | *x*    | *y*    | *z*    |
|-------|------|--------|--------|--------|
| Ba$_1$ | 4e   | 0.1617 | 0.6555 | 0.1633 |
| Ba$_2$ | 4e   | 0.1349 | 0.9574 | 0.6316 |
| W$_1$  | 4e   | 0.0825 | 0.1633 | 0.0836 |
| W$_2$  | 4e   | 0.0912 | 0.4609 | 0.6497 |
| O$_1$  | 4e   | 0.1078 | 0.0279 | 0.2876 |
| O$_2$  | 4e   | 0.1845 | 0.6029 | 0.7777 |
| O$_3$  | 4e   | 0.0490 | 0.6510 | 0.4746 |
| O$_4$  | 4e   | 0.2128 | 0.2676 | 0.0618 |
| O$_5$  | 4e   | 0.0579 | 0.2693 | 0.8186 |
| O$_6$  | 4e   | 0.1783 | 0.3319 | 0.5103 |
| O$_7$  | 4e   | 0.0198 | 0.3756 | 0.1829 |
| O$_8$  | 4e   | 0.0789 | 0.9201 | 0.9539 |

**Table IX:** Phase transition pressures and $BX_4/A$ ratios for some scheelite compounds.

| Compound | $BX_4/A$ ratio | $P_T$ (GPa) | Reference |
|---|---|---|---|
| $KIO_4$ | 1.39 | 6.5 | [58] |
| $RbIO_4$ | 1.25 | 5.3 | [59] |
| $AgReO_4$ | 1.90 | 13±1 | [147] |
| $TlReO_4$ | 1.84 | 10 | [179] |
| $KReO_4$ | 1.45 | 7.5 | [56] |
| $RbReO_4$ | 1.30 | 1.6 | [56] |
| $CaWO_4$ | 1.89 | 11±1 | [14, 17] |
| $SrWO_4$ | 1.76 | 10.5±2 | [17, 26, 51] |
| $EuWO_4$ | 1.76 | 8.5±1 | [32] |
| $PbWO_4$ | 1.66 | 6.5±2.5 | [19, 29, 52] |
| $BaWO_4$ | 1.47 | 7±0.5 | [15, 19, 28, 51] |
| $CdMoO_4$ | 2.03 | 12 | [67] |
| $CaMoO_4$ | 1.88 | 10±1.5 | [16, 85] |
| $SrMoO_4$ | 1.74 | 12.2±1 | [66, 134] |
| $PbMoO_4$ | 1.64 | 6.5±3.3 | [43, 52] |
| $BaMoO_4$ | 1.46 | 5.8 | [20, 27] |
| $CaZnF_4$ | 1.97 | 10 | [159] |
| $YLiF_4$ | 2.11 | 11±1.1 | [80] |
| $GdLiF_4$ | 2.01 | 11 | [117] |
| $LuLiF_4$ | 1.93 | 10.7 | [178] |
| $NaAlH_4$ | 2.40 | 14 | [180] |
| $NaSm(WO_4)_2$ | 1.88 | 10 | [181] |
| $NaTb(WO_4)_2$ | 1.90 | 10 | [181] |
| $NaHo(WO_4)_2$ | 1.93 | 10 | [181] |

**Table X:** Summary of the data plotted in **Figure 18**. The structure, A-O bond distance, cation formal charge, and bulk modulus are given.

| ABO$_4$ compound | Space Group | mean A-O bond distance [Å] | cation formal charge | B$_0$ [GPa] | Reference |
|---|---|---|---|---|---|
| HfGeO$_4$ | I4$_1$/a | 2.196 | 4 | 242(6) | [135] |
| ZrGeO$_4$ | I4$_1$/a | 2.203 | 4 | 238(6) | [135] |
| ZrSiO$_4$ | I4$_1$/a | 2.243 | 4 | 230(18) | [144] |
| ZrSiO$_4$ | I4$_1$/amd | 2.198 | 4 | 215(15) | [136, 144, 202, 203] |
| LaNbO$_4$ | I4$_1$/a | 2.505 | 3 | 111(3) | [154] |
| YVO$_4$ | I4$_1$/a | 2.387 | 3 | 138(9) | [139] |
| TbVO$_4$ | I4$_1$/amd | 2.369 | 3 | 149(5) | [204] |
| BiVO$_4$ | I4$_1$/a | 2.350 | 3 | 150(5) | [206] |
| DyVO$_4$ | I4$_1$/amd | 2.354 | 3 | 160(5) | [205] |
| YCrO$_4$ | I4$_1$/amd | 2.391 | 3 | 135(5) | [116] |
| YVO$_4$ | I4$_1$/amd | 2.348 | 3 | 130(3) | [139] |
| ErVO$_4$ | I4$_1$/amd | 2.341 | 3 | 136(9) | [207] |
| LuPO$_4$ | I4$_1$/amd | 2.306 | 3 | 166(9) | [208] |
| YLiF$_4$ | I4$_1$/a | 3.044 | 3 | 81(6) | [80] |
| BaSO$_4$ | Pnma | 2.879 | 2 | 58(5) | [209, 210] |
| BaWO$_4$ | I4$_1$/a | 2.678 | 2 | 57(4) | [15, 19] |
| BaMoO$_4$ | I4$_1$/a | 2.679 | 2 | 59(6) | [20] |
| PbWO$_4$ | I4$_1$/a | 2.579 | 2 | 64(2) | [19, 21, 60] |
| PbMoO$_4$ | I4$_1$/a | 2.576 | 2 | 64(2) | [60] |
| SrWO$_4$ | I4$_1$/a | 2.557 | 2 | 63(7) | [17] |
| EuWO$_4$ | I4$_1$/a | 2.557 | 2 | 71(6) | [32] |
| SrMoO$_4$ | I4$_1$/a | 2.556 | 2 | 73(5) | [134, 211] |
| NaY(WO$_4$)$_2$ | I4$_1$/a | 2.478 | 2 | 77(8) | [212] |
| CaMoO$_4$ | I4$_1$/a | 2.458 | 2 | 82(7) | [16, 60] |
| CaWO$_4$ | I4$_1$/a | 2.457 | 2 | 75(7) | [14, 17, 60, 68, 210] |
| SrSO$_4$ | Pnma | 2.452 | 2 | 82(5) | [211] |
| CdMoO$_4$ | I4$_1$/a | 2.419 | 2 | 104(2) | [60] |
| ZnWO$_4$ | P2/c | 2.112 | 2 | 140 | [70] |
| CdWO$_4$ | P2/c | 2.197 | 2 | 120(8) | [62] |
| KReO$_4$ | I4$_1$/a | 2.791 | 1 | 18(6) | [213] |
| TlReO$_4$ | Pnma | 2.765 | 1 | 26(4) | [57] |
| AgReO$_4$ | I4$_1$/a | 2.524 | 1 | 31(6) | [147] |
| NaAlH$_4$ | I4$_1$/a | 2.850 | 1 | 27(4) | [180] |

**Figure Captions**

**Figure 1:** Schematic view of the scheelite structure. The drawing was done using the structural parameters of CaWO$_4$ given in **Table I**. Large circles represent the Ca atoms, middle-size circles correspond to the W atoms, and the small circles are the O atoms. The unit cell, Ca-O bonds, and W-O bonds are also shown together with a WO$_4$ tetrahedron and a CaO$_8$ dodecahedron.

**Figure 2:** Selected x-ray powder patterns of CaWO$_4$ at different pressures. Differences between the spectra collected at 9.7 GPa and 11.3 GPa illustrate the occurrence of the scheelite-to-fergusonite transition.

**Figure 3:** Pressure dependence of the lattice parameters, volume and axial ratio of CaWO$_4$ and SrWO$_4$. Solid squares [14, 18], circles [17], and diamonds [68] correspond to data for the scheelite phase and open circles [17] and squares [14, 18] to data for the fergusonite phase. The stars [71] and crosses [44, 81] represent atmospheric pressure data. The solid lines represent the EOS described in the text and a fit to the pressure dependence of the axial ratios.

**Figure 4:** W-O and Pb-O distances in scheelite and fergusonite PbWO$_4$. Theoretical W-O (Pb-O) distances are represented by solid and empty squares (circles) in the scheelite and fergusonite phases, respectively. Experimental W-O (Pb-O) distances are represented by solid triangles (diamonds). The dashed line indicates the onset of the scheelite-to-fergusonite transition. The dotted line indicates the pressure where the W-O coordination becomes 4+2.

**Figure 5:** Experimental XANES spectra (W $L_3$-edge) of $PbWO_4$ at different pressures. A spectrum collected on pressure release is marked with d. The analysis of the spectra reveals a coordination change associated with a pressure-induced phase transition between 9 GPa and 10.5 GPa. Subtle changes in the spectra suggest a second phase transition in $PbWO_4$ at 16.7 GPa. Figure reproduced from Ref. [19].

**Figure 6:** Raman spectra in $PbWO_4$ at three different pressures. Arrows show the main peaks of the scheelite, fergusonite and $PbWO_4$-III structures. The dashed line indicates the position of a plasma line used for spectral calibration. The asterisk indicates the position of a Raman mode arising from the pressure-transmitting medium.

**Figure 7:** Total-energy versus volume *ab initio* calculations for $BaWO_4$. Different candidate structures are indicated in the plot. The inset extends the pressure range to the region where the $BaWO_4$-II structure becomes unstable. Figure reproduced from Ref. [19].

**Figure 8:** ADXRD pattern of $CaWO_4$ at 11.3 GPa which has been assigned to the fergusonite structure. The background was subtracted. Black symbols: experimental observations, solid line: refined model, and dotted line: the difference between the measured data and the refined profile. The bars indicate the calculated positions of the reflections.

**Figure 9:** Schematic view of fergusonite $PbWO_4$ at two different pressures. In order to illustrate the structural distortion induced by pressure, figure (a) shows the fergusonite structure at 7.9 GPa where W is tetrahedrally coordinated while figure (b) shows the fergusonite structure at 9.5 GPa where W is octahedrally coordinated [19]. Large circles represent the Pb atoms, middle-size circles correspond to the W atoms, and the small circles

are the O atoms. The unit cell, Pb-O bonds, and W-O bonds are also shown together with the W-O and Pb-O polyhedra.

**Figure 10:** Pressure dependence of the Raman active modes in the scheelite, fergusonite, and PbWO$_4$–III phases of PbWO$_4$.

**Figure11:** ADXRD pattern of BaWO$_4$ at 10.9 GPa which has been assigned to the BaWO$_4$-II structure. The background was subtracted. Black symbols: experimental observations, solid line: refined model, and dotted line: the difference between the measured data and the refined profile. The bars indicate the calculated positions of the reflections.

**Figure 12:** Schematic view of the structure of the BaWO$_4$-II-type phase. The drawing was done using the structural parameters of BaWO$_4$ at 9.3 GPa given in **Table VIII**. Large circles represent the Ba atoms, middle-size circles correspond to the W atoms, and the small circles are the O atoms. The unit cell, Ba-O bonds, and W-O bonds are also shown together with the W-O and Ba-O polyhedra.

**Figure 13:** Schematic HP-HT phase diagram of BaWO$_4$. The solid line was taken from Ref. [46] and the room temperature transition pressures from Ref. [19]. The dotted line is the melting curve and the dotted lines postulated phase boundaries. The existence of a new HP-HT phase is also indicated.

**Figure 14:** Phase transition pressure in several scheelites as a function of the BX$_4$/A ratio. The symbols correspond to the data summarized in **Table IX**, the solid lines correspond to the fitted linear relation and to its lower and higher deviations.

**Figure 15:** Updated Bastide's diagram for $ABX_4$ compounds. The dashed lines show the evolution of the ionic radii ratios with increasing pressure in a number of scheelite-structured compounds.

**Figure 16:** Evolution of structures with increasing pressure for $ABX_4$ compounds depending on their $r_A/r_X$ ratios. Arrows show the most likely phase transitions to occur in $ABX_4$ compounds.

**Figure 17:** Correlation between the spontaneous strain $\varepsilon_s$ and the Landau order parameter $\eta'$ in $CaWO_4$.

**Figure 18:** Values of the ambient-pressure bulk modulus of $ABO_4$ scheelite and scheelite-related compounds plotted against the value of the cation charge density of the $AO_8$ polyhedra.

**Figure 19:** Optical-absorption spectra of scheelite $PbWO_4$ single crystals for different pressures.

**Figure 20:** Pressure dependence of the absorption edge of $PbWO_4$.

**Figure 1**

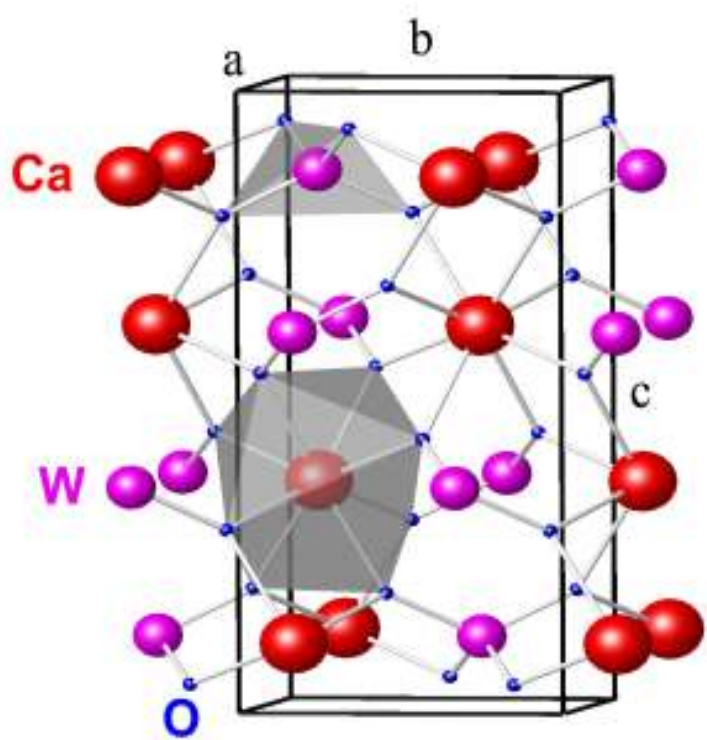

**Figure 2**

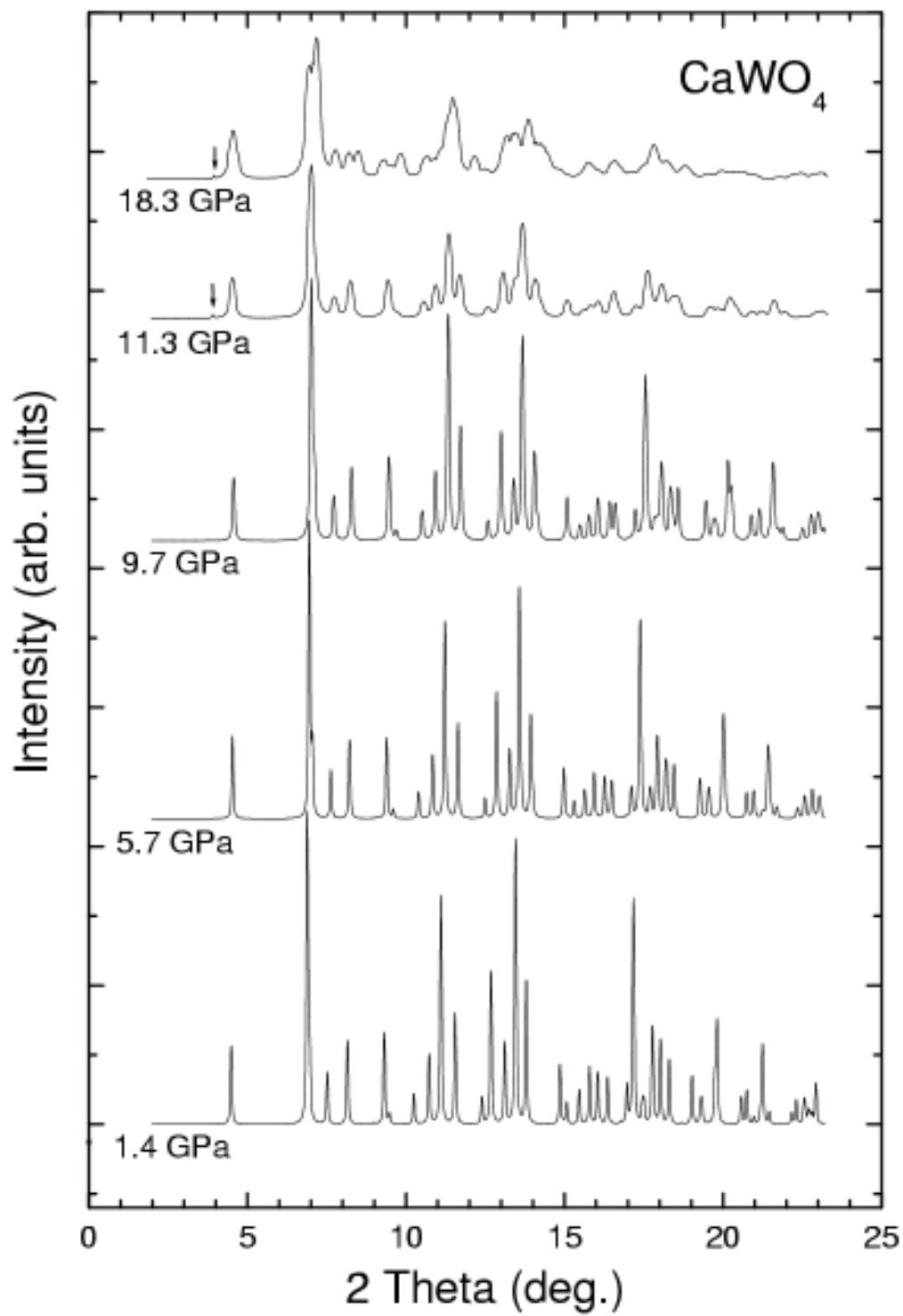

**Figure 3**

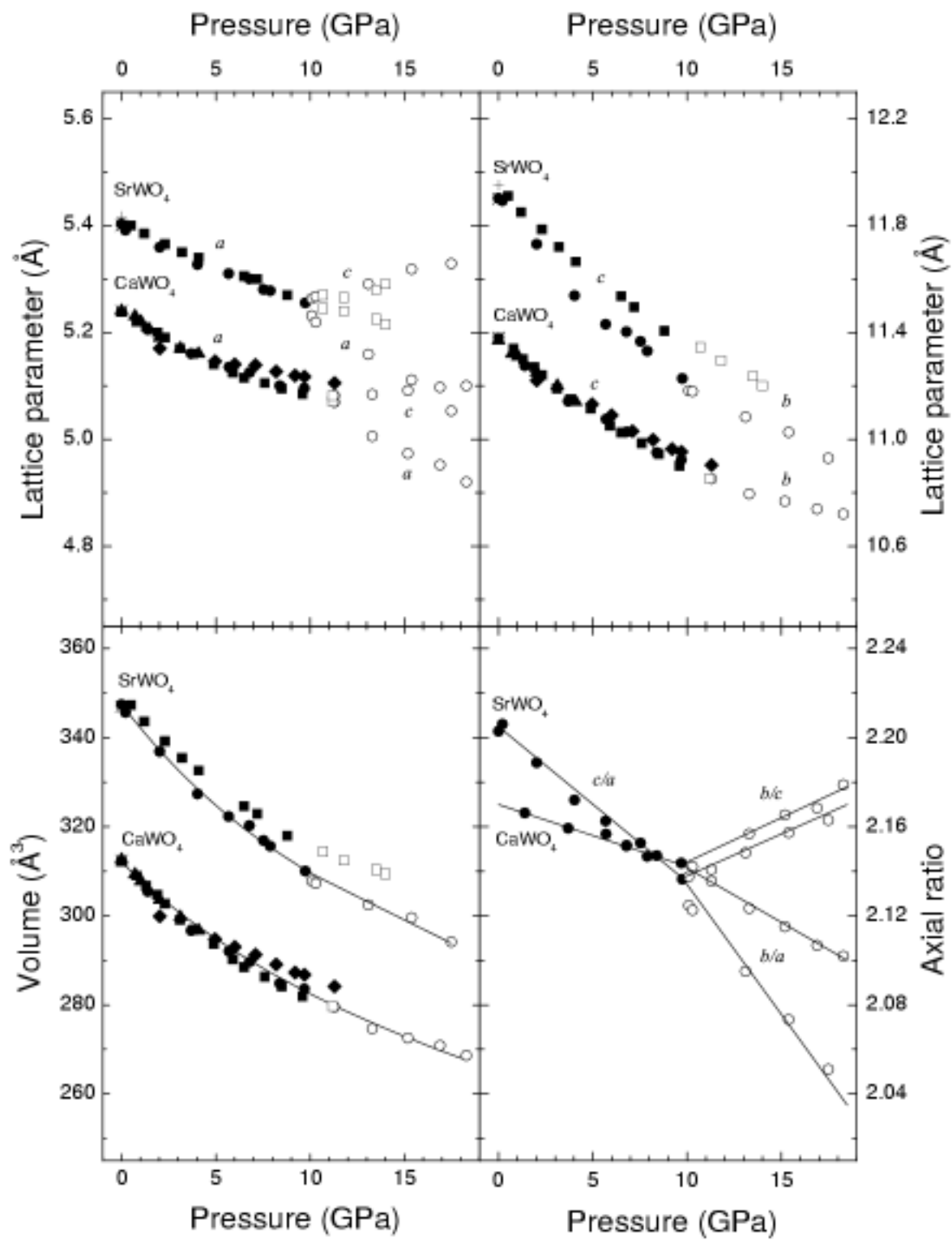

**Figure 4**

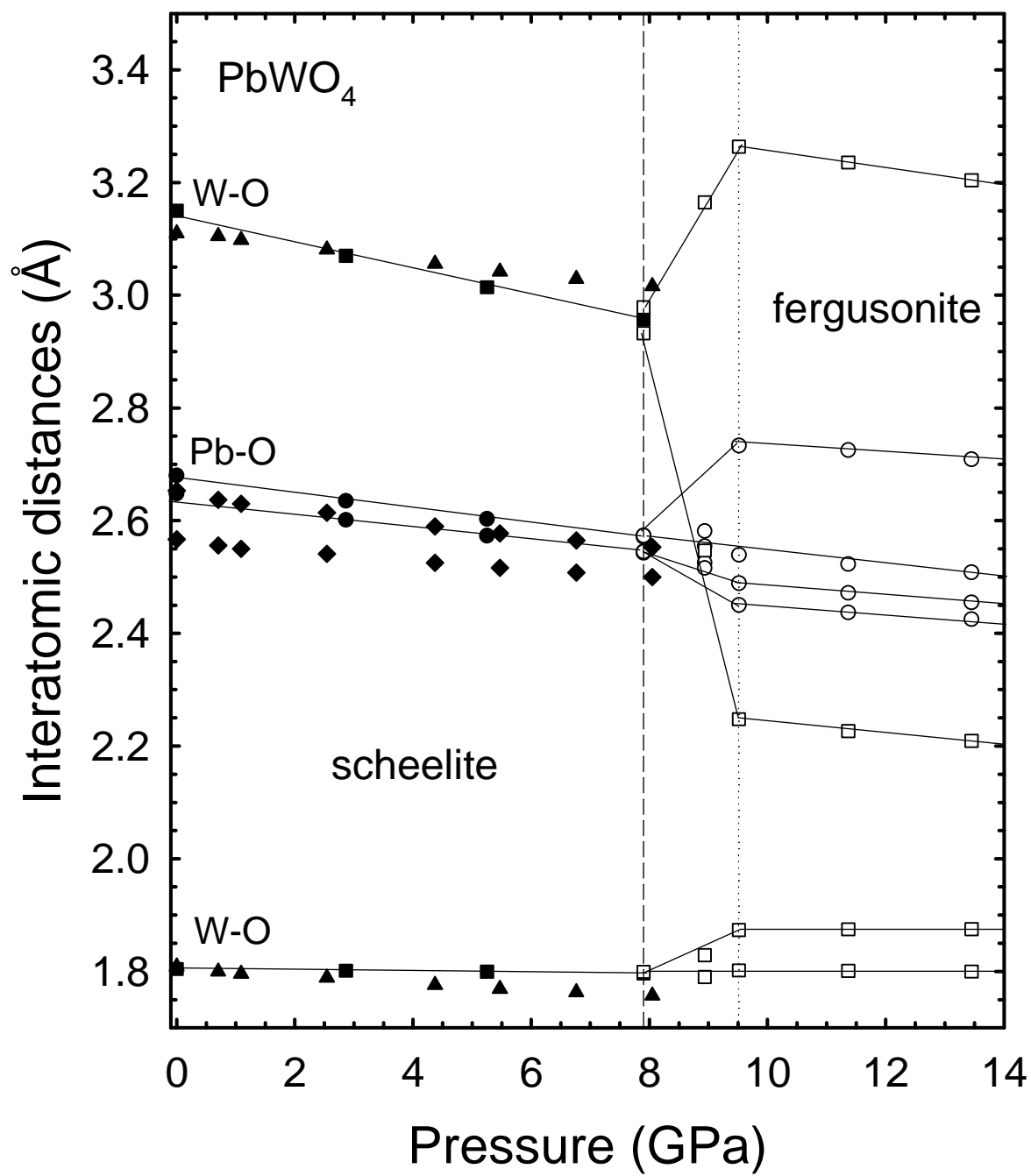

**Figure 5**

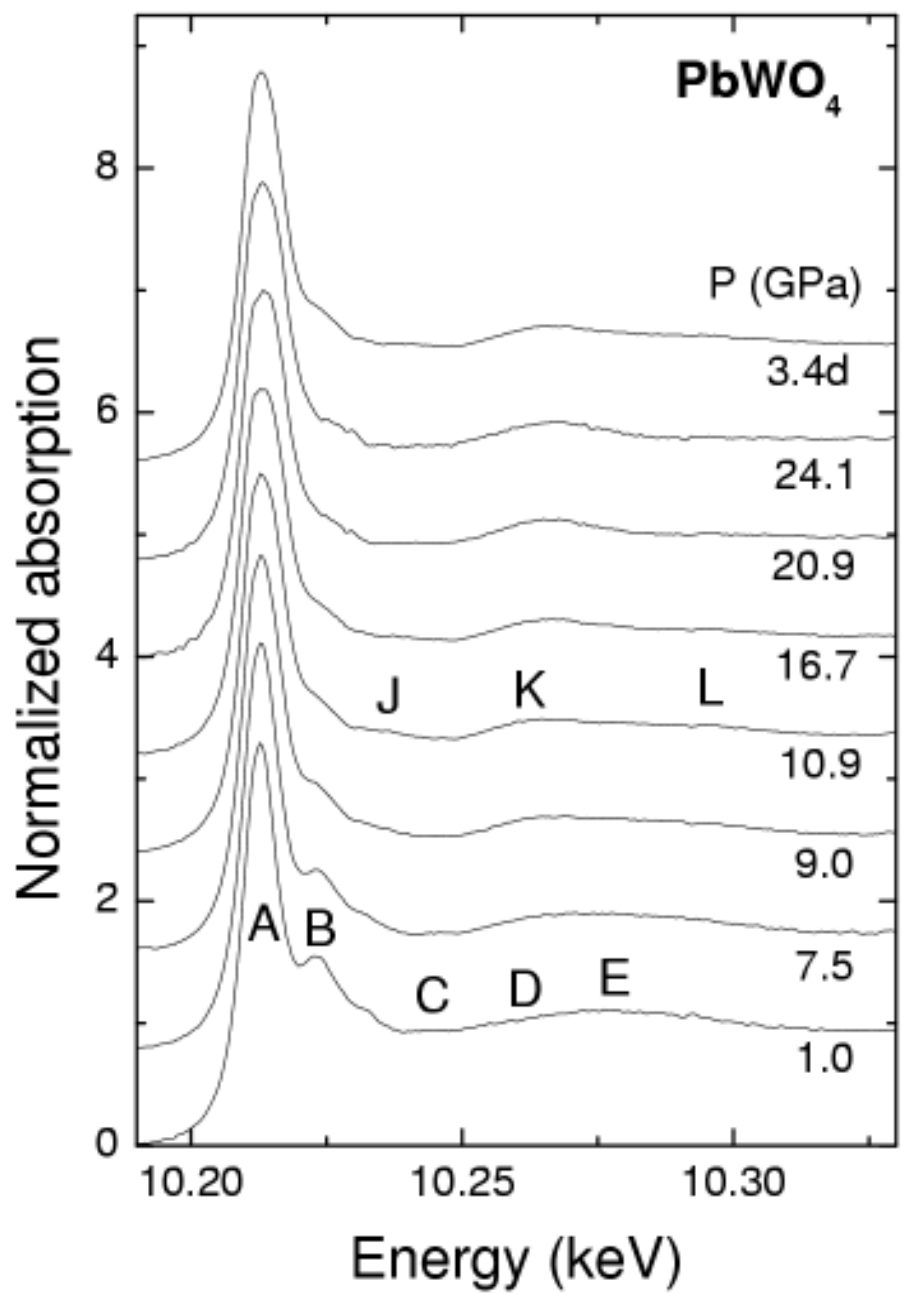

**Figure 6**

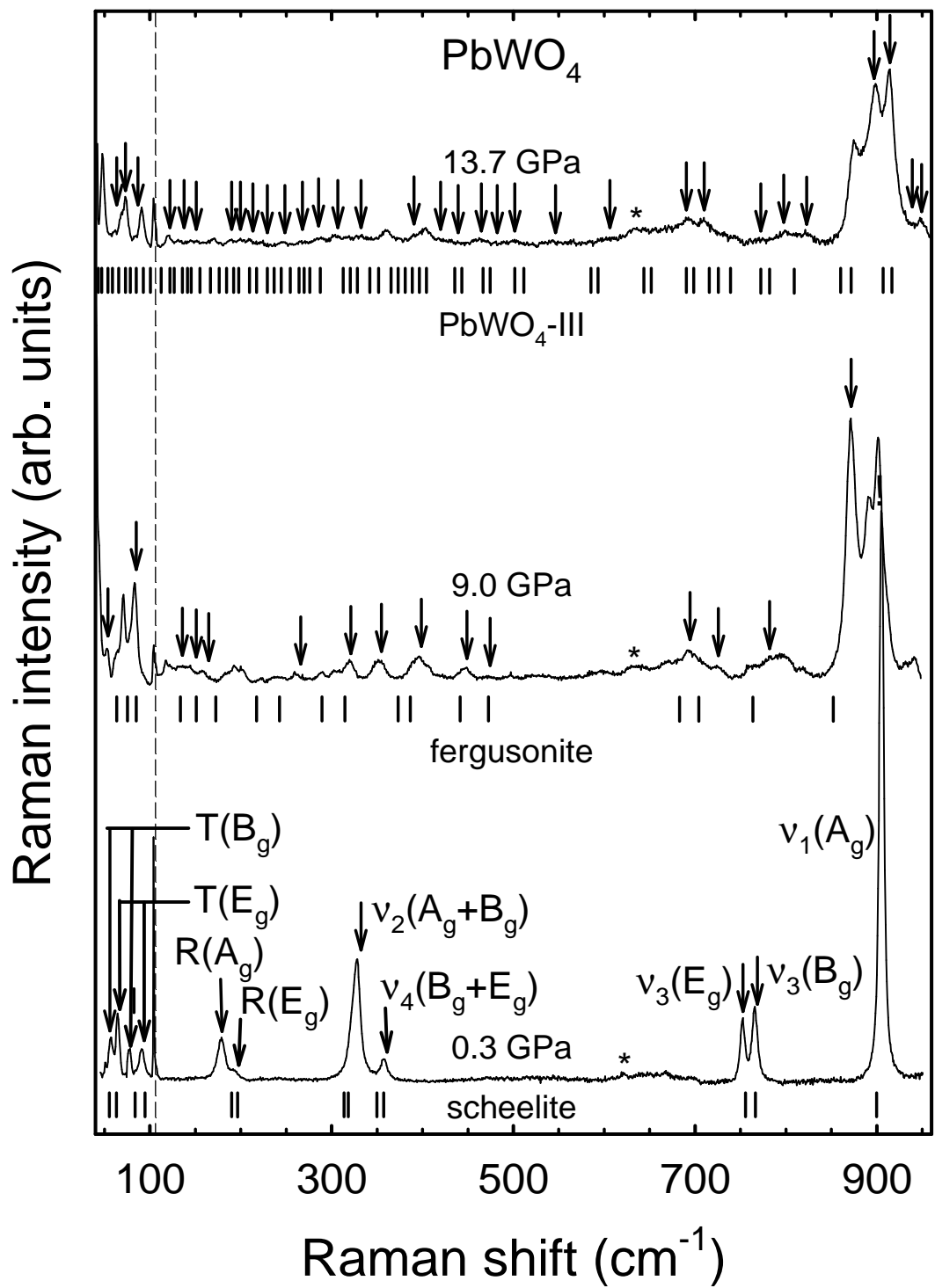

**Figure 7**

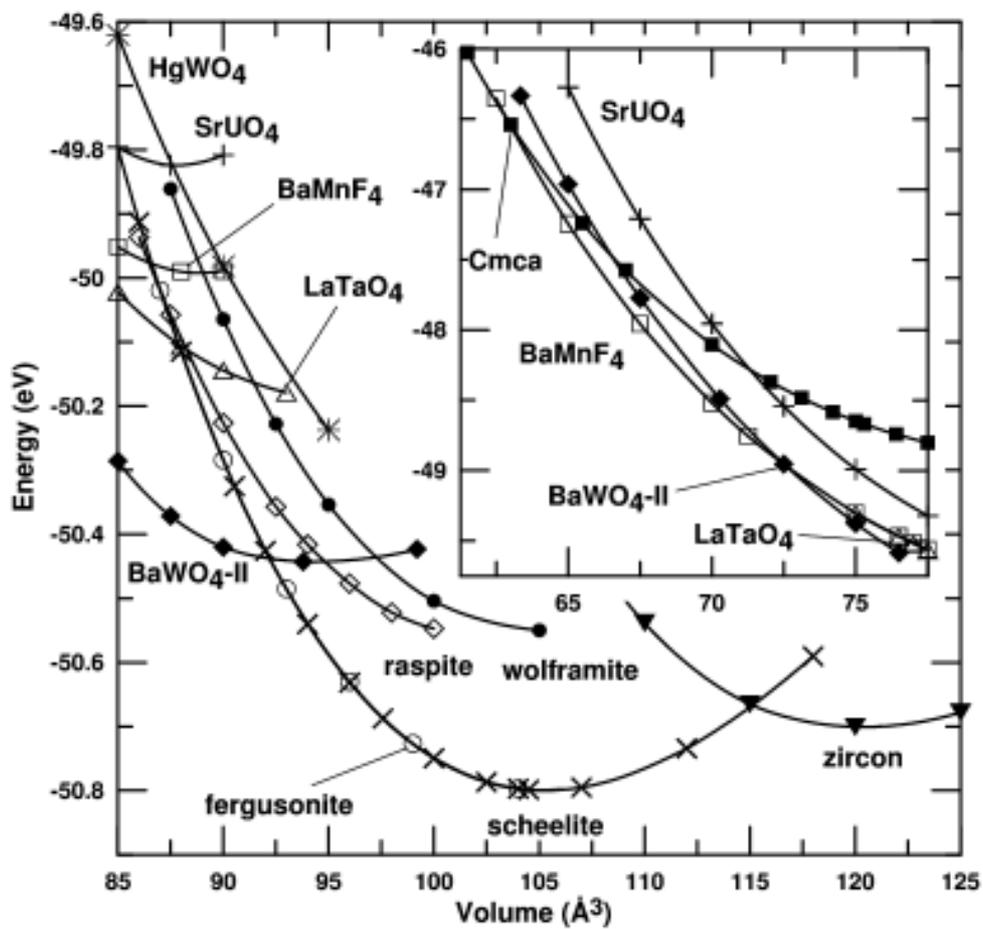

**Figure 8**

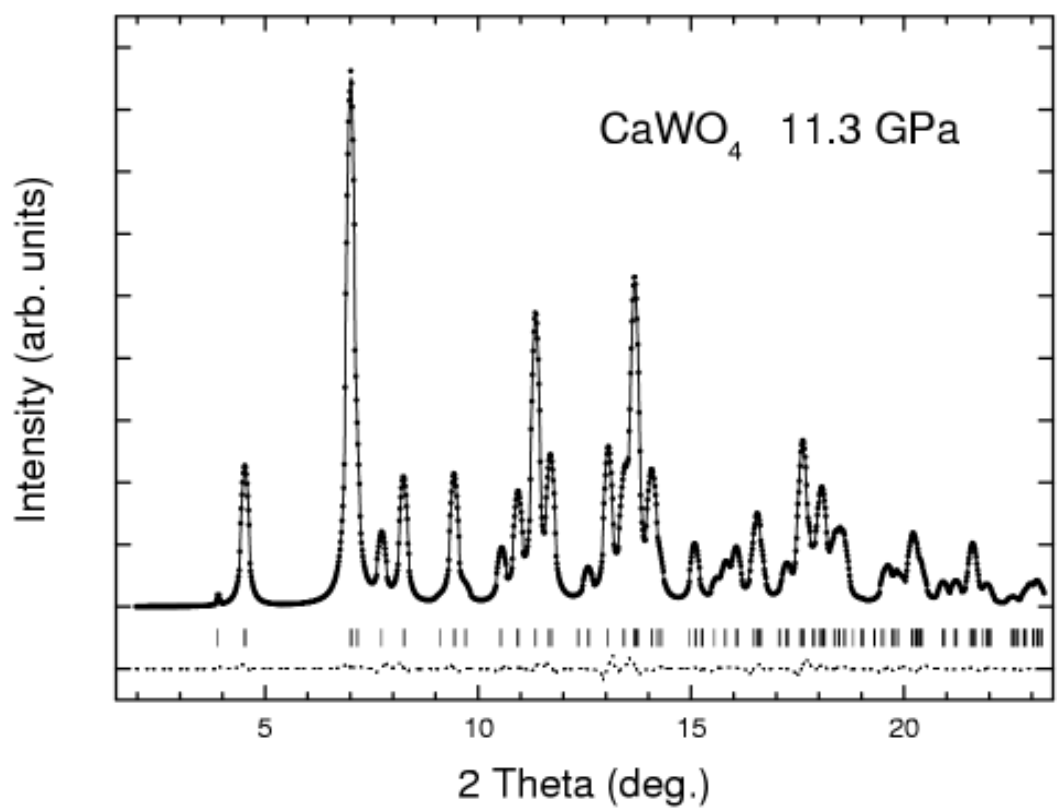

**Figure 9**

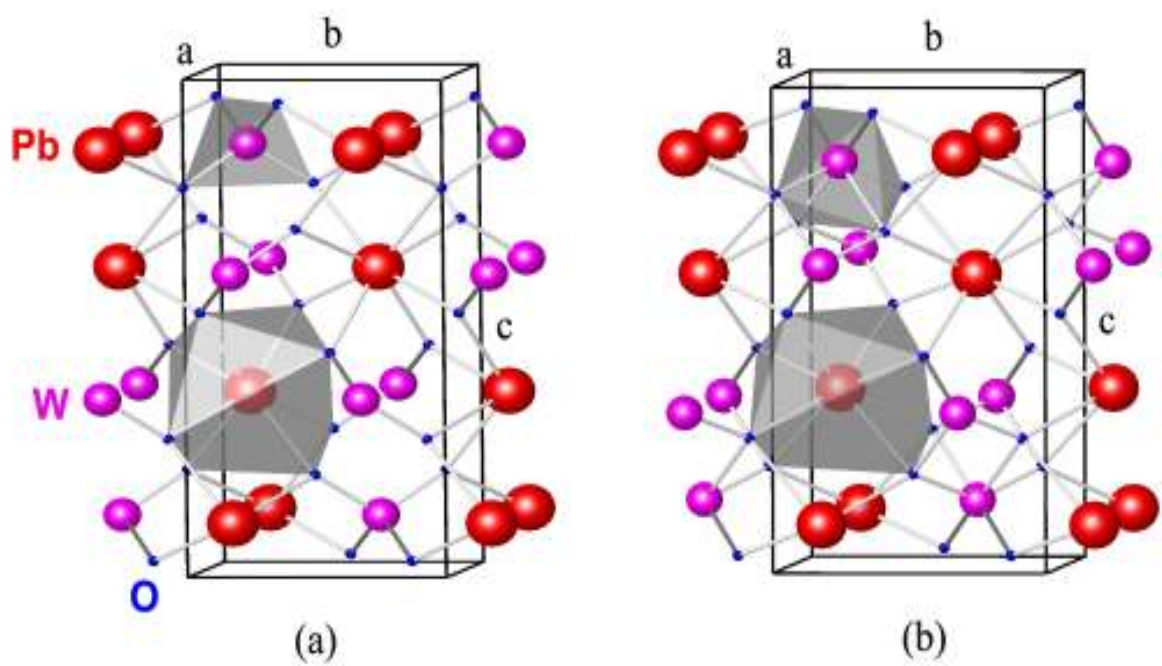

**Figure 10**

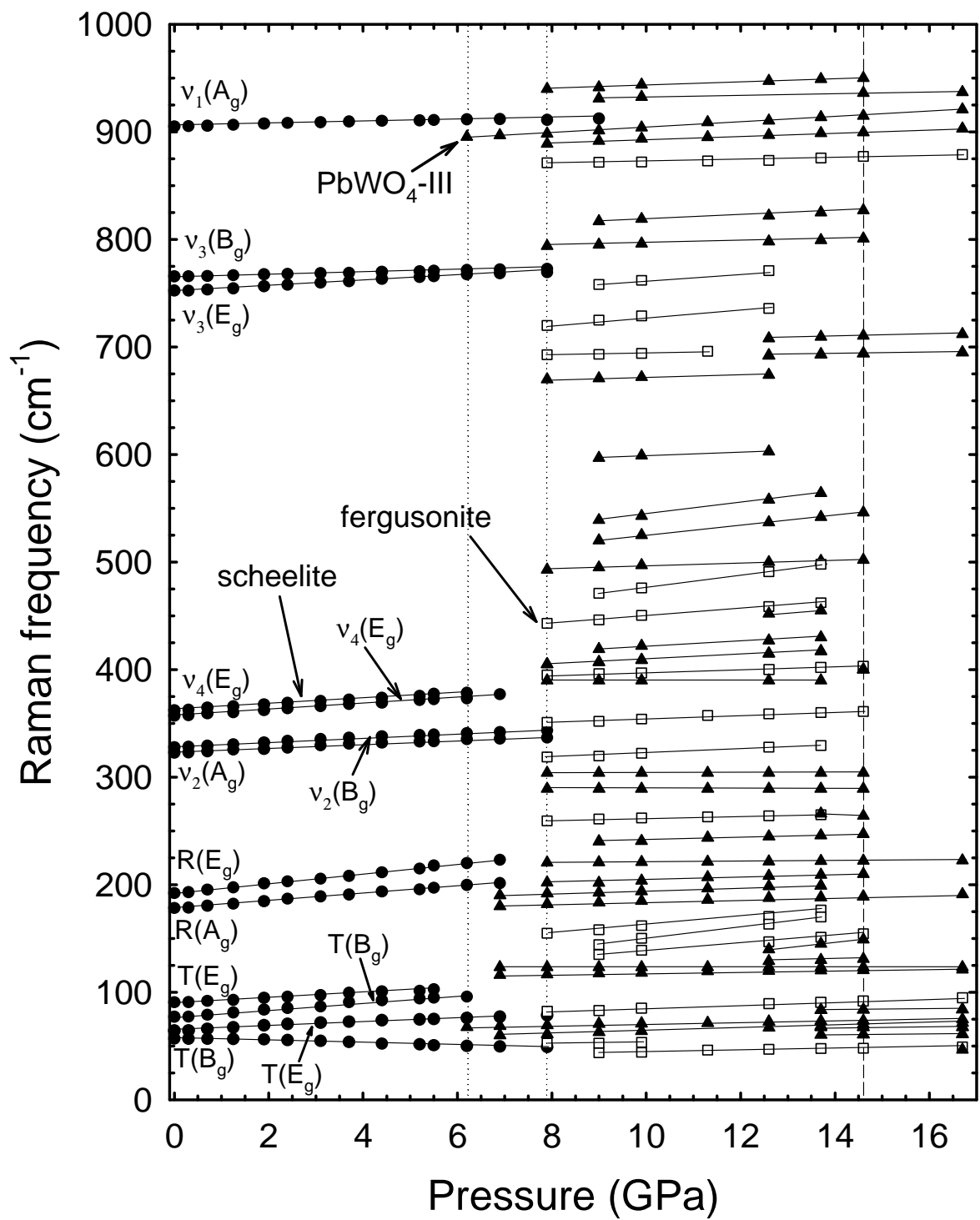

**Figure 11**

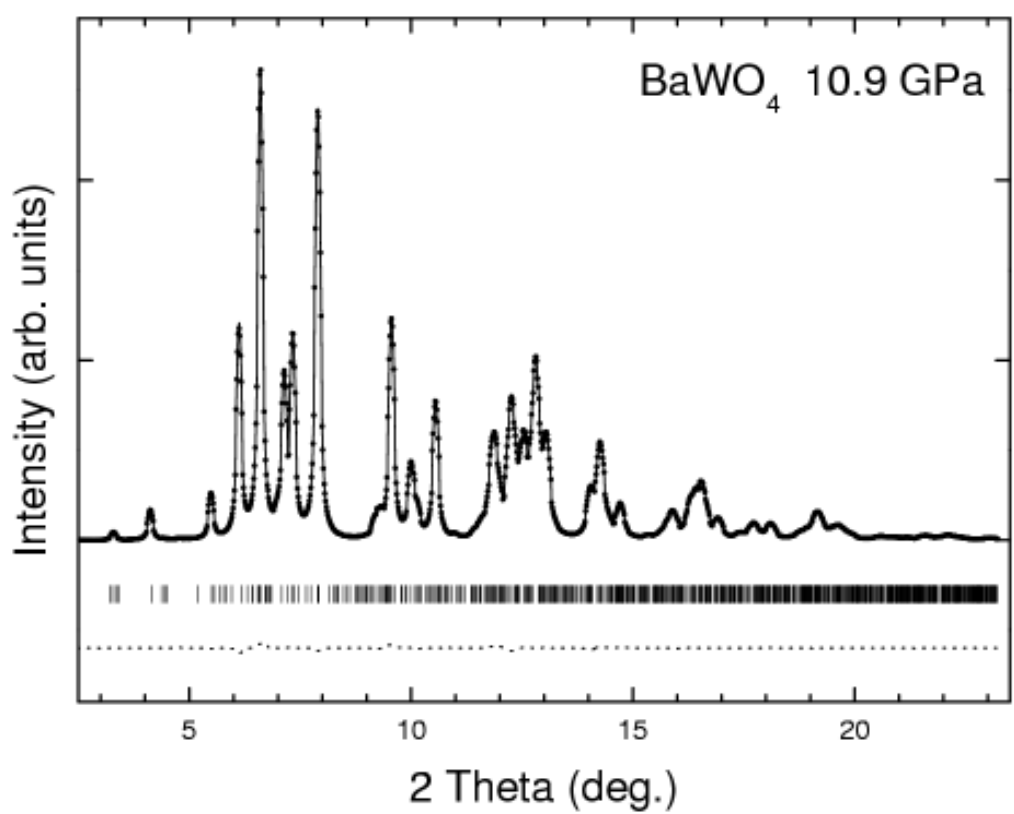

**Figure 12**

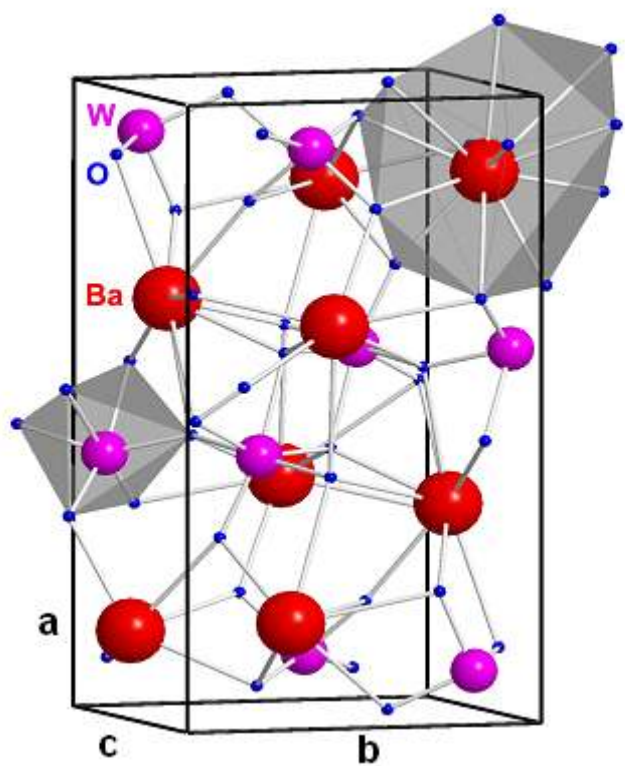

**Figure 13**

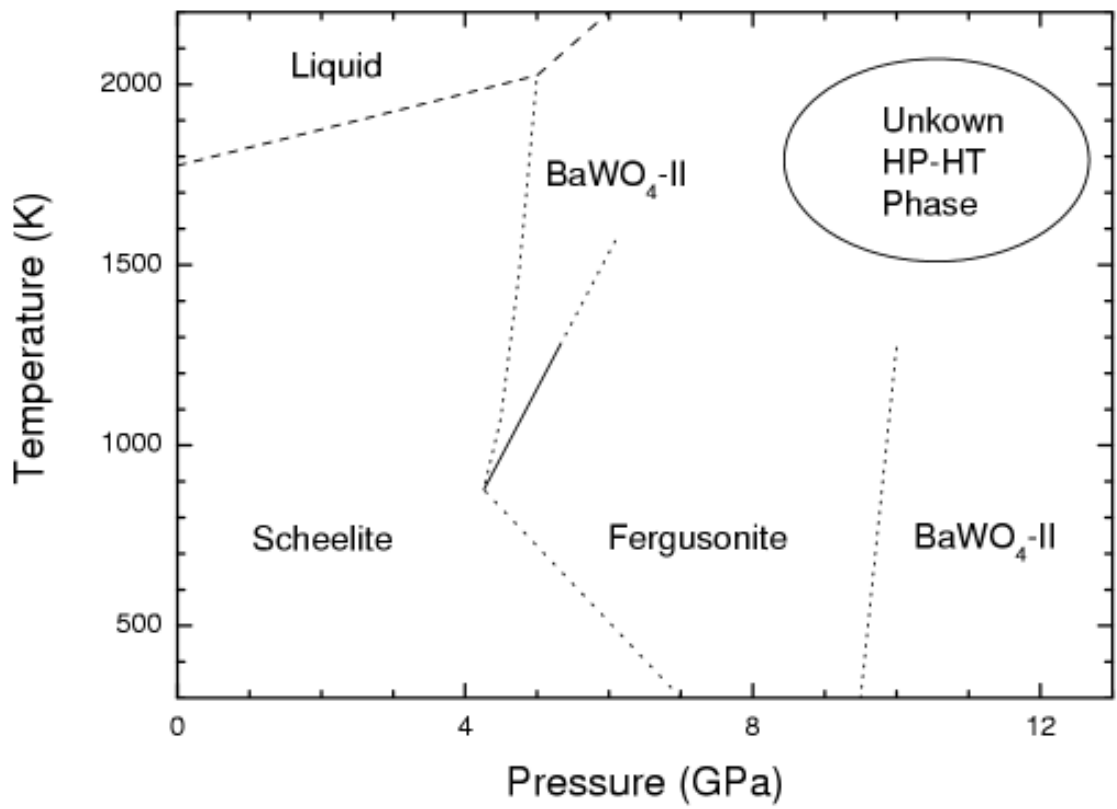

**Figure 14**

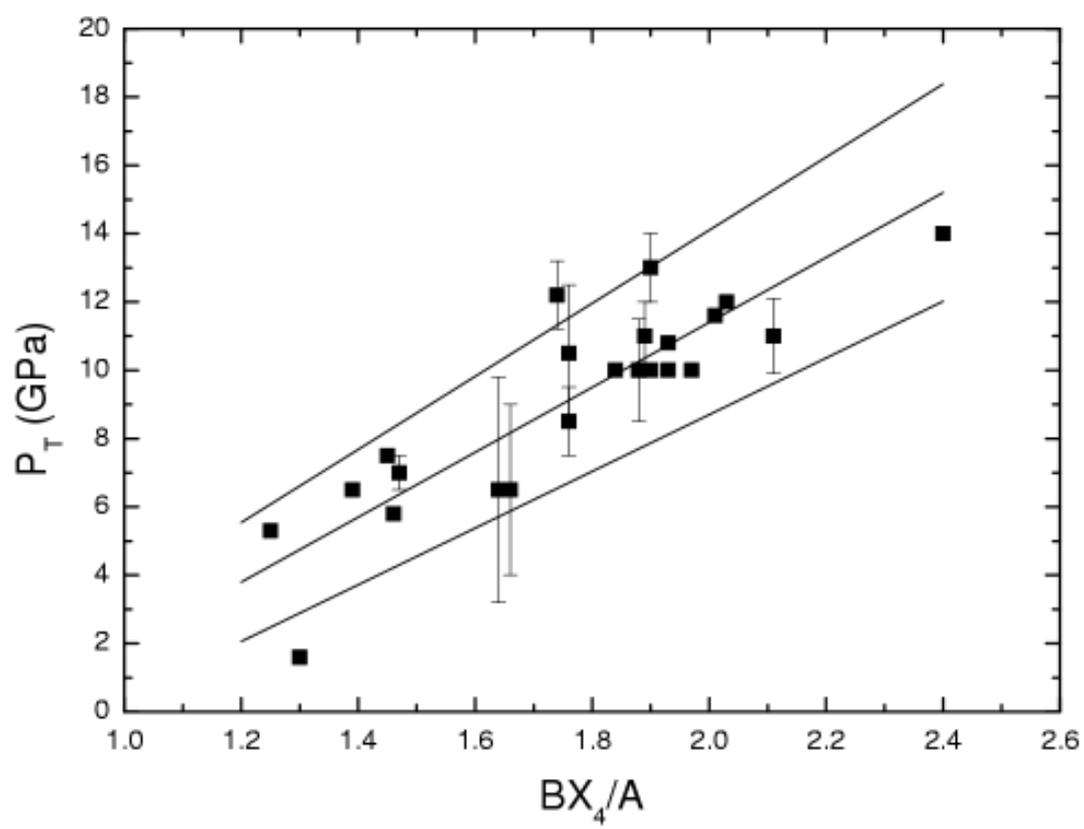

**Figure 15**

**Figure 16**

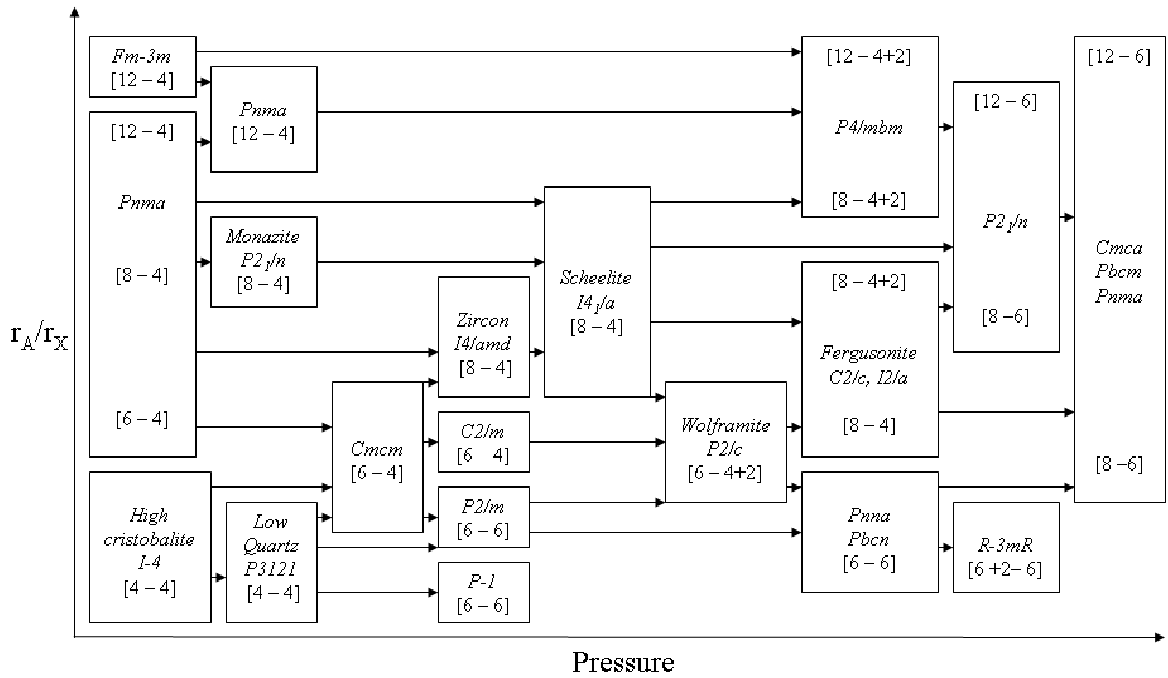

**Figure 17**

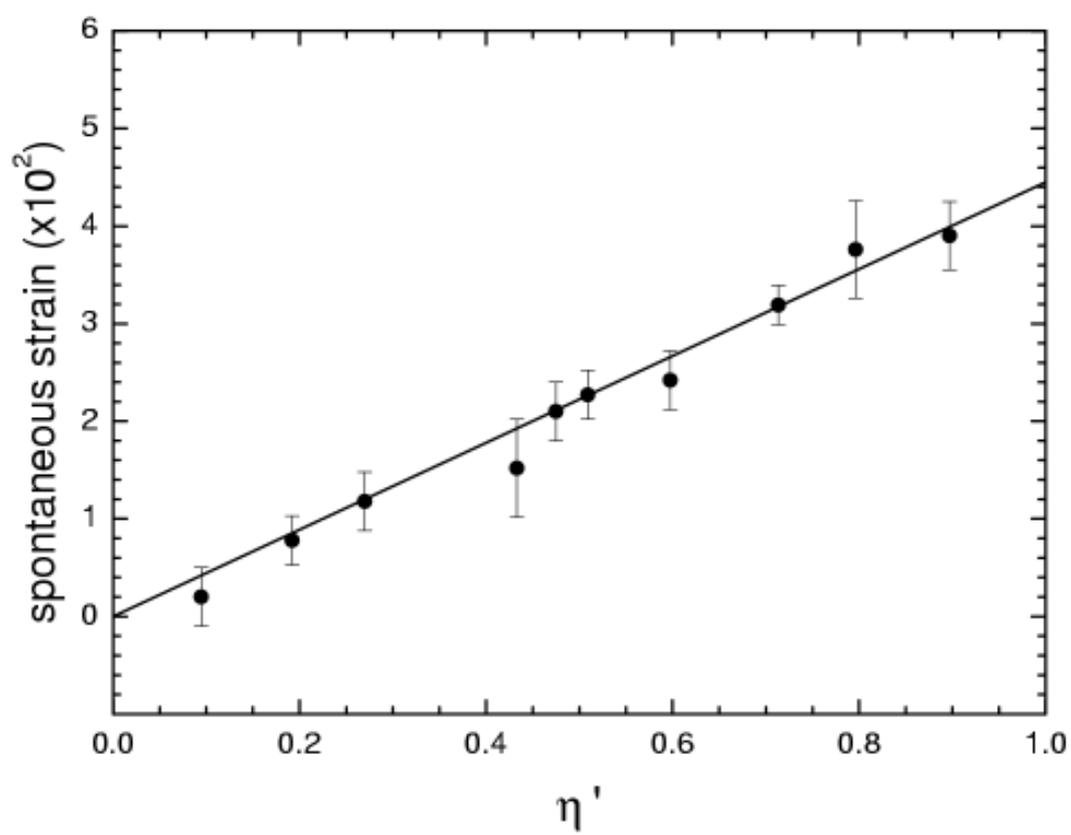

**Figure 18**

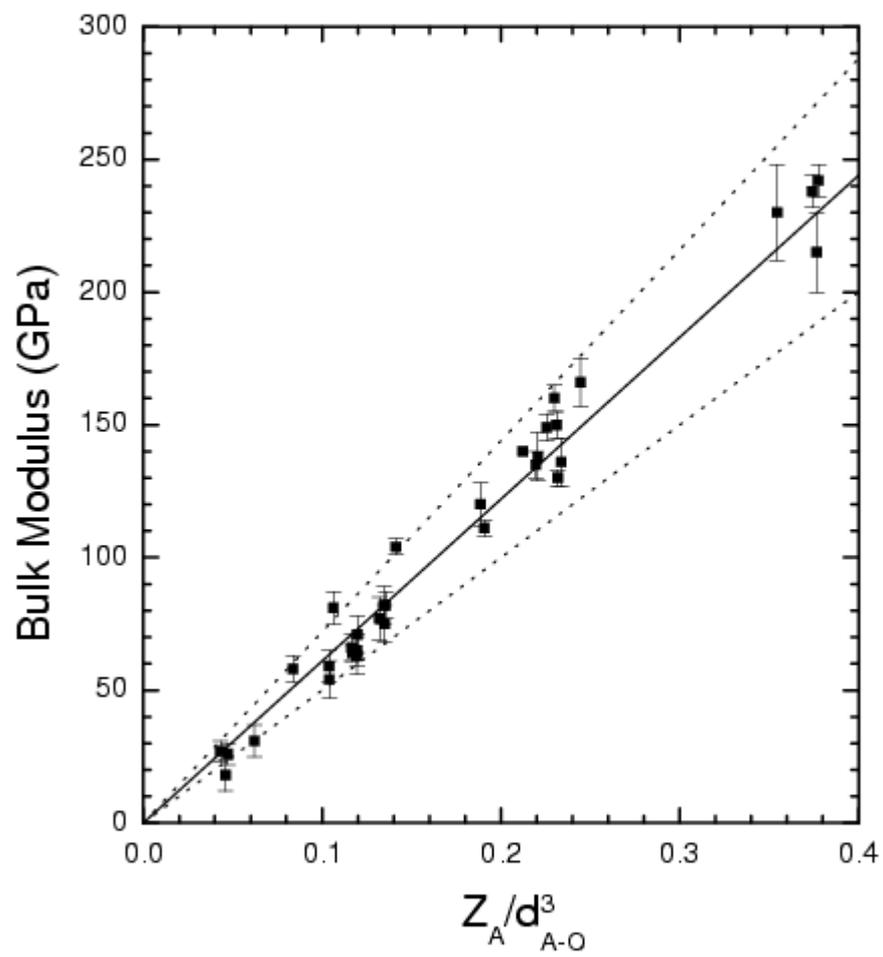

**Figure 19**

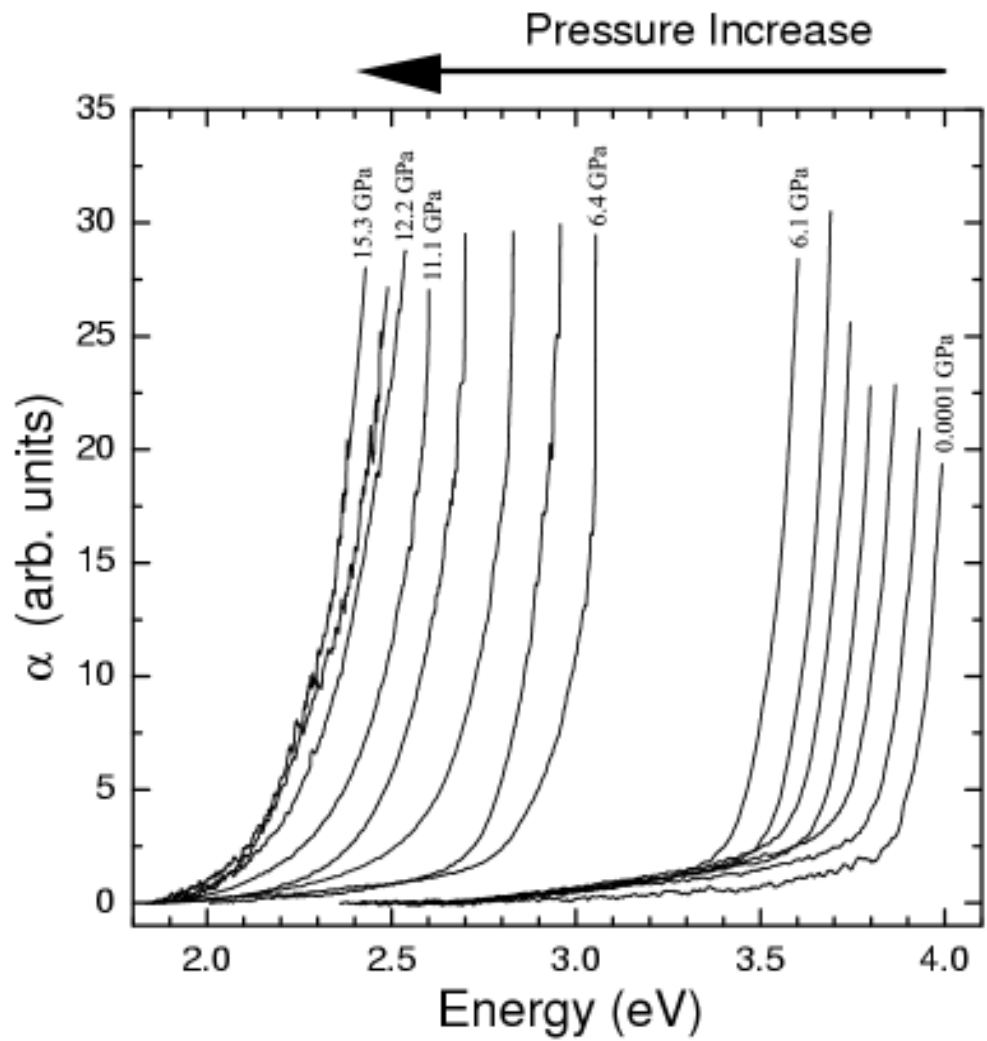

**Figure 20**

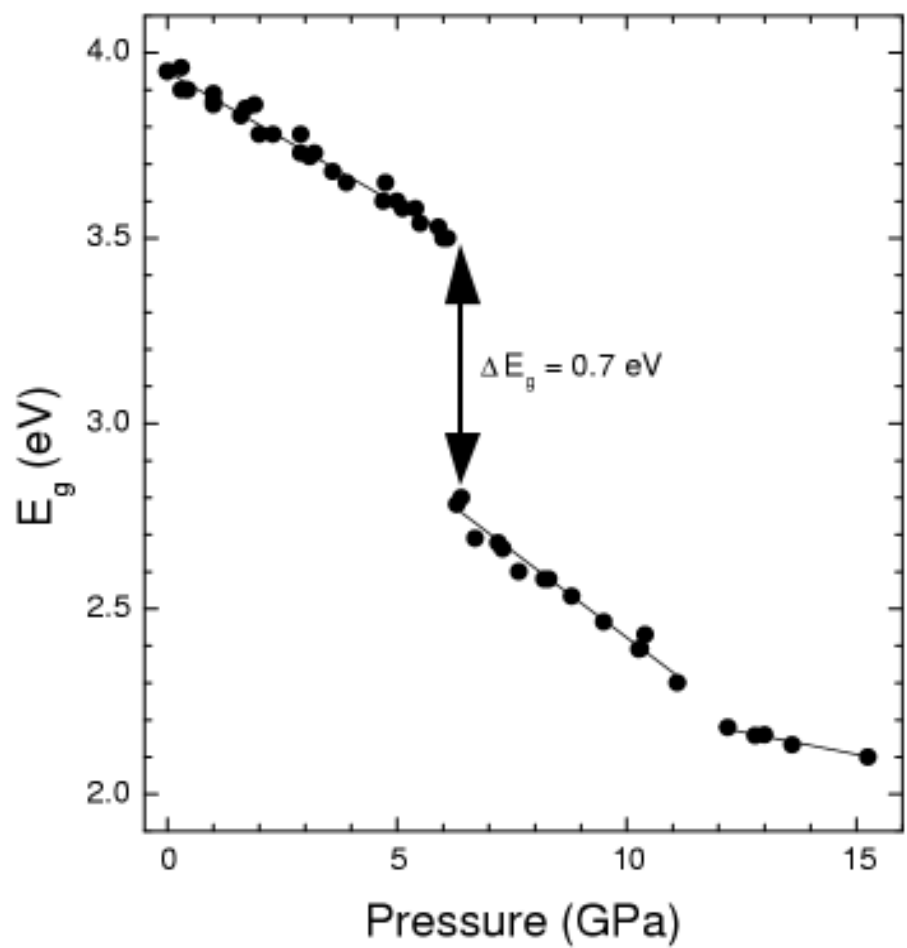